\documentclass[ reprint, floatfix, superscriptaddress, amsmath, amssymb, longbibliography ]{revtex4-1}
 \pdfoutput=1

\usepackage{graphicx}
\usepackage{amsmath}
\usepackage{appendix}
\usepackage{xcolor}
\usepackage{tikz}
\usepackage{tikzit}
\usetikzlibrary{quantikz}
\usepackage{braket}
\usepackage{subfig}
\usepackage{lipsum} 
\usepackage{floatrow}
\usepackage{hyperref}
\hypersetup{ 
    colorlinks=true,
    linkcolor=blue,
    filecolor=blue,      
    urlcolor=blue,
}
\usepackage{graphicx}

\usepackage{caption}
\newcommand{\sx}[1]{\textcolor{black}{#1}}

\input{zx.tikzdefs}

\tikzstyle{box}=[shape=rectangle, text height=1.5ex, text depth=0.25ex, yshift=0.5mm, fill=white, draw=black, minimum height=5mm, yshift=-0.5mm, minimum width=5mm, font={\small}]
\tikzstyle{Z dot}=[inner sep=0mm, minimum size=2mm, shape=circle, draw=black, fill={rgb,255: red,216; green,248; blue,216}, tikzit fill={rgb,255: red,216; green,248; blue,216}]
\tikzstyle{Z phase dot}=[minimum size=4.75mm, font={\footnotesize}, shape=rectangle, rounded corners=1.9mm, inner sep=0.1mm, outer sep=-2mm, scale=0.8, tikzit shape=circle, draw=black, fill={rgb,255: red,216; green,248; blue,216}, tikzit draw=blue, tikzit fill={rgb,255: red,216; green,248; blue,216}]
\tikzstyle{X dot}=[Z dot, shape=circle, draw=black, fill={rgb,255: red,232; green,165; blue,165}, tikzit fill={rgb,255: red,232; green,165; blue,165}]
\tikzstyle{X phase dot}=[Z phase dot, tikzit shape=circle, tikzit fill={rgb,255: red,232; green,165; blue,165}, fill={rgb,255: red,232; green,165; blue,165}, font={\footnotesize}, tikzit draw=blue]
\tikzstyle{XD dot}=[shape=XDdot, inner sep=2pt, draw=black]
\tikzstyle{XD phase dot}=[shape=XDdotphase, minimum size=4.75mm, font={\footnotesize}, inner sep=.1mm, outer sep=0mm, scale=0.8, tikzit shape=circle, rounded corners=1.9mm, draw=black]
\tikzstyle{zn}=[shape=zn, tikzit draw=black, draw=black, inner sep=2pt]
\tikzstyle{hadamard}=[fill={rgb,255: red,140; green,220; blue,248}, draw=black, shape=rectangle, inner sep=0.6mm, minimum height=1.5mm, minimum width=1.5mm, xslant=0.5]
\tikzstyle{hz}=[hadamard, fill={rgb,255: red,216; green,248; blue,216}, shape=rectangle, tikzit fill={rgb,255: red,216; green,248; blue,216}, minimum height=2 mm, minimum width=1.25 mm, tikzit draw=black]
\tikzstyle{hx}=[hadamard, fill={rgb,255: red,232; green,165; blue,165}, shape=rectangle, tikzit fill={rgb,255: red,232; green,165; blue,165}, minimum height=2 mm, minimum width=1.25 mm, tikzit draw=black]
\tikzstyle{vertex}=[inner sep=0mm, minimum size=1mm, shape=circle, draw=black, fill=black]
\tikzstyle{vertex set}=[inner sep=0mm, minimum size=1mm, shape=circle, draw=black, fill=white, font={\footnotesize\boldmath}]
\tikzstyle{meter}=[draw, fill=white, minimum width=2em, minimum height=1.5em, rectangle, path picture={\draw ([shift={(.1,.24)}]path picture bounding box.south west) to[bend left=50] ([shift={(-.1,.24)}]path picture bounding box.south east);\draw[-{Latex[scale=0.6]}] ([shift={(0,.1)}]path picture bounding box.south) -- ([shift={(.3,-.1)}]path picture bounding box.north);}, tikzit shape=rectangle]
\tikzstyle{white dot}=[Z dot]
\tikzstyle{gray dot}=[X dot]
\tikzstyle{white phase dot}=[Z phase dot]
\tikzstyle{gray phase dot}=[X phase dot]
\tikzstyle{red ket}=[fill={rgb,255: red,232; green,165; blue,165}, draw=black, shape=isosceles triangle, tikzit fill={rgb,255: red,232; green,165; blue,165}, tikzit draw=black, inner sep=0 mm, outer sep=2 mm]
\tikzstyle{tiny none}=[none, font={\tiny}]
\tikzstyle{green ket}=[fill={rgb,255: red,216; green,248; blue,216}, draw=black, shape=isosceles triangle, tikzit fill={rgb,255: red,216; green,248; blue,216}, tikzit draw=black, inner sep=0 mm, outer sep=2 mm]
\tikzstyle{filament}=[hadamard, fill=yellow, draw=none, minimum height=0.01mm]
\tikzstyle{unit circle}=[shape=circle, minimum size=42.5 mm, fill=none, draw={rgb,255: red,223; green,223; blue,223}, tikzit draw={rgb,255: red,223; green,223; blue,223}]
\tikzstyle{new style 0}=[shape=ellipse, minimum height=10 mm, minimum width=100 mm, fill=white, draw=black]
\tikzstyle{gate}=[box, minimum height=10mm, minimum width=10mm]
\tikzstyle{2 control}=[vertex set, draw=blue, inner sep=0.5pt]
\tikzstyle{1 control}=[2 control, draw=red]
\tikzstyle{0 control}=[2 control, draw=black]
\tikzstyle{small dot}=[vertex, minimum size=1 mm, draw=black, tikzit draw=black, tikzit fill=black, tikzit shape=circle]
\tikzstyle{tallbox}=[box, minimum height=12mm]
\tikzstyle{targ}=[vertex set, minimum size=0.5mm, inner sep=-0.5mm, tikzit shape=circle, shape=circle, tikzit draw=black]
\tikzstyle{hadamardbox}=[hadamard, xslant=0]

\tikzstyle{directedarrow}=[draw={rgb,255: red,223; green,223; blue,223}, ->, tikzit draw={rgb,255: red,223; green,223; blue,223}, line width=1 pt]
\tikzstyle{simple}=[-]
\tikzstyle{hadamard edge}=[-, color={rgb,255: red,0; green,100; blue,248}, dashed, dash pattern=on 2pt off 0.7pt, tikzit draw={rgb,255: red,0; green,100; blue,248}]
\tikzstyle{brace edge}=[-, tikzit draw=blue, decorate, decoration={brace,amplitude=1mm,raise=-1mm}]
\tikzstyle{gray}=[-, draw={rgb,255: red,223; green,223; blue,223}, line width=1 pt]
\tikzstyle{arrow}=[<-, draw={rgb,255: red,128; green,128; blue,128}]
\tikzstyle{double-arrow}=[draw={rgb,255: red,128; green,128; blue,128}, <->]
\tikzstyle{dashed edge}=[-, dashed, dash pattern=on 2pt off 0.5pt, draw=black]
\tikzstyle{diredge}=[->]
\tikzstyle{double edge}=[-, double, shorten <=-1mm, shorten >=-1mm, double distance=2pt]
\tikzstyle{thin}=[-, line width=0.05mm]
\tikzstyle{thin gray}=[-, draw={rgb,255: red,223; green,223; blue,223}, line width=0.05mm]
\tikzstyle{less thin}=[-, line width=0.1mm]
\tikzstyle{dashed gray edge}=[-, dashed edge, draw={rgb,255: red,128; green,128; blue,128}]
\tikzstyle{light right directed arrow}=[->, directedarrow, draw={rgb,255: red,223; green,223; blue,223}, line width=0.2mm]
\tikzstyle{diredge0.3}=[->, line width=0.3 mm]
\tikzstyle{less thin gray}=[-, draw={rgb,255: red,223; green,223; blue,223}]
\tikzstyle{dashed thin purple}=[-, dashed, line width=0.1mm, draw={rgb,255: red,128; green,106; blue,219}]
\tikzstyle{hadamardedge}=[-, color={rgb,255: red,100; green,200; blue,248}, dashed, dash pattern=on 2pt off 0.7pt, tikzit draw={rgb,255: red,120; green,220; blue,248}]

\begin{document}

\title{Emulating two qubits with a four-level transmon qudit \\ for variational quantum algorithms}

\author{Shuxiang Cao}
 \email{shuxiang.cao@physics.ox.ac.uk} 
 \affiliation{Department of Physics, Clarendon Laboratory, University of Oxford, OX1 3PU, UK}
 \author{Mustafa Bakr}
 \affiliation{Department of Physics, Clarendon Laboratory, University of Oxford, OX1 3PU, UK}
 \author{Giulio Campanaro}
 \affiliation{Department of Physics, Clarendon Laboratory, University of Oxford, OX1 3PU, UK}
 \author{Simone D. Fasciati}
 \affiliation{Department of Physics, Clarendon Laboratory, University of Oxford, OX1 3PU, UK}
 \author{James Wills}
 \thanks{Present address: Oxford Quantum Circuits. Reading  RG2 9LH, UK}
 \affiliation{Department of Physics, Clarendon Laboratory, University of Oxford, OX1 3PU, UK}
\author{Deep Lall}
\affiliation{National Physical Laboratory, Teddington London, TW11 0LW, UK}
\affiliation{Department of Materials, University of Oxford, Parks Road, Oxford, OX1 3PH, UK}
 \author{Boris Shteynas}
 \thanks{Present address: Oxford Quantum Circuits. Reading RG2 9LH, UK}
 \affiliation{Department of Physics, Clarendon Laboratory, University of Oxford, OX1 3PU, UK}
 \author{Vivek Chidambaram}
 \affiliation{Department of Physics, Clarendon Laboratory, University of Oxford, OX1 3PU, UK}
\author{Ivan Rungger}
\affiliation{National Physical Laboratory, Teddington London, TW11 0LW, UK}
\author{Peter Leek}
 \email{peter.leek@physics.ox.ac.uk} 
 \affiliation{Department of Physics, Clarendon Laboratory, University of Oxford, OX1 3PU, UK}

\date{\today}

\begin{abstract}

Using quantum systems with more than two levels, or qudits, can scale the computational space of quantum processors more efficiently than using qubits, which may offer an easier physical implementation for larger Hilbert spaces. However, individual qudits may exhibit larger noise, and algorithms designed for qubits require to be recompiled to qudit algorithms for execution. In this work, we implemented a two-qubit emulator using a 4-level superconducting transmon qudit for variational quantum algorithm applications and analyzed its noise model. The major source of error for the variational algorithm was readout misclassification error and amplitude damping. To improve the accuracy of the results, we applied error-mitigation techniques to reduce the effects of the misclassification and qudit decay event. The final predicted energy value is within the range of chemical accuracy. 
\end{abstract}

\maketitle

\section{Introduction}





%

Quantum computing is widely considered a promising computing model due to its exponentially large Hilbert space~\cite{Nielsen2000QuantumInformation}. Most current quantum processors use two-level quantum systems, or qubits, as the basic building block for their ease of implementation, inheriting from the successful experience with binary classical computers. For a qubit processor made of $N$ elementary building blocks, the size of the Hilbert space is $2^N$. Current research focuses on scaling up the number of qubits $N$, but this approach faces challenges such as wiring \cite{TAMATE2022TowardComputers}, expensive control electronics \cite{Stefanazzi2022TheDetectors}, and frequency crowding \cite{Theis2016SimultaneousShapes,Schutjens2013Single-qubitSystems}. A complementary strategy is to increase the size of each building block by replacing the qubit with a qudit (d-level system),  which yields a Hilbert space of size $d^N$. In current quantum processors, each building block requires a separate set of control electronics to operate, resulting in complex and costly systems that scale linearly with the number of building blocks. However, by using qudit-based quantum processors, the number of required building blocks can be significantly reduced while maintaining the same size of Hilbert space. This also reduces the number of required control electronics, thereby simplifying the system and reducing financial costs. As a result, qudit-based approaches are a more practical solution for scaling up quantum processors in the near term.  


The experimental use of qudits as the fundamental building blocks for quantum processors has been investigated in various systems, including photonic systems \cite{Slussarenko2019PhotonicReview,Fickler2012QuantumMomenta,Malik2016Multi-photonDimensions,Wang2018MultidimensionalOptics,Lu2020QuantumPhoton,Kues2017On-chipControl}, ion traps \cite{Bruzewicz2019Trapped-ionChallenges,Ringbauer2022AIons} and nuclear magnetic resonance \cite{Vandersypen2004NMRComputation,Jones2011QuantumNMR,Dogra2014DeterminingQutrit,Dogra2015ImplementationEmulator,Dogra2016ExperimentalQutrit,Gedik2015ComputationalQudit}. Superconducting circuits, which are among the leading platforms for quantum computing, have also been explored for this purpose~\cite{Krantz2019AQubits, Wallquist2009CircuitQubits, PhysRevApplied.14.014072}. The transmon is the most widely used building block in superconducting quantum processors~\cite{Koch2007Charge-insensitiveBox}, and it has more than two energy levels that can potentially be utilized for computation. However, these higher levels are more susceptible to charge noise and spontaneous decay, which presents a significant challenge~\cite{Peterer2015CoherenceQubit, Bianchetti2010ControlAtom, Dong2022SimulationQudit,PhysRevA.79.064305}. In recent years, transmons have been used as qutrits (three-level systems) in several studies, including the implementation of qutrit-based quantum algorithms \cite{2211.06523}, simulation of topological Maxwell metal bands~\cite{Tan2018TopologicalQutrit}, the dynamics of quantum information in strongly interacting systems~\cite{Blok2021QuantumProcessor}, and Tensor Monopoles~\cite{Tan2021ExperimentalQudit}. While these works used transmons with up to three levels of computational space, the results already demonstrate the potential of using transmons qudits as the fundamental building blocks for quantum processors. Previous work has also explored the use of higher levels of transmons beyond qutrits, and demonstrated quantum gates on 4-level transmons \cite{Dong2021, PhysRevLett.125.170502, Zheng2022,2304.11159}. 
Other applications on 4-level transmons have also been showcased, including enhancement of qubit readout fidelity~\cite{PhysRevX.10.011001} and encoding information for autonomous error correction \cite{2302.06707}. 

The implementation of qudit-based quantum processors to execute quantum algorithms has two main challenges. The first challenge is that using qudits as building blocks requires a higher system quality, which is more difficult to fabricate and control \cite{Cozzolino2019High-DimensionalChallenges,PhysRevX.9.041042,PhysRevA.79.064305,PhysRevA.104.052416}. For example, transmons have lower lifetimes at higher energy levels and exhibit larger charge dispersion. Implementing the readout of qudit typically presents more challenges due to the need to discriminate between more than two states from a single readout pulse. Therefore, a more careful design of the processor and signal processing is required to enable qudit readout. Another challenge for qudit-based quantum processors is that algorithms designed for qubit-based systems must be modified and adapted to fully exploit the increased Hilbert space provided by qudits. The algorithms must be compiled using a qudit gate set, which differs from the qubit gate set~\cite{Wang2020QuditsComputing, Luo2014UniversalQudits, Brylinski2001UniversalGates,Di2011ElementaryCircuit}. Some well-known quantum algorithms, such as the Deutsch-Jozsa algorithm \cite{Fan2007ALogic,Mogos2007TheN-Qudits}, Bernstein-Vazirani algorithm \cite{Krishna2016ASystems}, Grover’s algorithm~\cite{Ivanov2012Time-efficientQudits}, Quantum Fourier Transform~\cite{Cao2011QuantumSystem, Pavlidis2021Quantum-Fourier-transform-basedQudits}, and Shor's algorithm \cite{Bocharov2017FactoringArchitectures}, have direct generalizations of the qubit counterparts, which maintain the same principles but change the positional notation from the base-2 numeral system to the base-$d$ numeral system. However, it is challenging to directly generalize an arbitrary quantum algorithm into a qudit version simply by changing the positional notation. Certain near-term applications are specifically designed to run on qubit-based quantum processors~\cite{Cerezo2021VariationalAlgorithms, Tilly2021ThePractices}. For example, variational quantum algorithms for chemistry applications decompose the molecule Hamiltonian into a sum of tensor products of qubit Pauli operators. Generalizing these variational algorithms to work with qudits requires an innovative encoding mechanism. \sx{Upon submission, we noticed a parallel work demonstrates implementing variational algorithms that are modified for qudits \cite{PhysRevX.13.021028}. }

The goal of this work is to address these challenges and provide solutions for the development of future qudit-based quantum processors. To tackle the challenge of qudit readout, we demonstrate a high-coherence transmon device capable of utilizing its lowest four levels for information processing and enabling single-shot readout. For the challenge of developing qudit algorithms, we implement emulations of qubit systems using qudits. When the Hilbert space of $N_d$ qudit is equivalent to that of an $N_b$-qubit system ($2^{N_b} = d^{N_d}$), the qudit can be used to emulate the $N$-qubit system. For example, a qudit with $d=4$ can emulate a two-qubit system, as proposed in \cite{Kiktenko2015MultilevelInequalities, Dong2021, PhysRevLett.125.170502}. By using this emulator, qubit algorithms can be executed directly on qudit-based processors without modification. These solutions bring us closer to developing practical and scalable qudit-based quantum processors.


In this paper, we present an implementation of a two-qubit emulator on a high-coherence transmon. Our transmon is designed to have the optimal operating parameters as a ququart (qudit with $d=4$), and it is tuned to have single-shot readout capabilities that can distinguish the lowest four levels. To create the emulator, we compiled two-qubit operations into sequences of 4-level intrinsic operations that include only neighbouring transitions and virtual-Z gates. We characterized the performance of both the intrinsic and emulated gates using randomized benchmarking and gate-set tomography. We also implemented active resets on the qudit to prepare a high-fidelity initial state. We demonstrated the efficacy of a two-qubit variational quantum eigensolver algorithm on the emulator. We also analyzed the primary sources of error and implemented error mitigation techniques to obtain more accurate results.

\section{Implementation of the emulator}

\begin{figure}[!t]
    \centering
    \includegraphics[width=\linewidth]{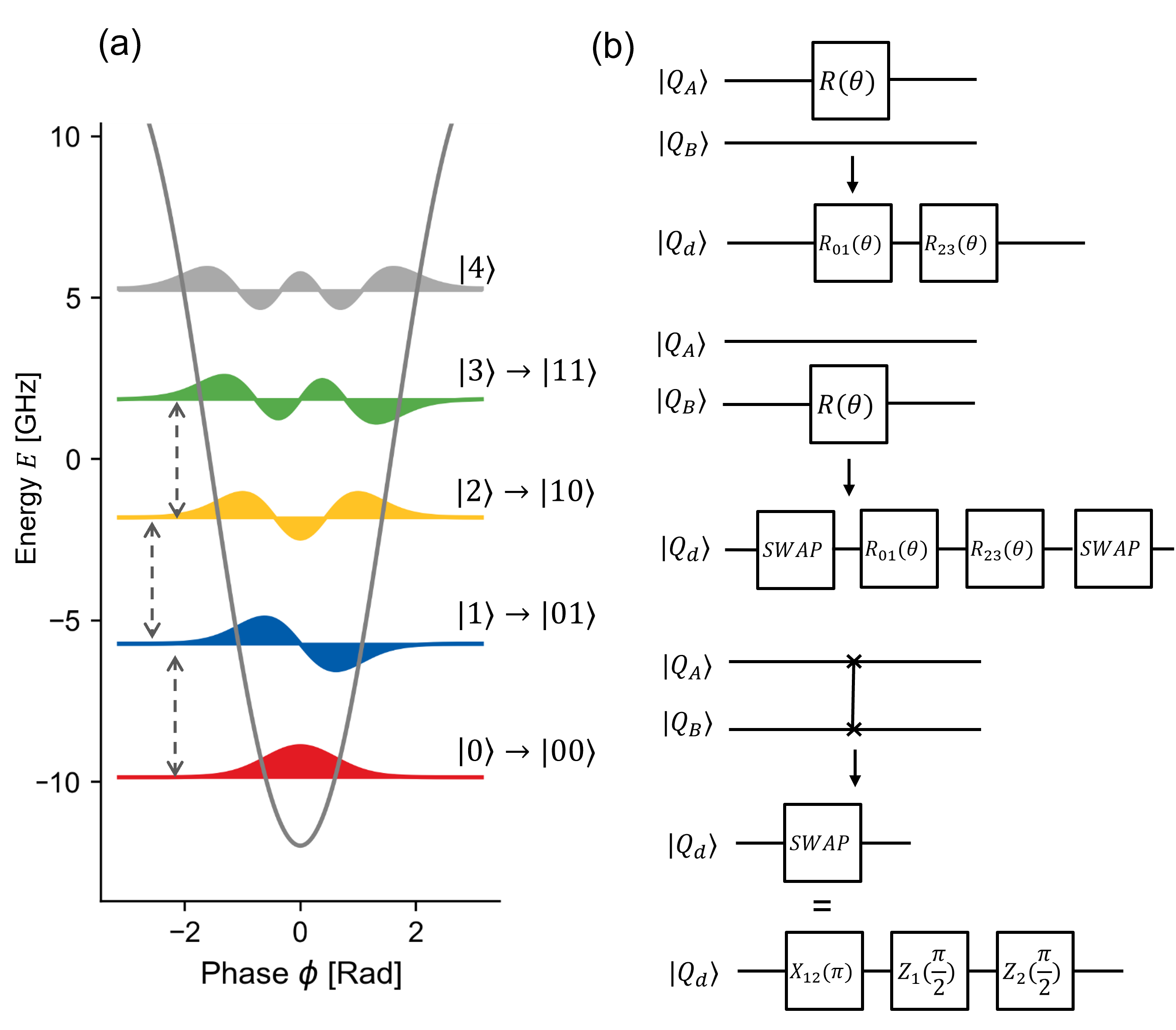}
    \caption{(a) The lowest four levels of the transmon are used as a qudit for quantum information processing. (b) The two-qubit operations are mapped to sequences of single qudit operations\sx{.} \sx{$R(\theta)$ denotes an arbitrary single-qubit rotation operation, and $R_{ij}$ denotes the equivalent single-qubit operation in the $\{\ket{i},\ket{j}\}$ subspace. $X_{ij}$ and $Y_{ij}$ and $Z_i$ are analogs of the single qubit $X, Y, Z$ rotation in the $SU(4)$ space, please refer to Appendix A for more details.}}
    \label{fig:emulator}
\end{figure}

\begin{figure}
	\sidesubfloat[]{
		\includegraphics[width=.97\linewidth]{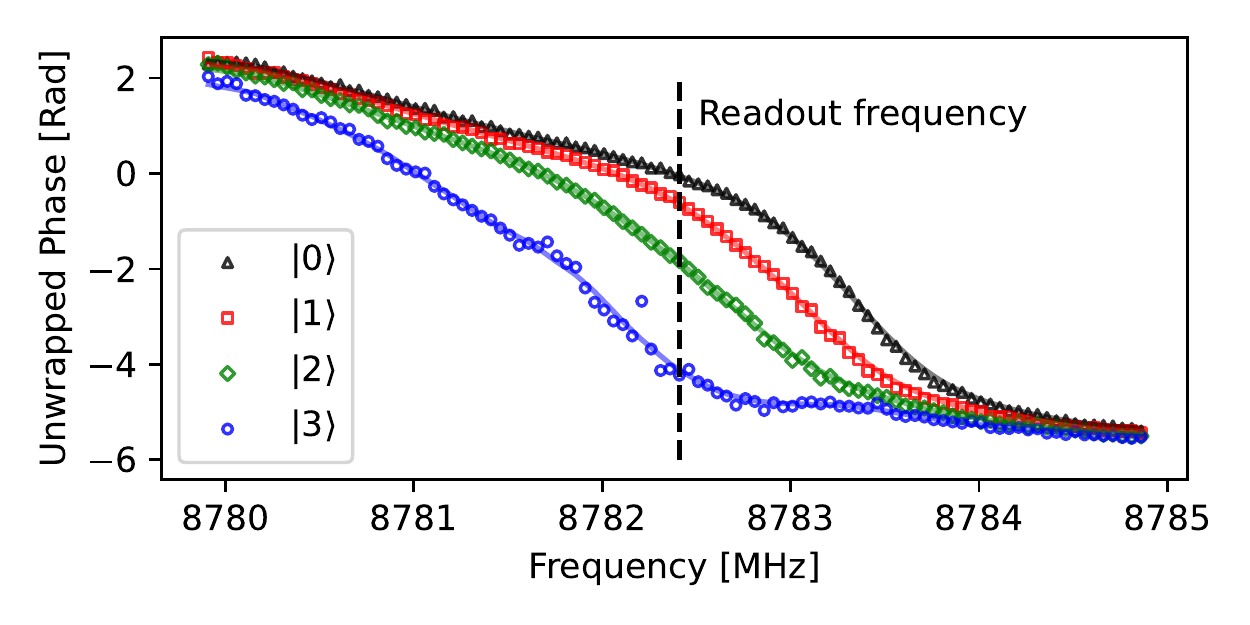}
	}\\
	\sidesubfloat[]{
		\includegraphics[width=.95\linewidth]{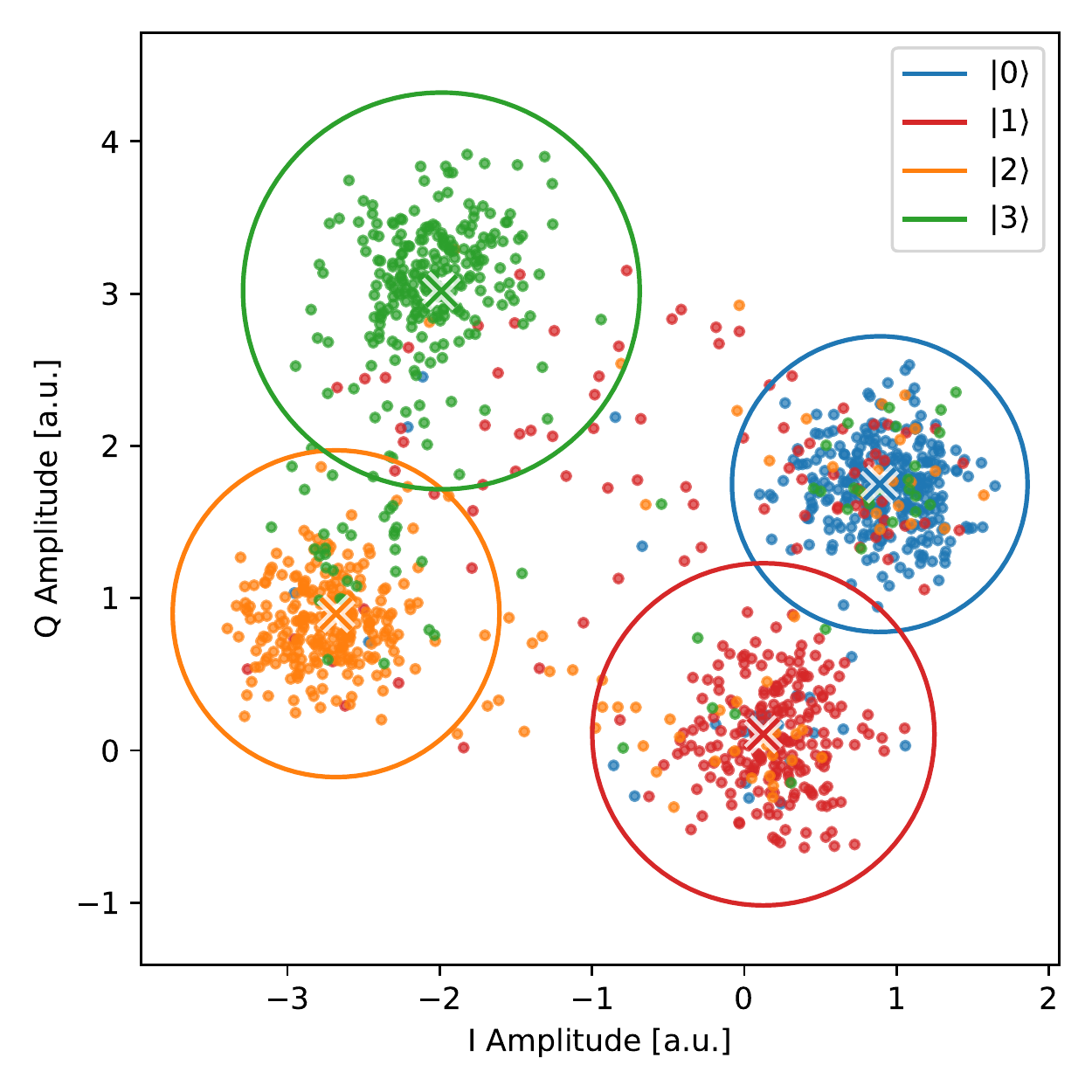}
	}
    \caption{(a) Phase response of the readout resonator while the transmon is excited to the $\ket{0}$, $\ket{1}$, $\ket{2}$, and $\ket{3}$ states, respectively. (b) IQ plane of the single-shot readout. The states are prepared in the $\ket{0}$, $\ket{1}$, $\ket{2}$, and $\ket{3}$ states, respectively, denoted by the color. The four different levels were classified by a spherical Gaussian mixture model, where the cross denotes the centre of the Gaussian distribution, and the radius of the circle denotes three times the standard deviation. \sx{See appendix \ref{app:spherical_gmm} for more details.}}
    \label{fig:readout}
\end{figure}

The two-qubit emulator is built on a single transmon with a coaxial geometry and off-chip wiring \cite{Spring2022,Rahamim2017Double-sided}, where the lowest 4 quantum states of the transmon are used as the computational space. The same device has also been previously used as a qutrit \cite{2210.04857}. The two-qubit emulator maps the physical states $\ket{0}$, $\ket{1}$, $\ket{2}$, $\ket{3}$ of the transmon to the $\ket{00}$, $\ket{01}$, $\ket{10}$, $\ket{11}$ states of a virtual two-qubit device, see Figure \ref{fig:emulator}(a).  Single qubit gates for the first virtual qubit $Q_A$ are performed by driving transitions in the $\{\ket{0},\ket{1}\}$ and $\{\ket{2},\ket{3}\}$ subspaces in sequence. The single qubit operation on $Q_B$ can be implemented with a similar approach as for $Q_A$ by driving the two-photon transitions in the $\{\ket{0},\ket{2}\}$ and $\{\ket{1},\ket{3}\}$ subspace. However, it requires higher input power to drive the two-photon transition and \sx{tracking} more parameters to implement the virtual Z gate. \sx{For simplicity, the gates are implemented by first performing an emulated SWAP gate, followed by applying the virtual qubit operations on $Q_A$, and then apply another SWAP gate, as shown in Figure \ref{fig:emulator}(b). The emulated SWAP gate is implemented by performing qubit-like $X$ gate in the ${\ket{1},\ket{2}}$ subspace followed by two virtual Z gates to correct the phases. The iSWAP gate can be implemented with a similar scheme by performing qubit-like $Y$ gate.}  Two-qubit gates are decomposed into single qubit gates and $ZZ$ interactions, which are implemented using virtual $Z$ gates. See the supplementary materials 
for further discussion of the qudit virtual Z gates.

The transmon is capacitively coupled to a resonator for dispersive readout. The readout pulse is a 10 $\mathrm{\mu s}$ square pulse with 1 $\mathrm{\mu s}$ hyperbolic-tangent rise and fall. 4-level single-shot readout in this device is close to optimal due to the use of a readout resonator with 
linewidth \sx{$\kappa/2\pi = 0.524~\mathrm{MHz}$ and state-dependent frequency shift $\chi/2\pi=0.288~\mathrm{MHz}$, where $ \kappa \approx 2\chi$. }
Further details can be found in the supplementary materials
. In the following experiments, we choose the single-shot readout frequency at 8782.41 MHz, see Figure \ref{fig:readout}(a). This choice provides a larger separation between the $\ket{2}$ and $\ket{3}$ states than between the $\ket{0}$ and $\ket{1}$ states, due to the larger standard deviation of the Gaussian distribution for the former pair of states, see Figure \ref{fig:readout}(b). This configuration is optimal for maximizing the signal-to-noise ratio. The frequency of the $\ket{0}\leftrightarrow\ket{1}$, $\ket{1}\leftrightarrow\ket{2}$, and $\ket{2}\leftrightarrow\ket{3}$ transitions are $4134.33\mathrm{~MHz}$, $3937.66\mathrm{~MHz}$, and $3721.58\mathrm{~MHz}$, respectively. For the intrinsic quantum gates, only the transitions between neighbouring states are used. 

We measure energy relaxation times of 
$T_1^{(01)} = 189 \pm 41~\mathrm{\mu s}$, $T_1^{(12)} = 119 \pm 21~\mathrm{\mu s}$, and $T_1^{(23)} = 87 \pm 23~\mathrm{\mu s}$, and the coherence time $T_{2Echo}^{(01)}=118\pm 21  \mathrm{~\mu s}$, $T_{2Echo}^{(12)}=76\pm 27 \mathrm{~\mu s}$, and $T_{2Echo}^{(23)}=35\pm14  \mathrm{~\mu s}$.
We observe a charge dispersion of $20\mathrm{~kHz}$ on the $\ket{1}\leftrightarrow \ket{2}$ transition,
which is significantly lower than the Rabi rate of a single qudit pulse (which is \sx{$10$} MHz for a $50$ ns long $\pi$ pulse). This implies that the charge noise contribution to the error is not detrimental to the implementation of quantum algorithms. See
 the supplementary materials
 for more details.
\section{Benchmarking}

The performance of the emulator is evaluated using gate-set tomography (GST) \cite{Greenbaum2015IntroductionTomography,Nielsen2021GateTomography} and randomized benchmarking (RB) \cite{PhysRevLett.126.210504,PRXQuantum.3.020357}. GST is a method for characterizing quantum gates in detail. It can reconstruct the full Pauli transfer matrix (PTM) of the gate, as well as the initial state density operator and measurement operators that describe the state preparation and measurement (SPAM) errors. We utilize the pyGSTi tool \cite{Nielsen_2020} to implement GST on the two-qubit emulator. GST reconstructs the PTM and SPAM operators and shows that $X_{01}(\pi/2)$, $X_{12}(\pi/2)$, and $X_{23}(\pi/2)$ have infidelities of $4.976 \pm 1.14 \times 10^{-3}$,	
$2.966 \pm 0.052\times 10^{-3}$,
$2.906 \pm 0.068\times 10^{-3}$, respectively. More details and discussions can be found in  the supplementary materials
. 

\begin{figure}[!t] 
    \centering
	\sidesubfloat[]{
    \includegraphics[width=.43\linewidth]{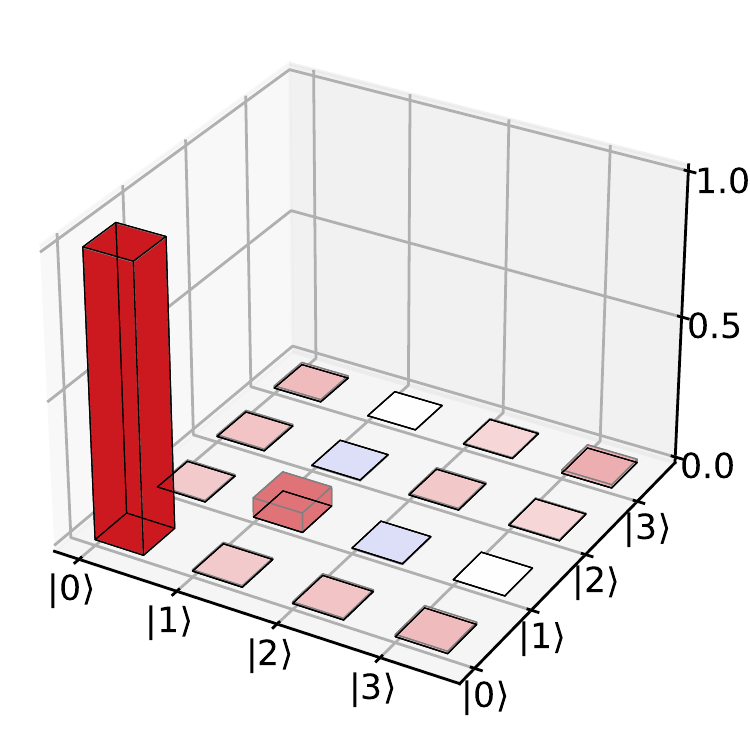}
	}
	\sidesubfloat[]{
        \includegraphics[width=.43\linewidth]{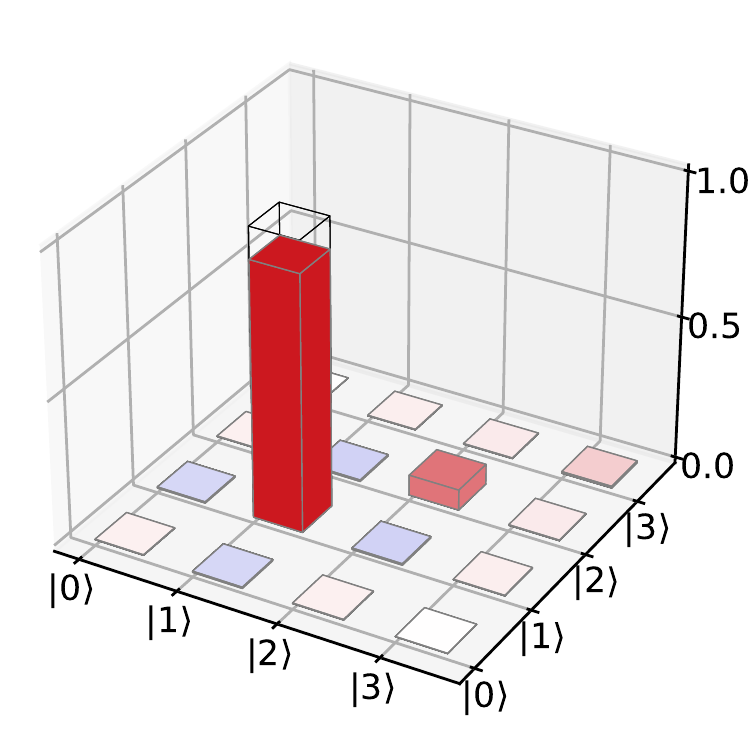}
	}\\
	\sidesubfloat[]{
        \includegraphics[width=.43\linewidth]{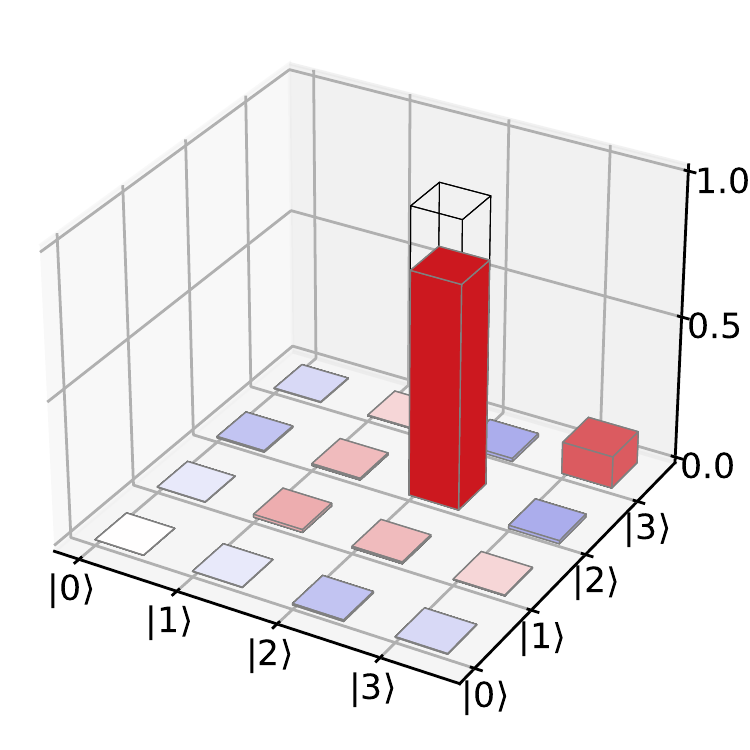}
	}
	\sidesubfloat[]{
        \includegraphics[width=.43\linewidth]{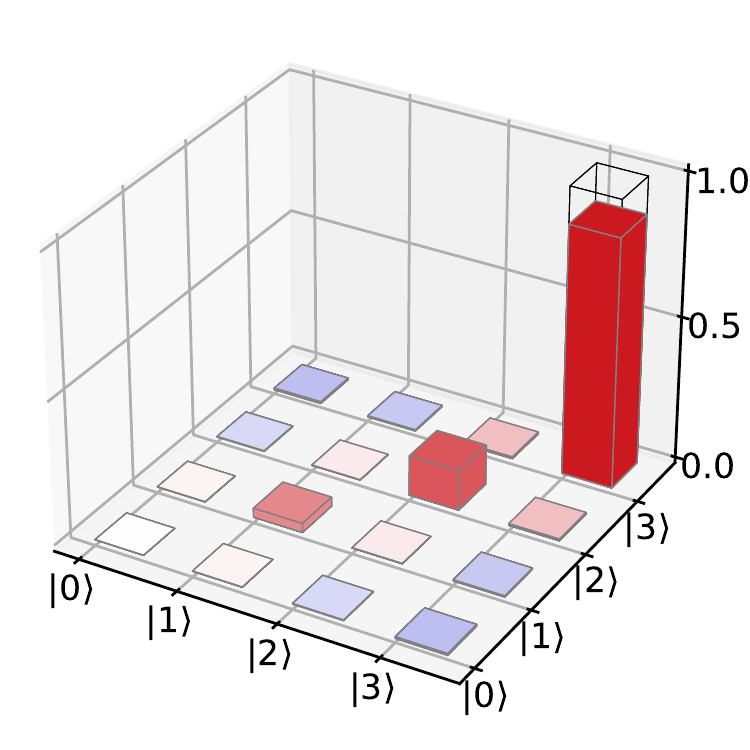}
	}\\
	\sidesubfloat[]{
        \includegraphics[width=.43\linewidth]{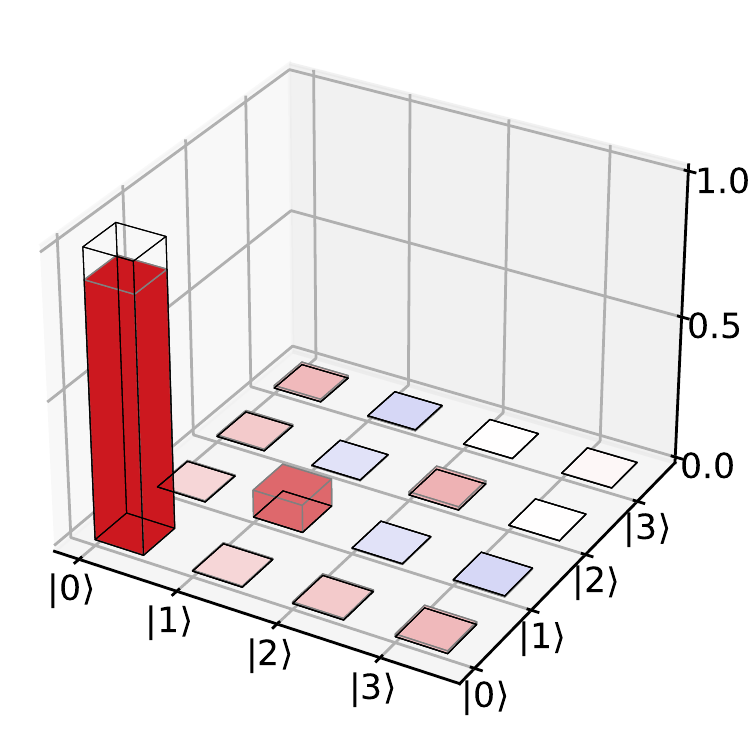}
	}
	\sidesubfloat[]{
        \includegraphics[width=.43\linewidth]{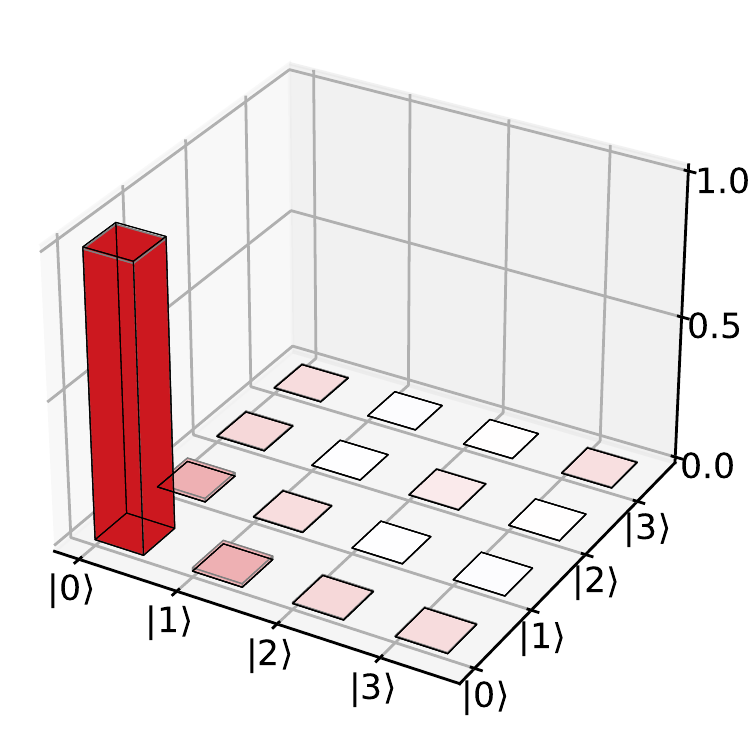}
	}
    \caption{(a-d) Real part of the measurement operators reconstructed from gate-set tomography (GST), which give the expectation value of transmon population in $\ket{0}$, $\ket{1}$, $\ket{2}$, and $\ket{3}$ states, respectively. The infidelity of these operators are $0.14 \pm 0.24 \times 10^{-2}$, $10.98 \pm 0.25 \times 10^{-2}$, $21.55 \pm 0.03 \times 10^{-2}$, $12.85 \pm 0.23 \times 10^{-2}$, respectively. (e-f) Real part of the density operator of the initial state without (e) and with (f) the active reset protocol enabled, respectively. The infidelity of the initial states are $ 9.9 \pm 1.1 \times 10^{-2}$ and $0.68 \pm 0.13 \times 10^{-2}$, without and with the active reset enabled, respectively.
}
    \label{fig:GST_SPAM}
\end{figure}

After characterizing the initial state density operator, we observe a non-negligible population in the excited state, which prompts us to implement a qudit active reset protocol to prepare a high-fidelity initial state. We achieve this by utilizing the programmed FPGA to distinguish all four states using the nearest neighbor method within 40 ns, and sending a conditional sequence of $\pi$ pulses in the $\{\ket{2},\ket{3}\}$, $\{\ket{1},\ket{2}\}$ and $\{\ket{0},\ket{1}\}$ subspaces to reset the state back to the ground state. The effectiveness of the active reset protocol is verified by characterizing the initial state operator using GST, see Figure \ref{fig:GST_SPAM}. We find that the active reset significantly improves the initial state fidelity from $0.900\pm 0.011$ to $0.9932 \pm 0.0013$, with twice repeated active reset being optimal. Prior to each following experiment, we apply the active reset protocol twice to ensure the preparation of a high-fidelity initial state.

\begin{figure}[!t]
    \centering
	\includegraphics[width=\linewidth]{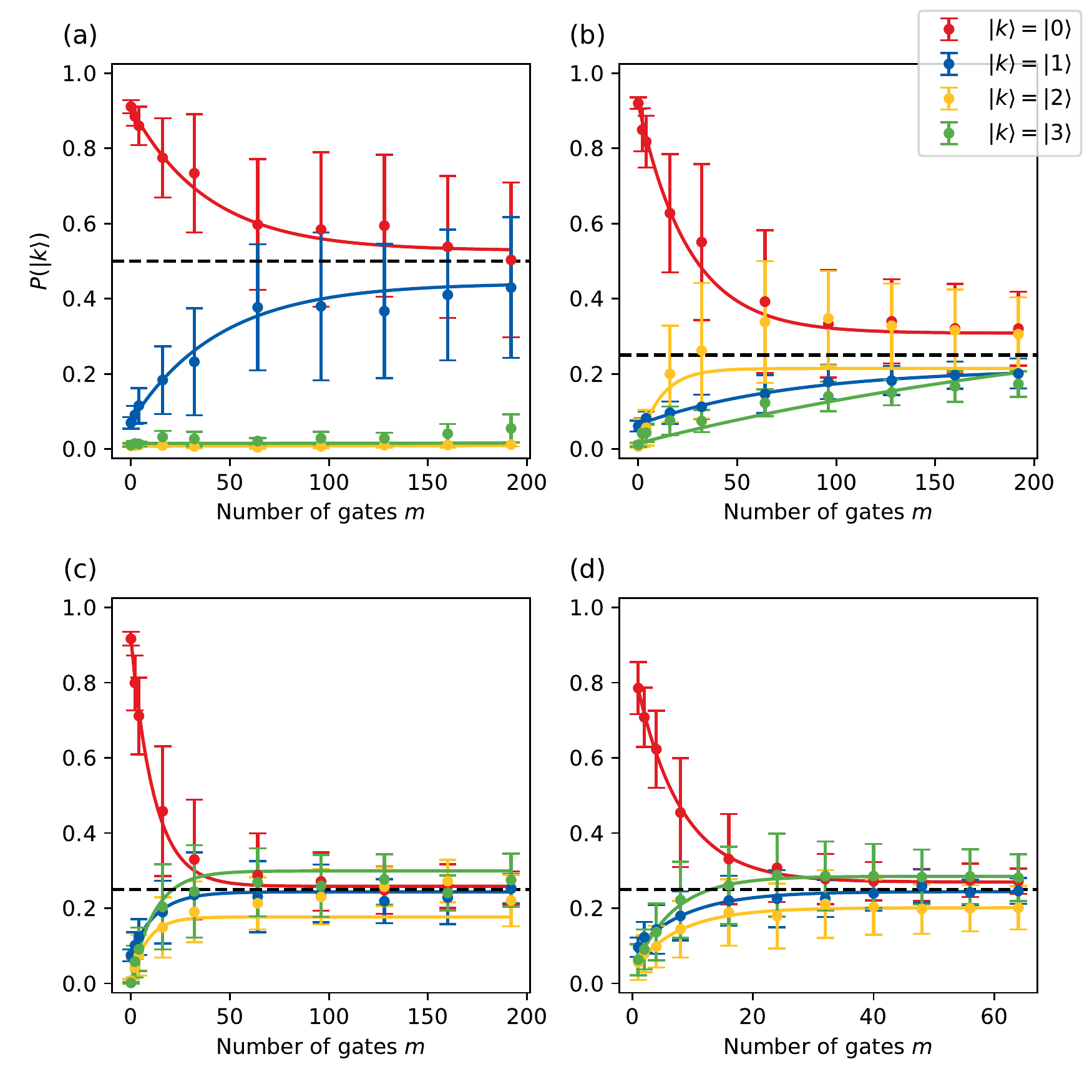}
    \caption{ (a-b) Results of single-qubit randomized benchmarking on the virtual qubits A and B of the emulator, yielding average infidelity per single-qubit Clifford gate $1.91 \pm 0.23 \times 10^{-2}$ and $2.89 \pm 0.31 \times 10^{-2}$, respectively. (c) Simultaneous randomized benchmarking of single-qubit gates on both virtual qubits, with gate sequences executed in series. The average infidelity per single-qubit Clifford is extracted to be $3.14 \pm 0.26 \times 10^{-2}$ (d) Results of full 2Q randomized benchmarking on the emulator, and the average infidelity per two-qubit Clifford gate is $9.51 \pm 0.71 \times 10^{-2}$.
    }
    \label{fig:randomized_benchmarking}
\end{figure}

To evaluate the performance of the emulator, we conduct RB on the virtual qubits A and B which allowes us to investigate the cost overhead of the emulation and determine its effectiveness. The result is shown in Figure \ref{fig:randomized_benchmarking}, showing that emulation causes fidelity loss. We generate single-qubit randomized Clifford gate sequences and convert them to the native qudit gate using the simulator's mapping rule. The RB sequences are run on the emulator, and the success probability $P$ is defined as the population landing on the emulated state $\ket{00}$. We fit the RB model to $P = A p^m + B$, where $A$, $B$, and $p$ are fitting parameters, and $m$ is the number of Clifford gates. We calculate the average infidelity per Clifford as $r = (1-p)(d-1)/d$, where $d=2$ for single-qubit RB and $d=4$ for two-qubit RB. The resulting average single-qubit Clifford gate infidelity for virtual qubits A and B are $1.91 \pm 0.23 \times 10^{-2}$ and $2.89 \pm 0.31 \times 10^{-2}$, respectively. Notably, performing single-qubit RB on virtual qubit A only drives the population within the ${\ket{0},\ket{1}}$ subspace. Although transitions within the $\{\ket{2},\ket{3}\}$ subspace are driven, a negligible population exists in this subspace. Therefore, we expect $P$ to converge to $0.5$, unlike the rest of the RB experiments that converges to $0.25$. We also perform RB on both virtual qubits using the emulated single-qubit gates, yielding an average infidelity per Clifford of $3.14 \pm 0.26 \times 10^{-2}$. Finally, we implement full two-qubit Clifford randomized benchmarking. \sx{The two-qubit gates in this setup are randomly selected from the two-qubit Clifford group, which contains 11,520 two-qubit gates. Each Clifford gate is then decomposed into a sequence consisting of single-qubit gates acting on individual qubits, followed by a fixed two-qubit gate, and then another set of single-qubit gates acting on each qubit again.  For this experiment, the fixed two-qubit gate is designed to execute the unitary operation $U_{ZX}=\mathrm{exp}(-i(\pi/4)ZX)$, where $ZX$ represents the tensor product of the Pauli Z and Pauli X operators. This operation is commonly realizable through cross-resonance interaction, which aligns with the conventional decomposition in fixed-coupling quantum processors.} Then, these decomposed gates are mapped to the qudit intrinsic gate set. \sx{The single qubit gates are mapped with the method described in Fig.\ref{fig:emulator}, and the $U_{ZX}$ is implemented by driving a X gate on the $\{\ket{2},\ket{3}\}$ subspace followed by a $X(-\frac{\pi}{2})$ gate on the first virtual qubit.} We extract the average emulated two-qubit Clifford gate infidelity to be $9.51 \pm 0.71 \times 10^{-2}$. Our results suggest that emulating two qubits on a single qudit results in fidelity loss, indicating the need for optimal mappings of the operations on the emulated qubits to the qudit to further improve performance. 
\section{Variational Quantum Eigensolver}

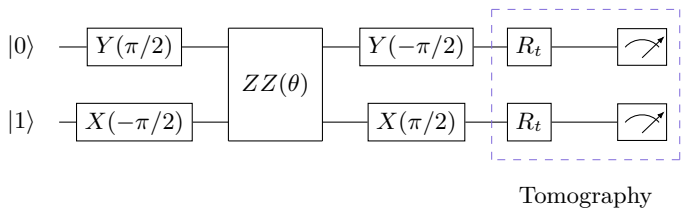
\begin{figure}
    \centering
    \begin{tikzpicture}
	\begin{pgfonlayer}{nodelayer}
		\node [style=none] (0) at (-8, 1) {};
		\node [style=none] (1) at (-8, -1) {};
		\node [style=box] (2) at (-6, 1) {$Y(\pi/2)$};
		\node [style=box] (3) at (1.5, -1) {$X(\pi/2)$};
		\node [style=none] (4) at (-3.5, 1.5) {};
		\node [style=none] (5) at (-1, 1.5) {};
		\node [style=none] (6) at (-1, -1.5) {};
		\node [style=none] (7) at (-3.5, -1.5) {};
		\node [style=none] (8) at (-2.25, 0) {$ZZ(\theta)$};
		\node [style=none] (9) at (-3.5, 1) {};
		\node [style=none] (10) at (-3.5, -1) {};
		\node [style=none] (11) at (-1, -1) {};
		\node [style=none] (12) at (-1, 1) {};
		\node [style=box] (13) at (-6, -1) {$X(-\pi/2)$};
		\node [style=box] (14) at (1.5, 1) {$Y(-\pi/2)$};
		\node [style=box] (15) at (4.5, 1) {$R_t$};
		\node [style=box] (16) at (4.5, -1) {$R_t$};
		\node [style=none] (17) at (3.5, 2) {};
		\node [style=none] (18) at (3.5, -2) {};
		\node [style=none] (19) at (8.5, -2) {};
		\node [style=none] (20) at (8.5, 2) {};
		\node [style=meter] (21) at (7.5, 1) {};
		\node [style=meter] (22) at (7.5, -1) {};
		\node [style=none] (23) at (6, -3) {Tomography};
		\node [style=none] (24) at (-9, 1) {$\ket{0}$};
		\node [style=none] (25) at (-9, -1) {$\ket{1}$};
	\end{pgfonlayer}
	\begin{pgfonlayer}{edgelayer}
		\draw (0.center) to (2);
		\draw (4.center) to (5.center);
		\draw (5.center) to (6.center);
		\draw (7.center) to (6.center);
		\draw (7.center) to (4.center);
		\draw (2) to (9.center);
		\draw (11.center) to (3);
		\draw (14) to (15);
		\draw (3) to (16);
		\draw (15) to (21);
		\draw (16) to (22);
		\draw [style=dashed thin purple] (17.center) to (20.center);
		\draw [style=dashed thin purple] (20.center) to (19.center);
		\draw [style=dashed thin purple] (19.center) to (18.center);
		\draw [style=dashed thin purple] (18.center) to (17.center);
		\draw (12.center) to (14);
		\draw (1.center) to (13);
		\draw (13) to (10.center);
	\end{pgfonlayer}
\end{tikzpicture}
    \caption{Gate sequence for implementing the variational ansatz used to approximate the binding energy of a hydrogen molecule. \sx{$R_t$ are fiducial gates for implementing tomography. } The gates are mapped to the qudit native operations under the rules described in Fig. \ref{fig:emulator}, and the $ZZ(\theta)$ gate is implemented by applying two virtual Z gates $Z_1(-\theta)$ and $Z_2(-\theta)$. More details can be found in  the supplementary materials
    .}
    \label{fig:VQE_circuit}
\end{figure}

\newcommand{\ivqe}[1]{\begin{tabular}{c} \includegraphics[width=4cm,height=2.5cm]{#1}\end{tabular}}

\begin{figure*}[!t]
    \centering
    \begin{tabular}{c|c|c|c|c}
                     & (a) Raw value &  (b) Outlier removed & (c) Measurement mitigation & (d) Simulation \\\hline  
         \begin{tabular}{c}IZ\\\end{tabular} & \ivqe{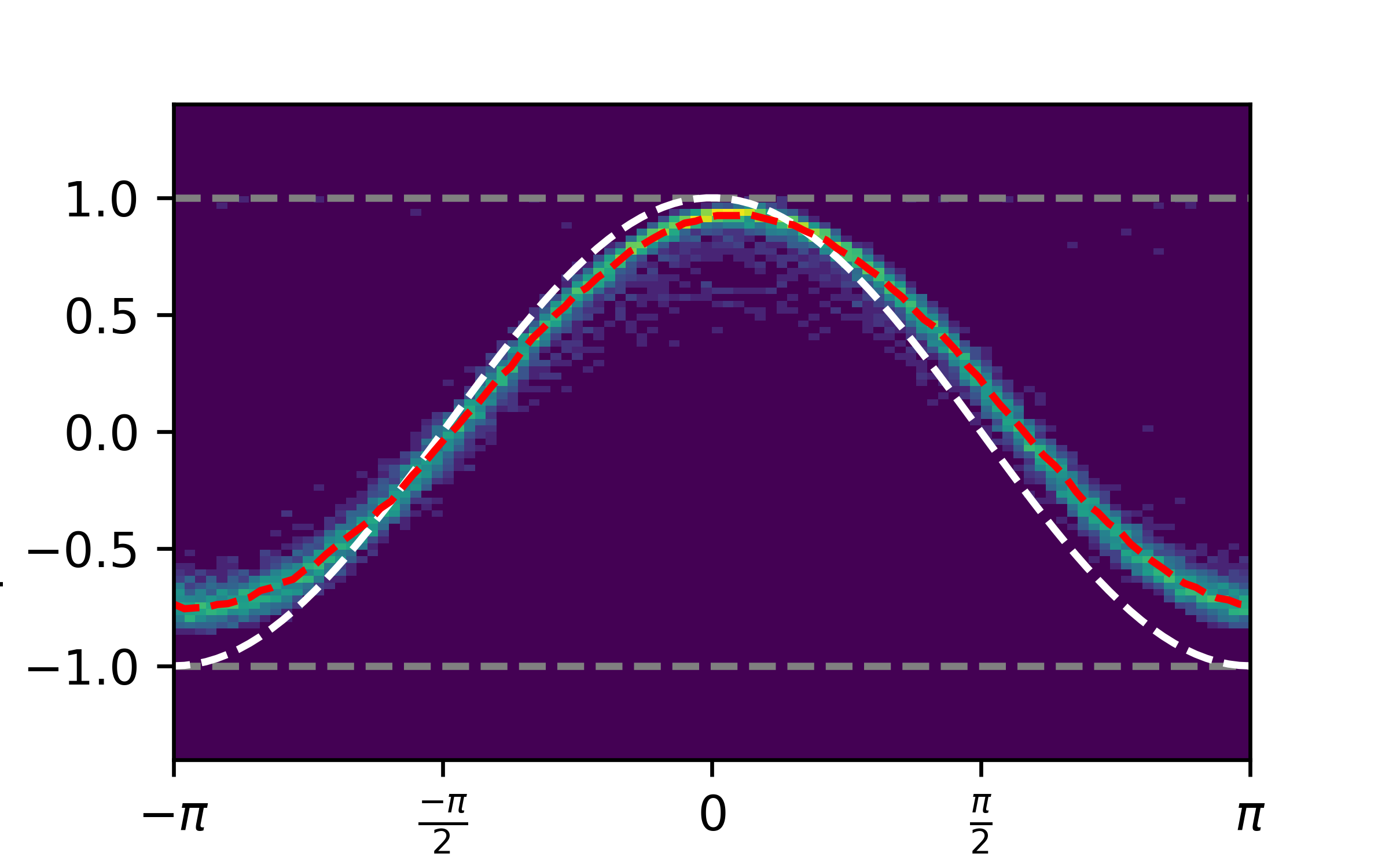} & \ivqe{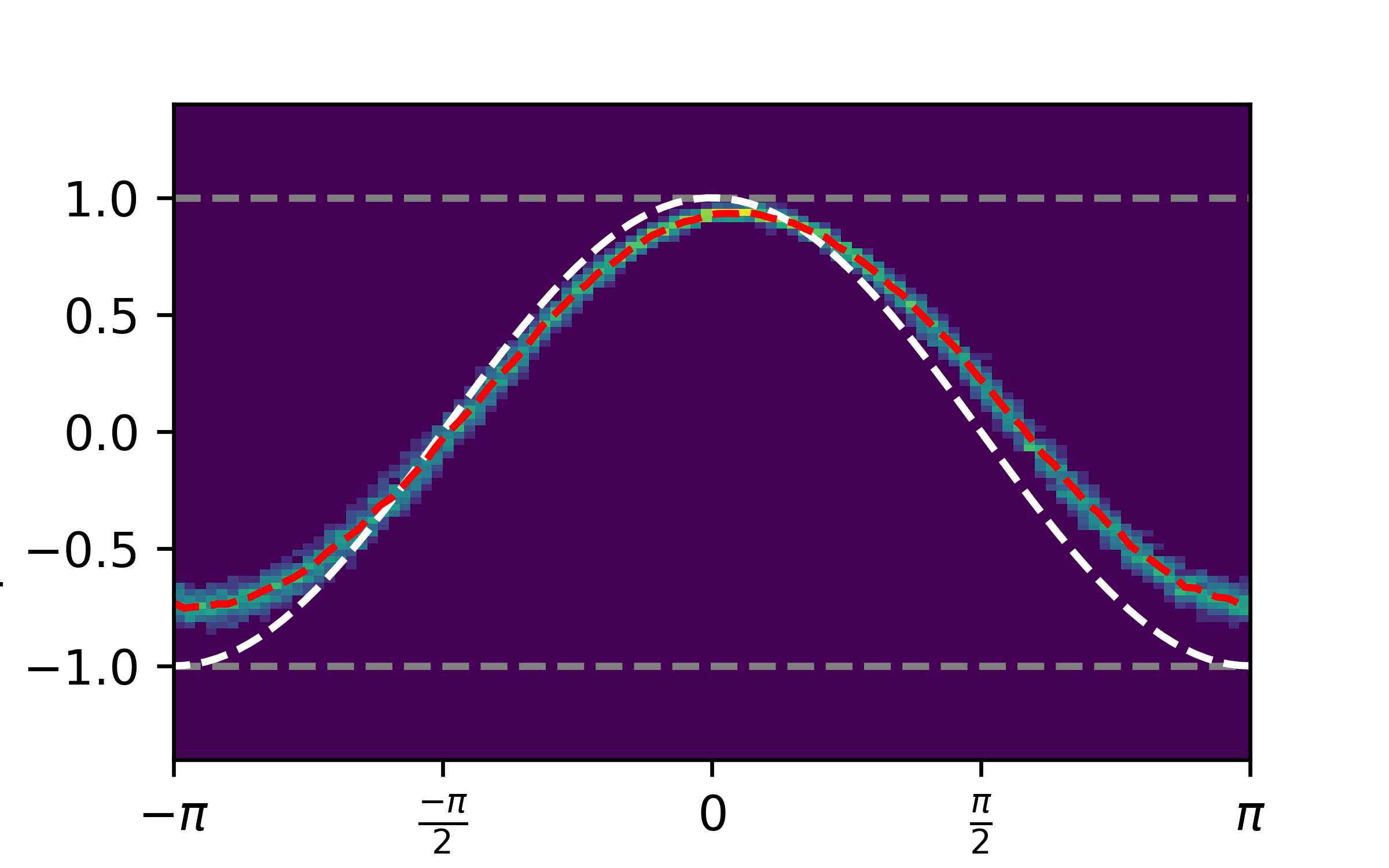} & \ivqe{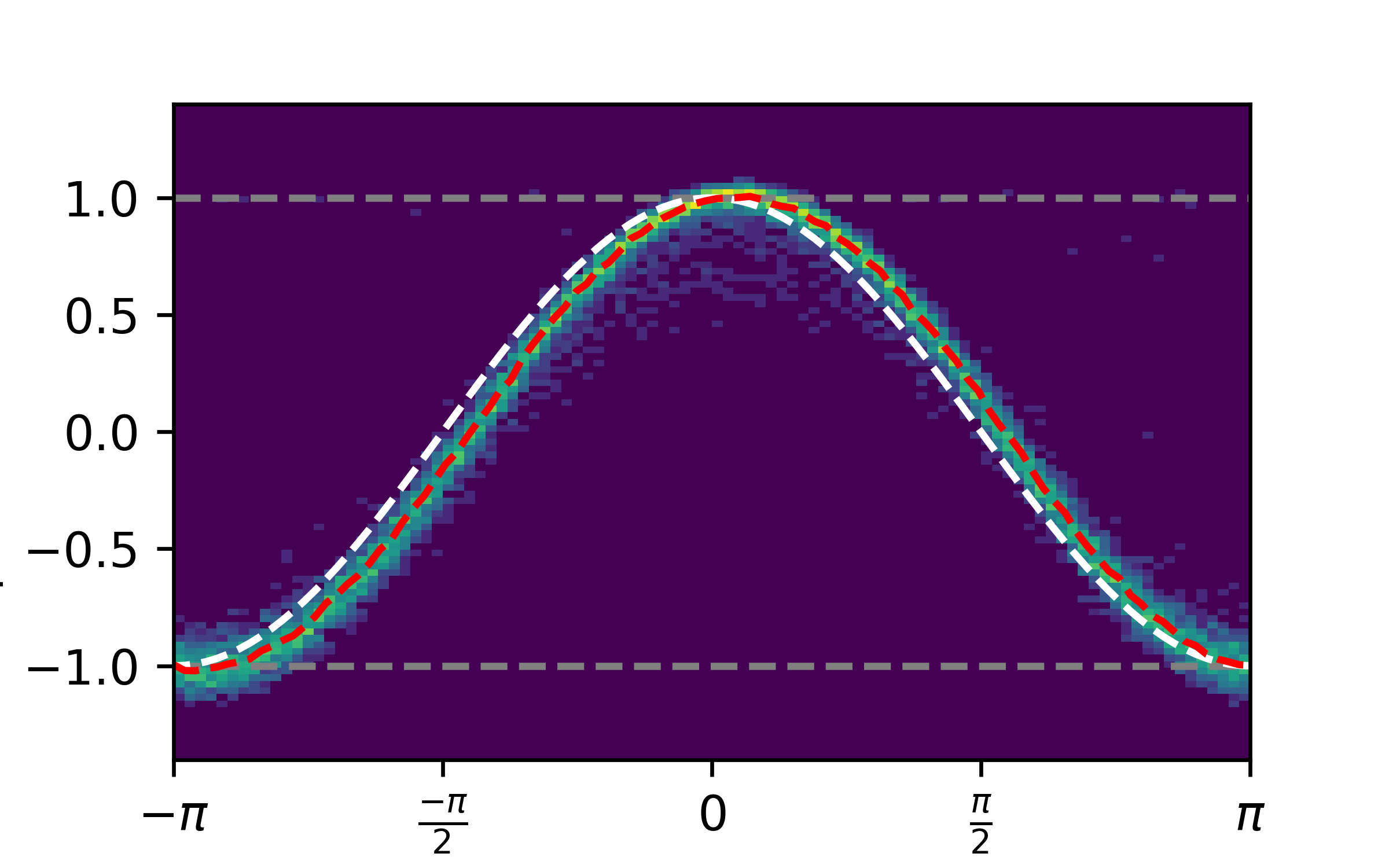}& \ivqe{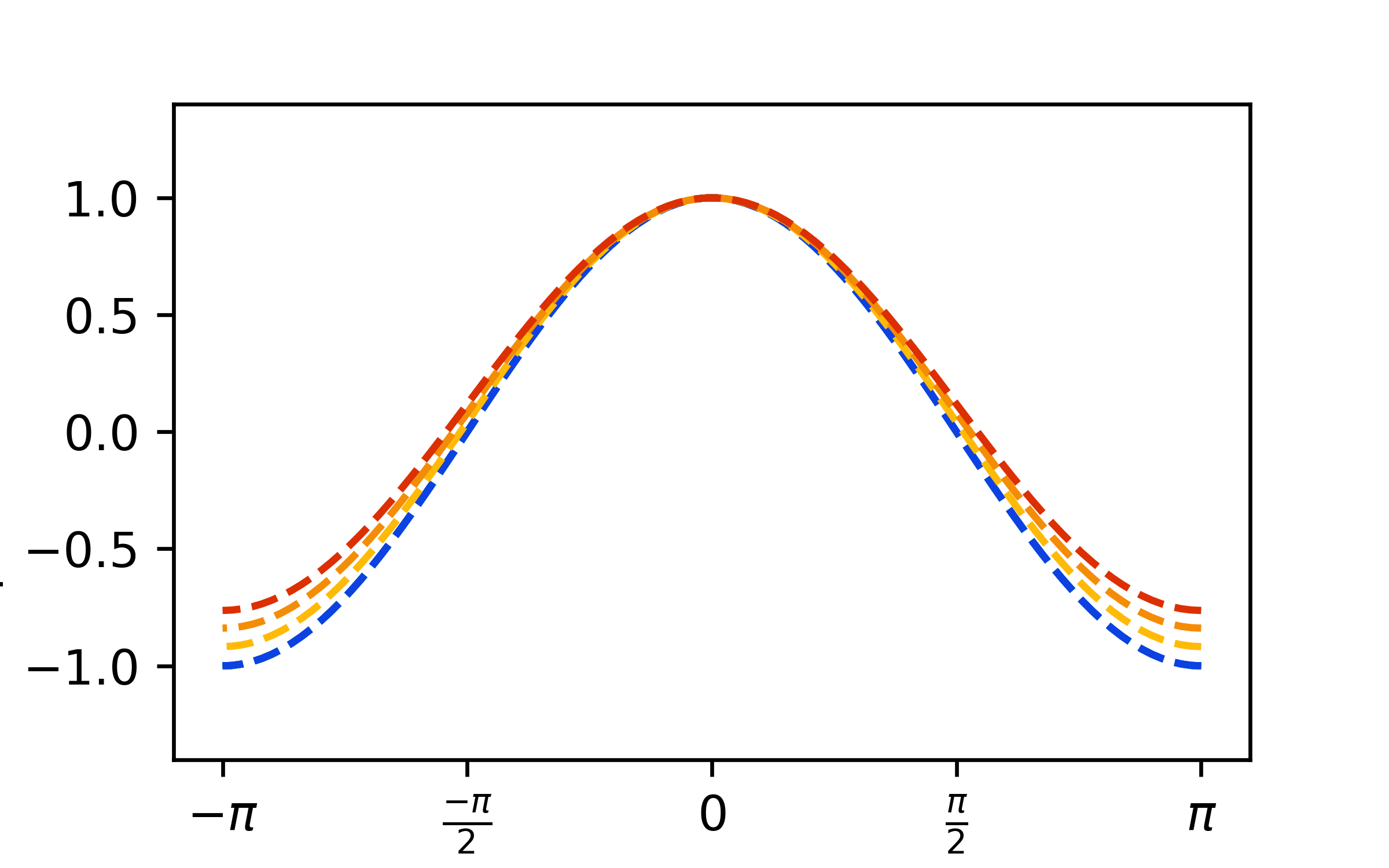} \\ \hline
         \begin{tabular}{c}ZI\\\end{tabular} & \ivqe{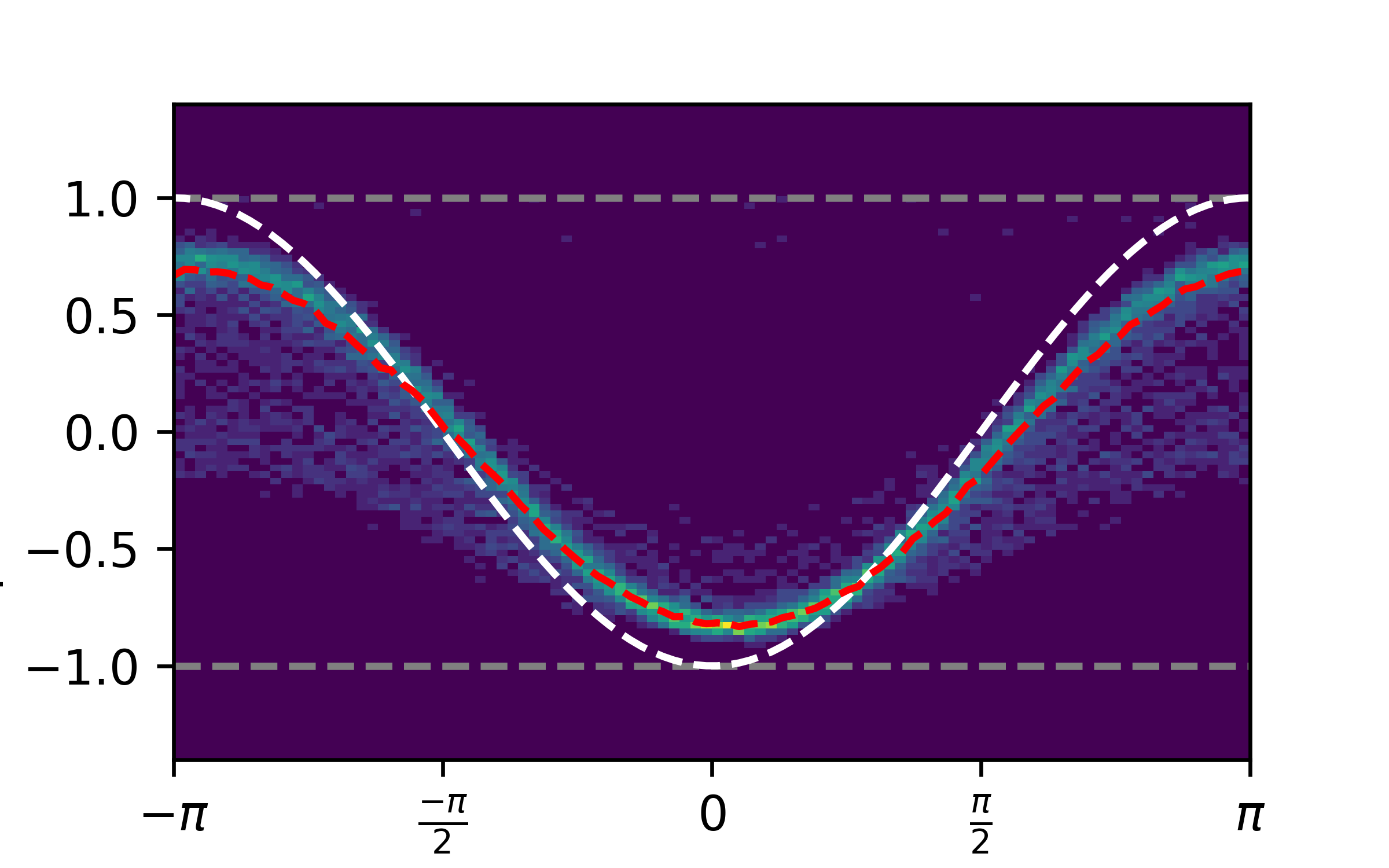} & \ivqe{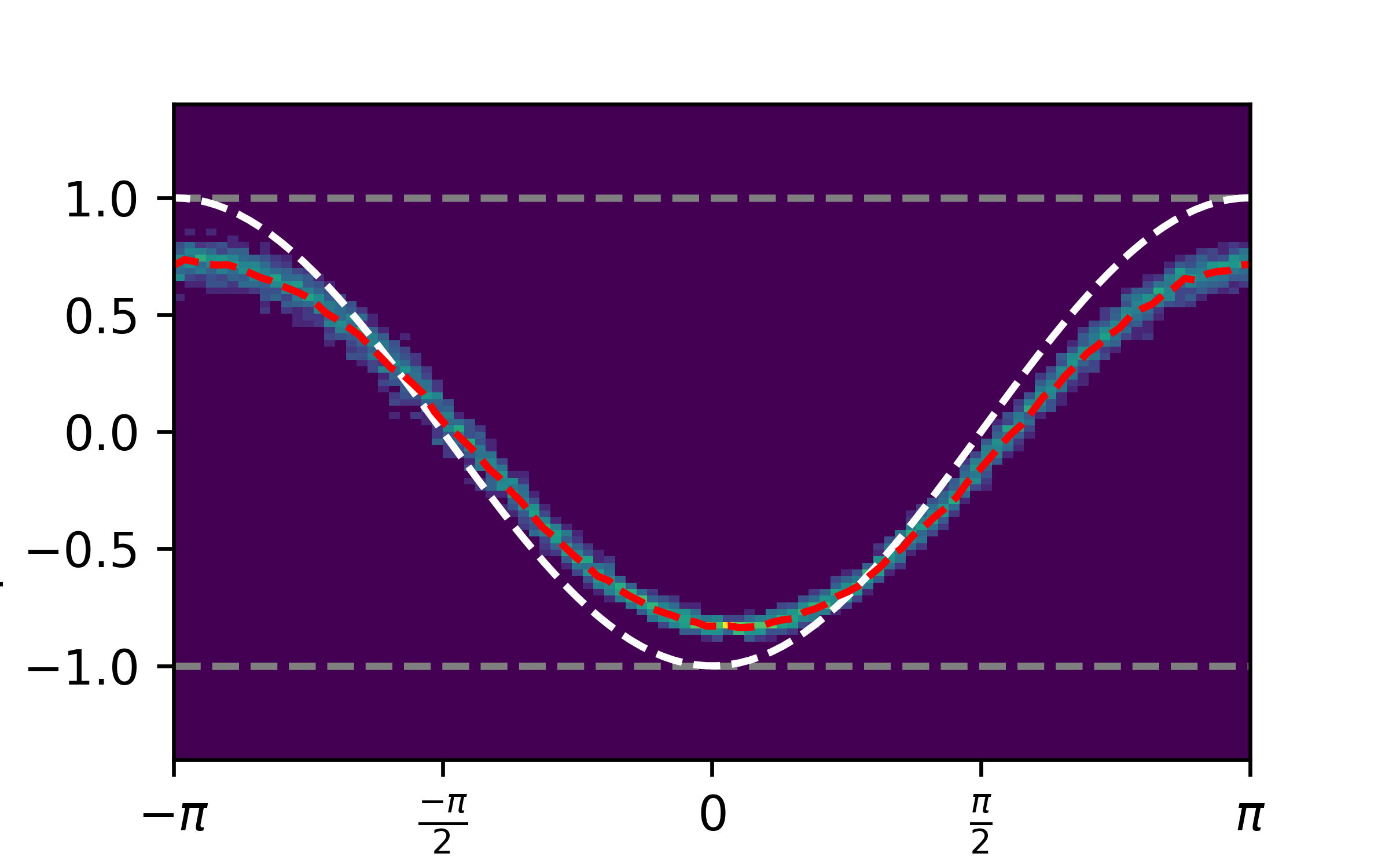} & \ivqe{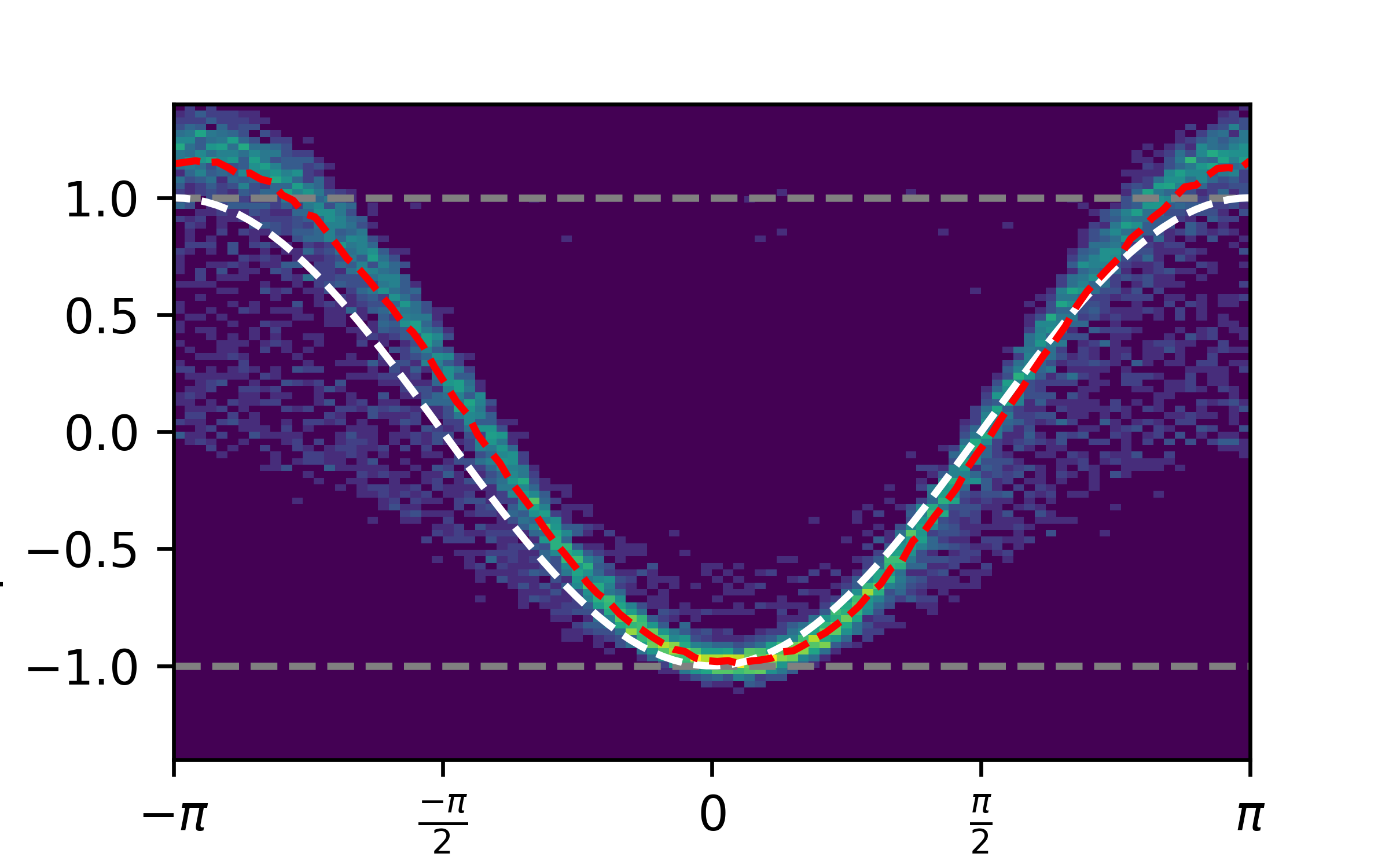}& \ivqe{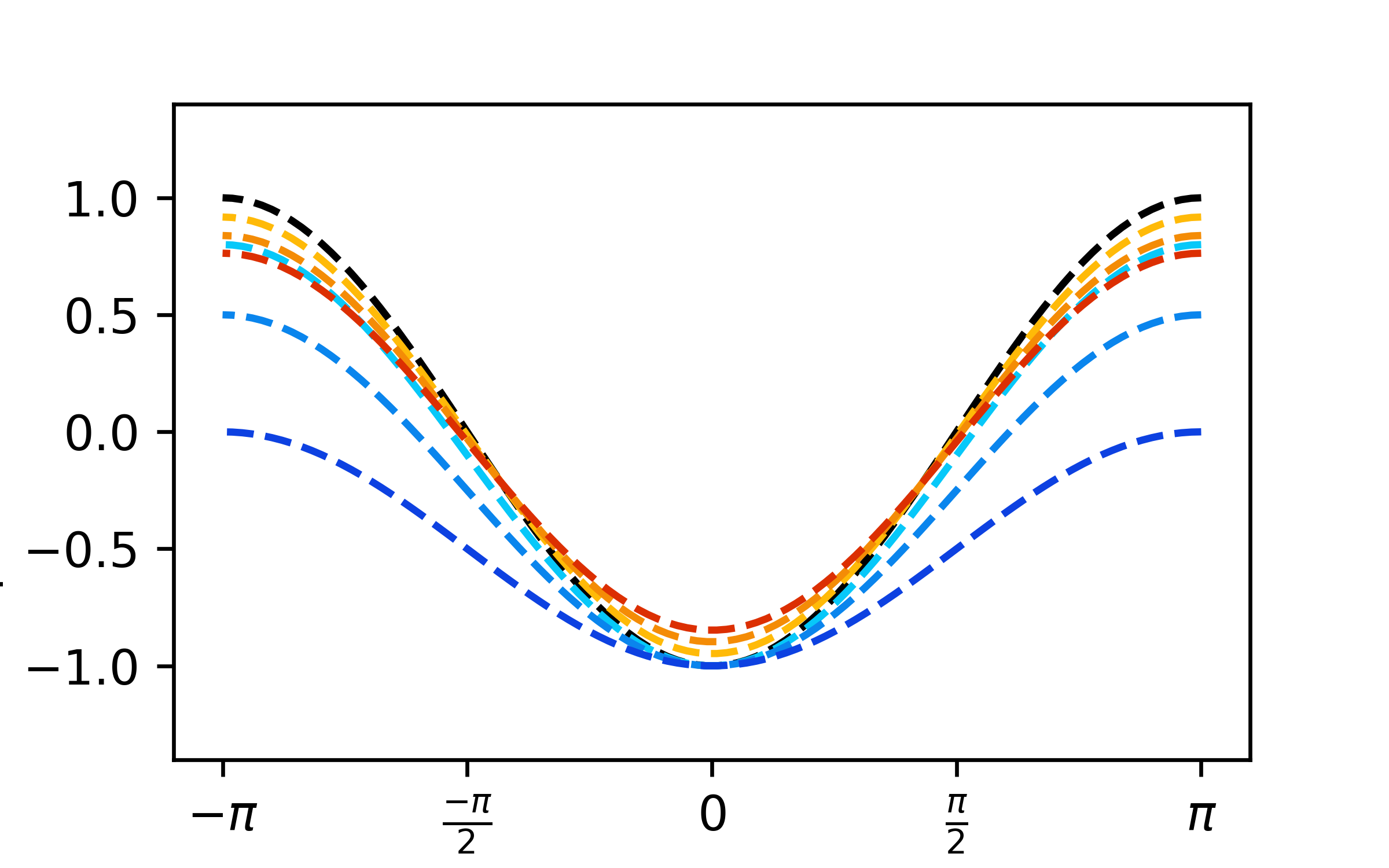} \\ \hline
         \begin{tabular}{c}ZZ\\\end{tabular} & \ivqe{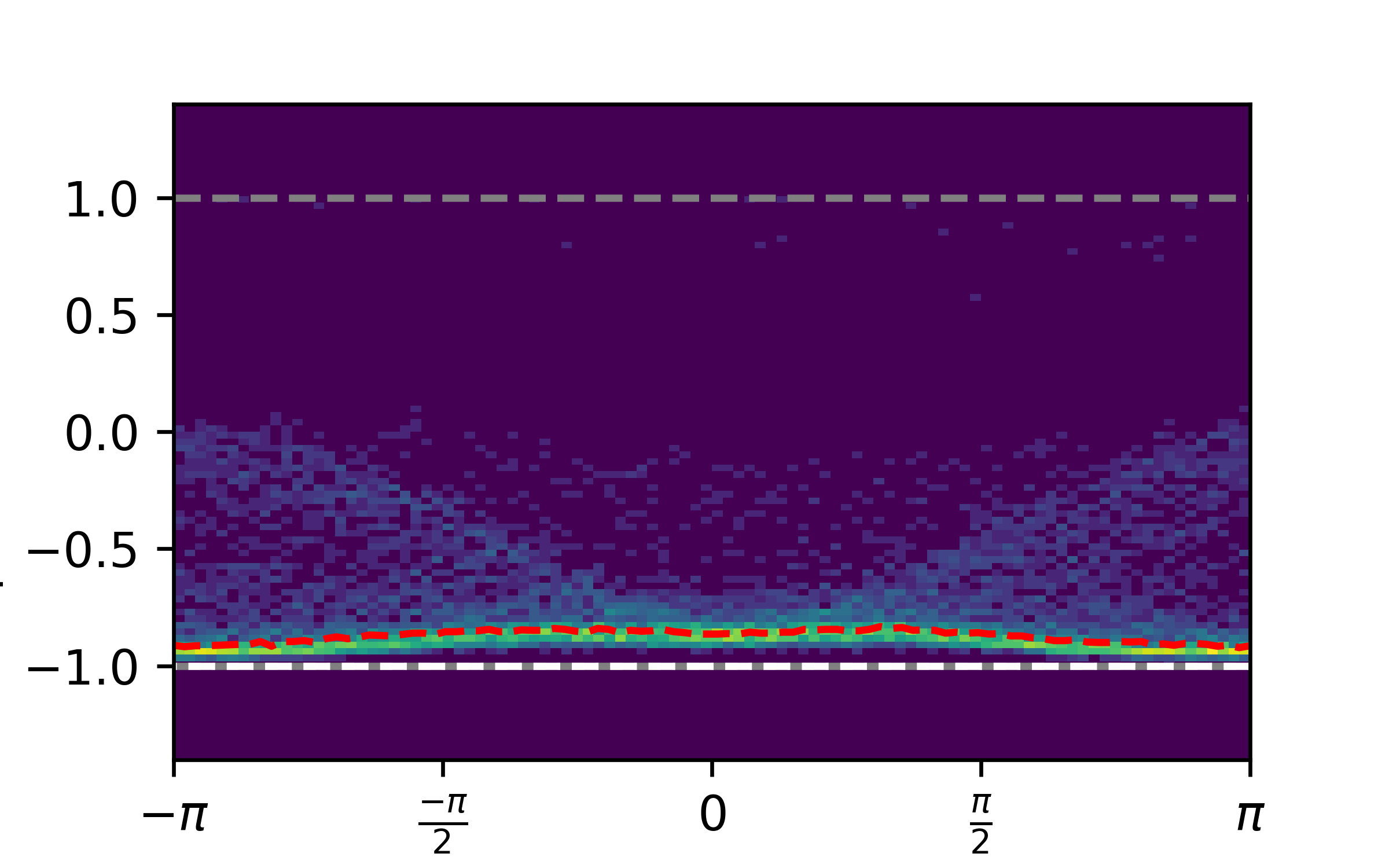} & \ivqe{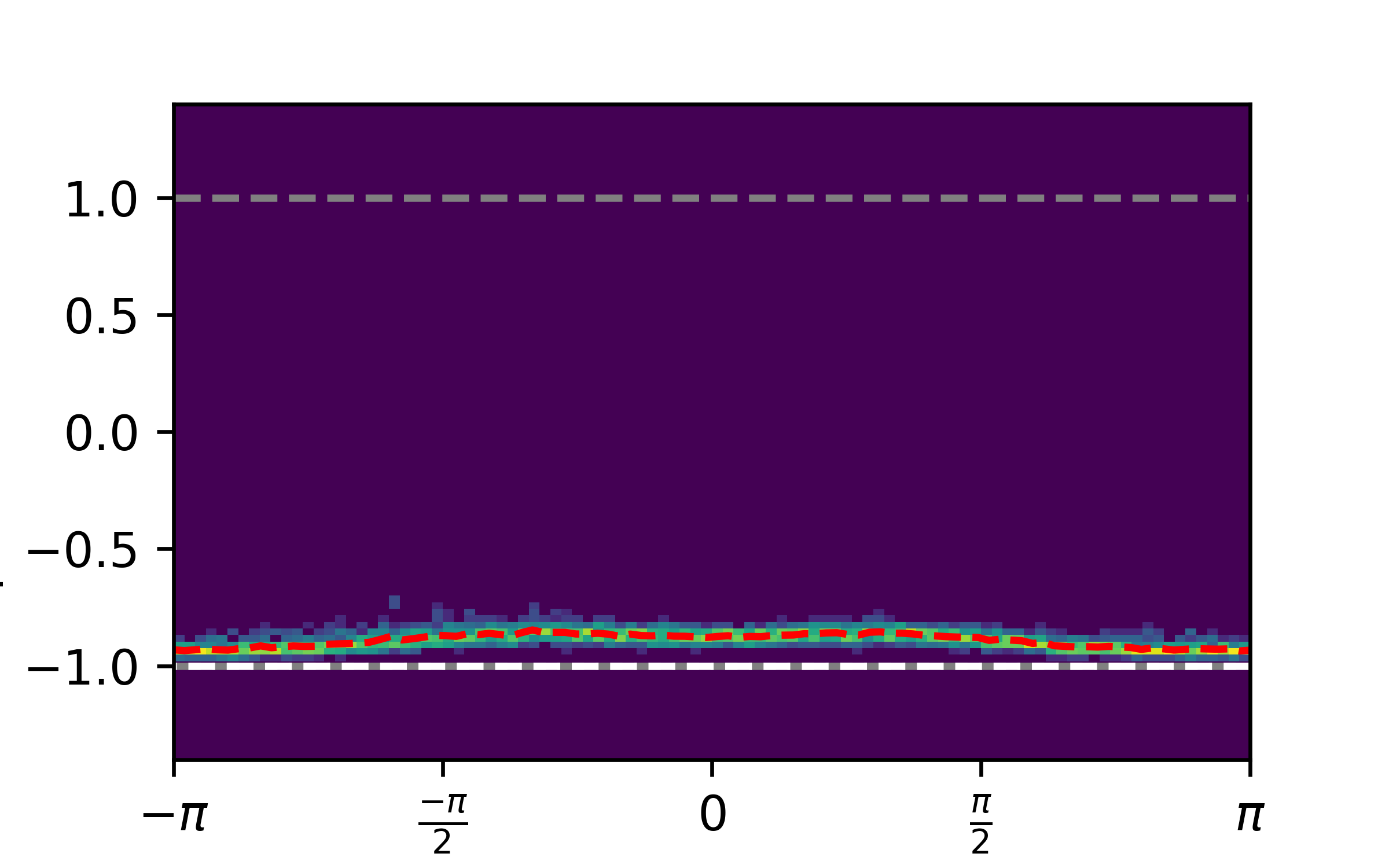} & \ivqe{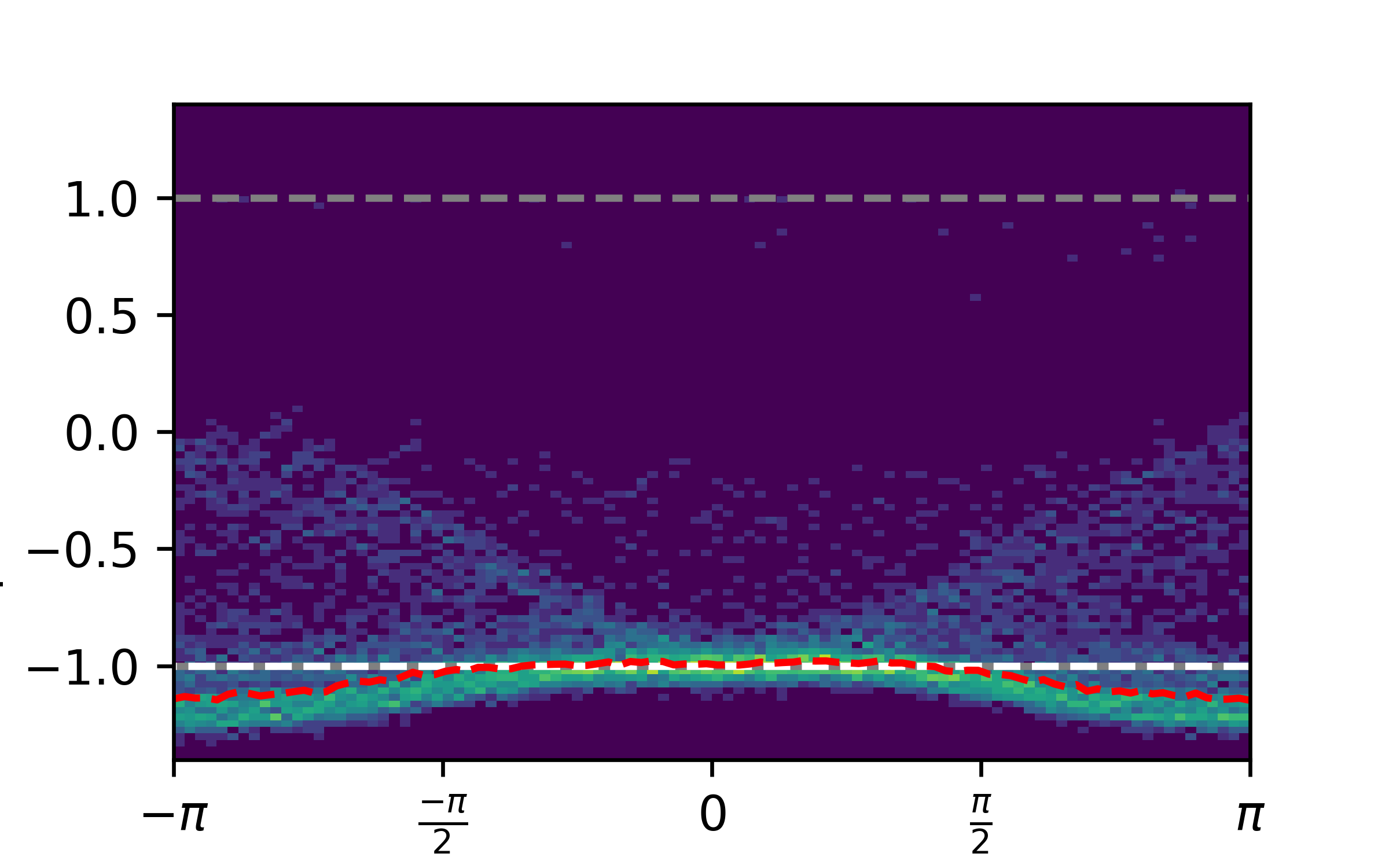}& \ivqe{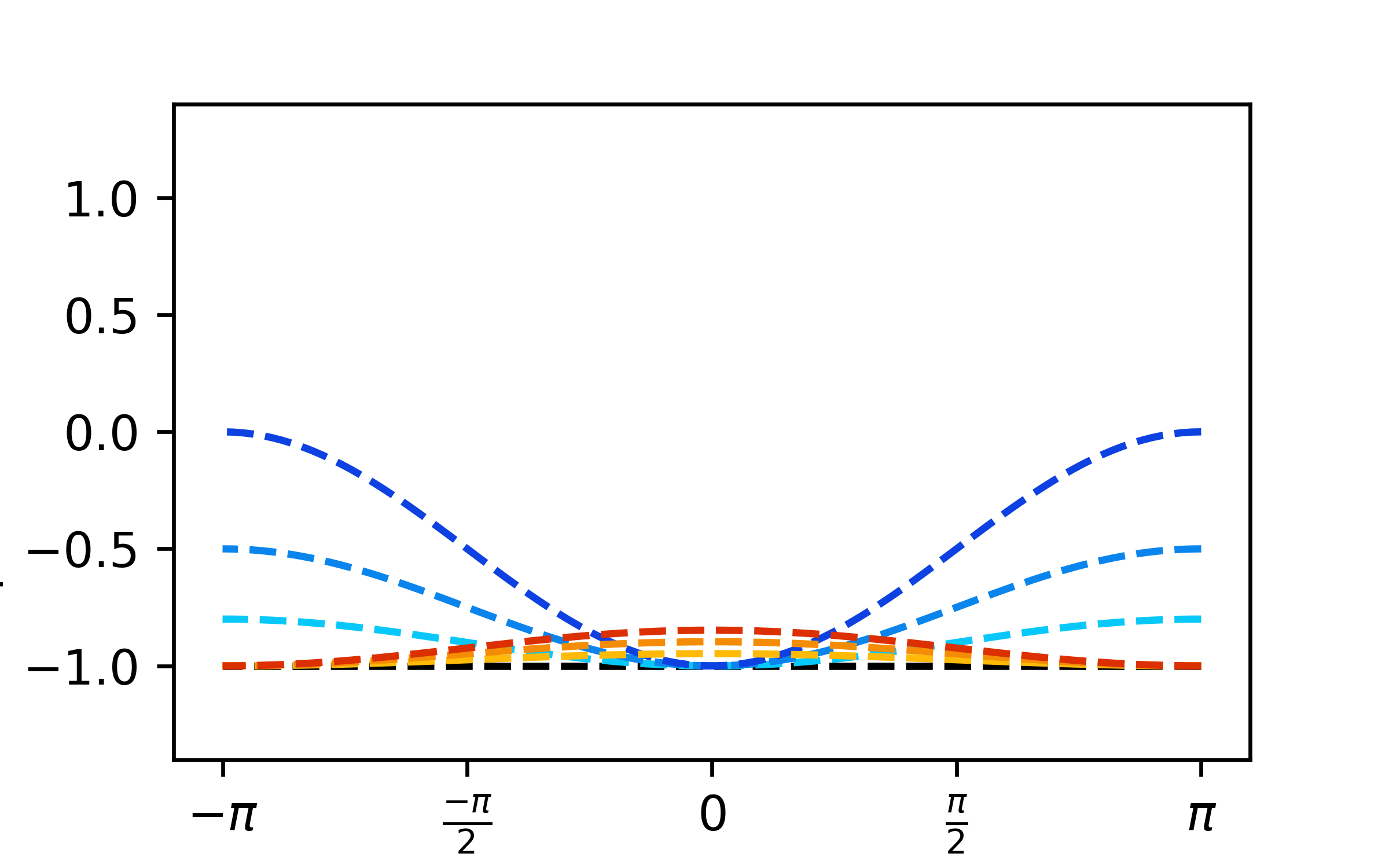} \\ \hline
         \begin{tabular}{c}XX\\\end{tabular} & \ivqe{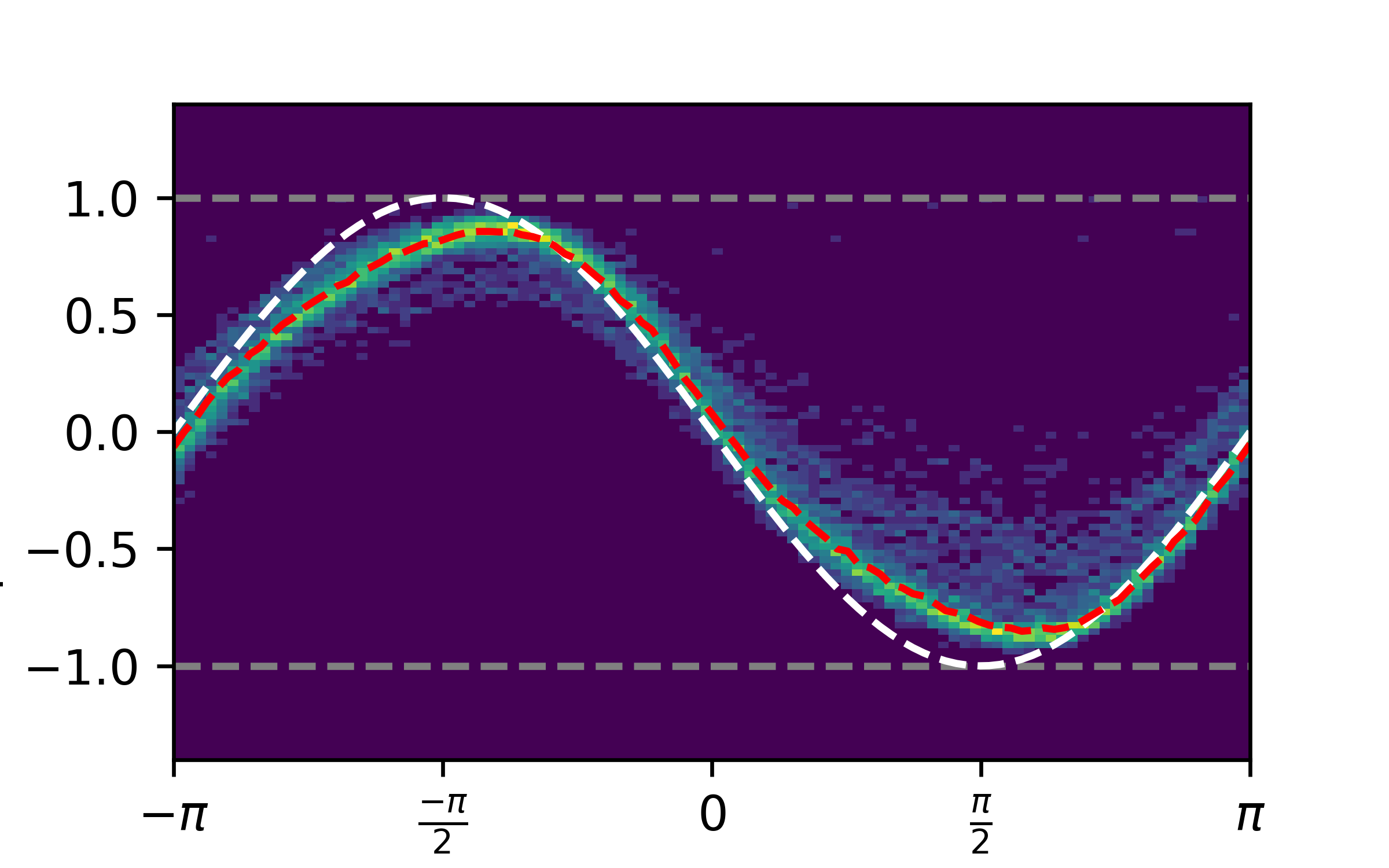} & \ivqe{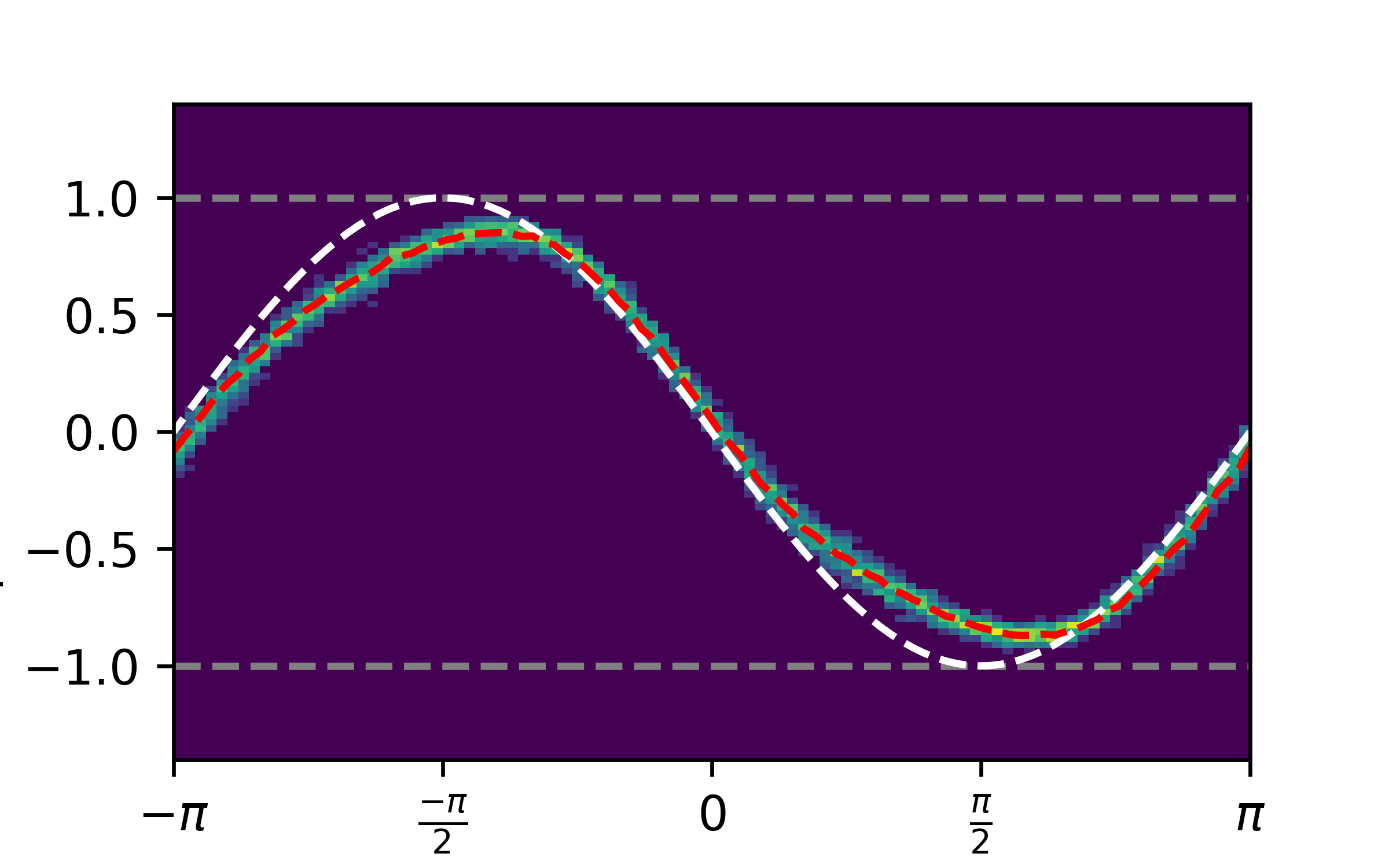} & \ivqe{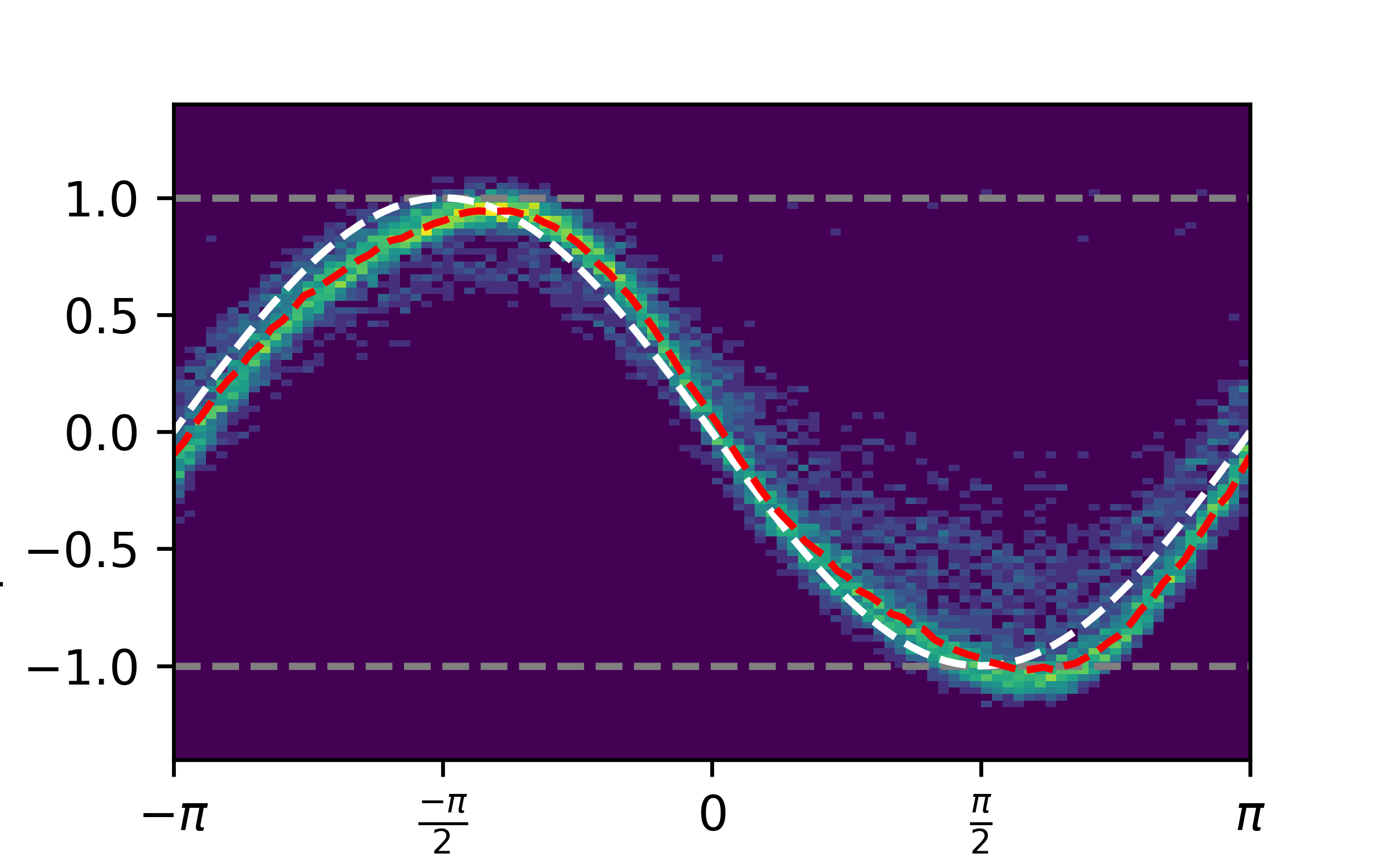}& \ivqe{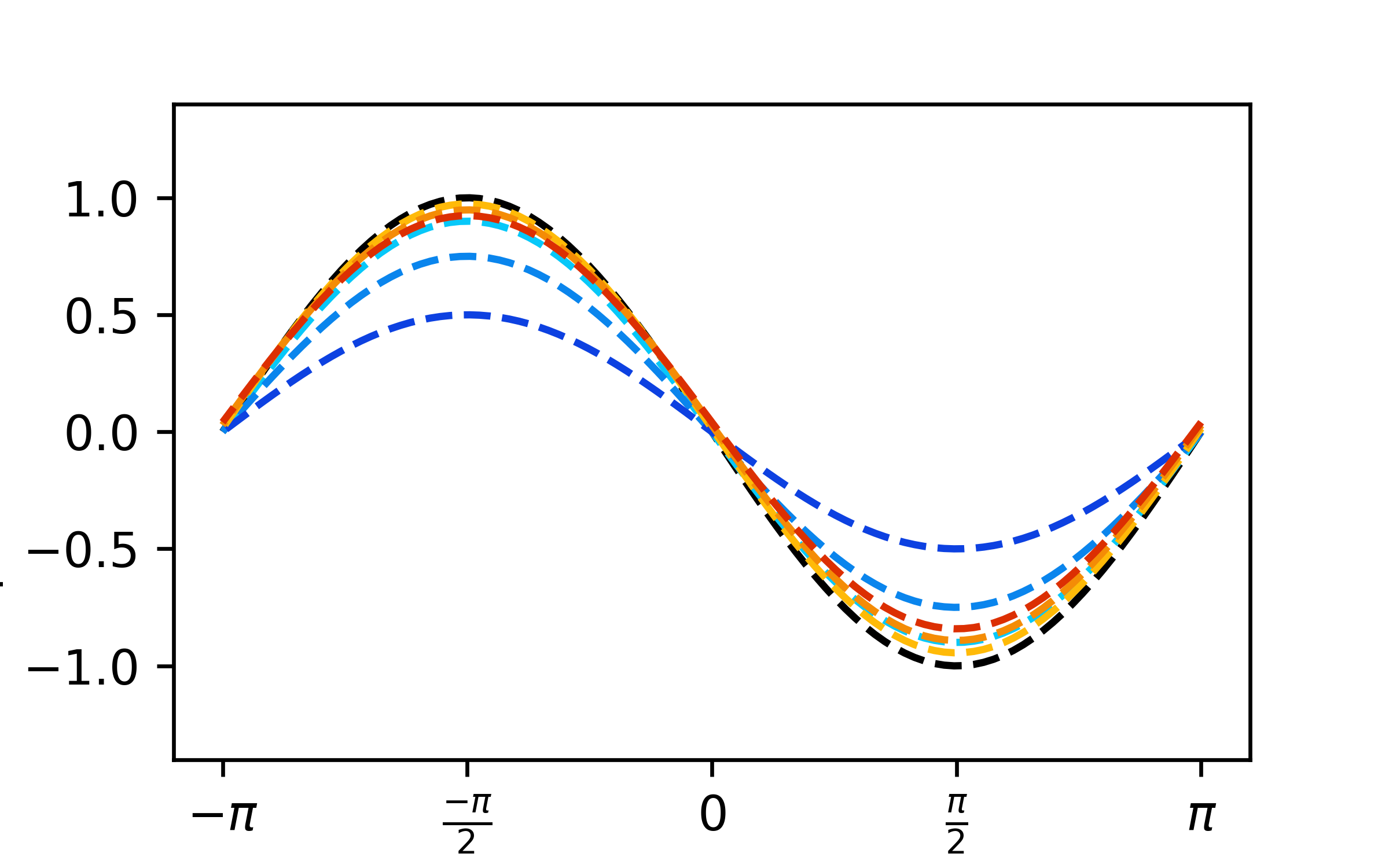} \\ \hline
         \begin{tabular}{c}YY\\\end{tabular} & \ivqe{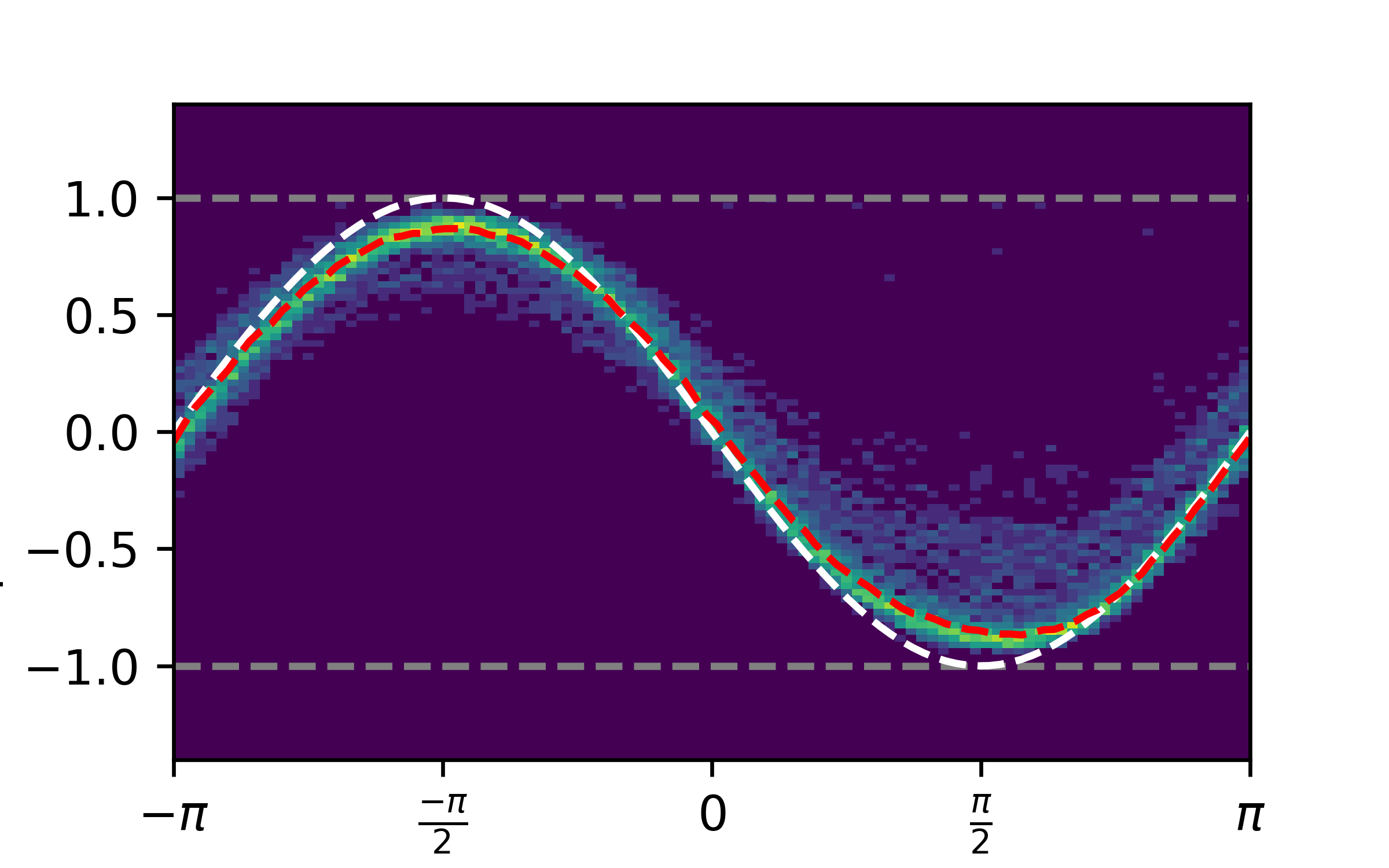} & \ivqe{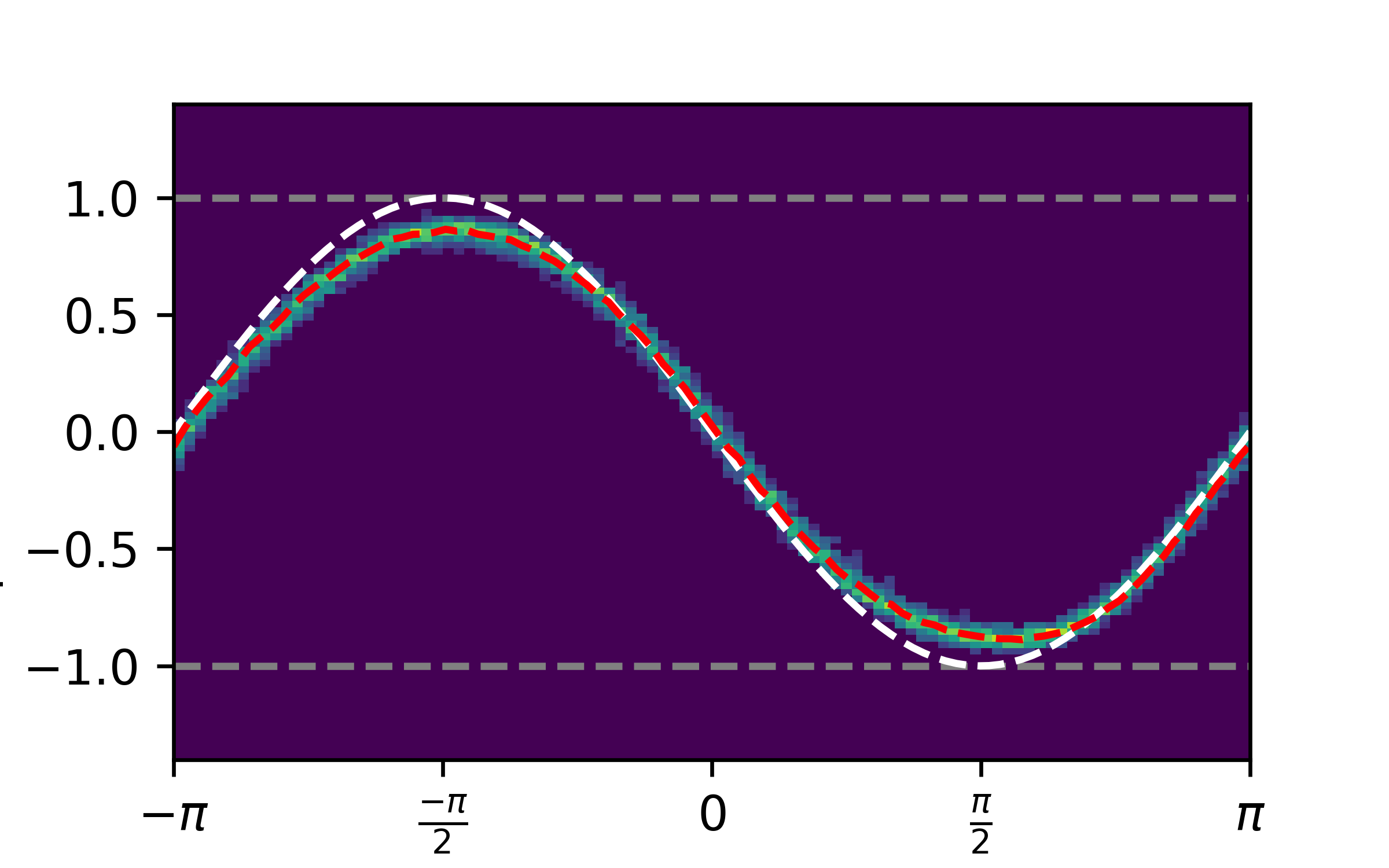} & \ivqe{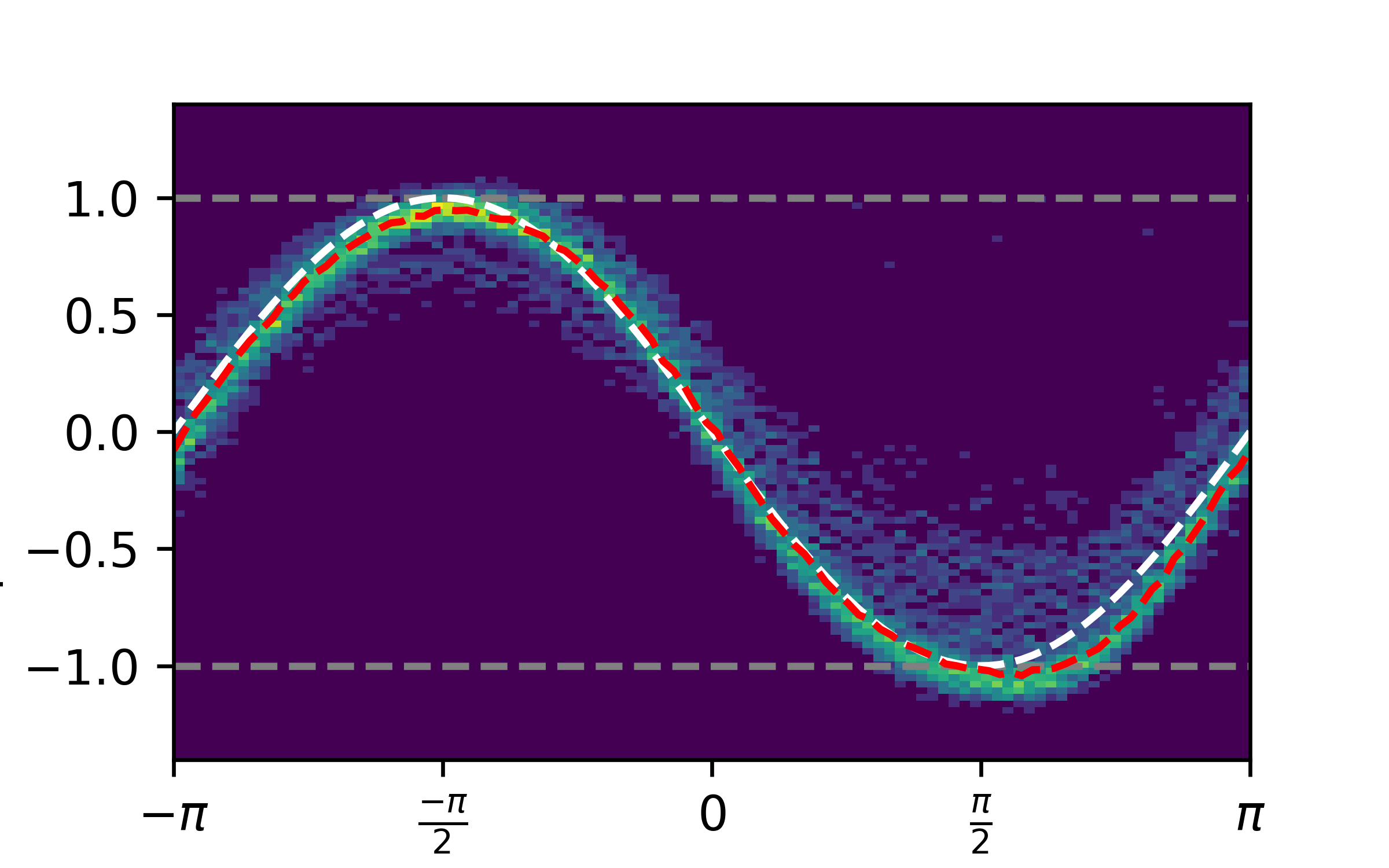}& \ivqe{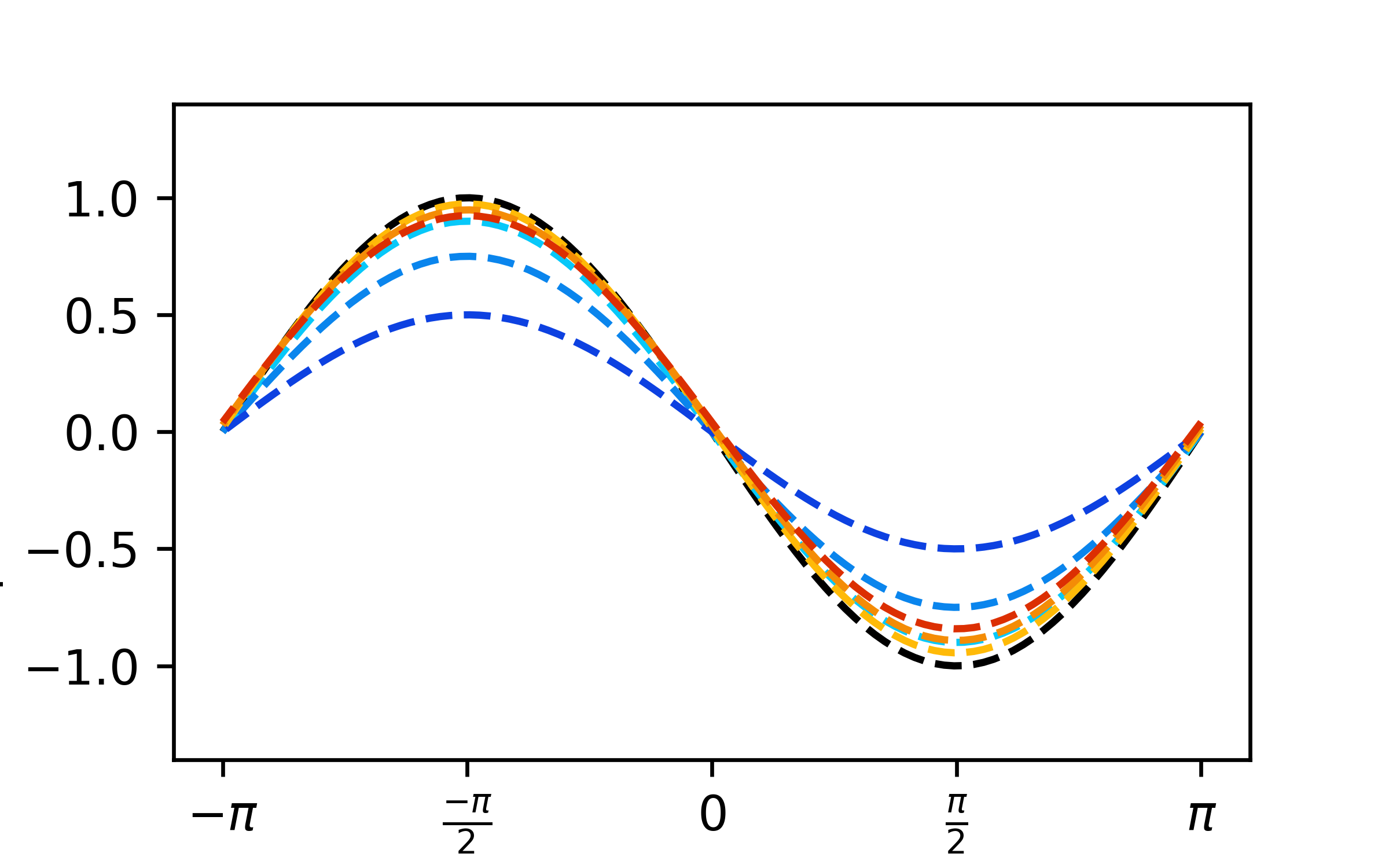} \\
    \end{tabular}
    \caption{Experimental results are shown for the expectation values of each Pauli operator as a function of the ansatz parameter in the VQE used to determine the binding energy of a hydrogen molecule. The VQE ansatz is implemented using the gate sequence shown in Fig. \ref{fig:VQE_circuit}. The white dashed curve represents the theoretically expected ground truth obtained by simulating the ansatz classically.  The red curve represents the average expectation value of the distribution.  Column (a) shows the raw experimental data. In column (b), the experimental data has been processed to remove $50\%$ of the population using Gaussian distribution outlier detection to eliminate misclassification events. Column (c) shows the expectation values after applying assignment mitigation. Column (d) displays the simulated expectation values versus the parameter under the amplitude damping noise (red) and misclassification noise (blue). The colours from light to dark denote the strength of the error. The black line denotes the simulated expectation value without noise.} 
    \label{fig:vqe}
\end{figure*}

\begin{figure}[b]
    \centering
	\sidesubfloat[]{
    \includegraphics[width=.4\linewidth]{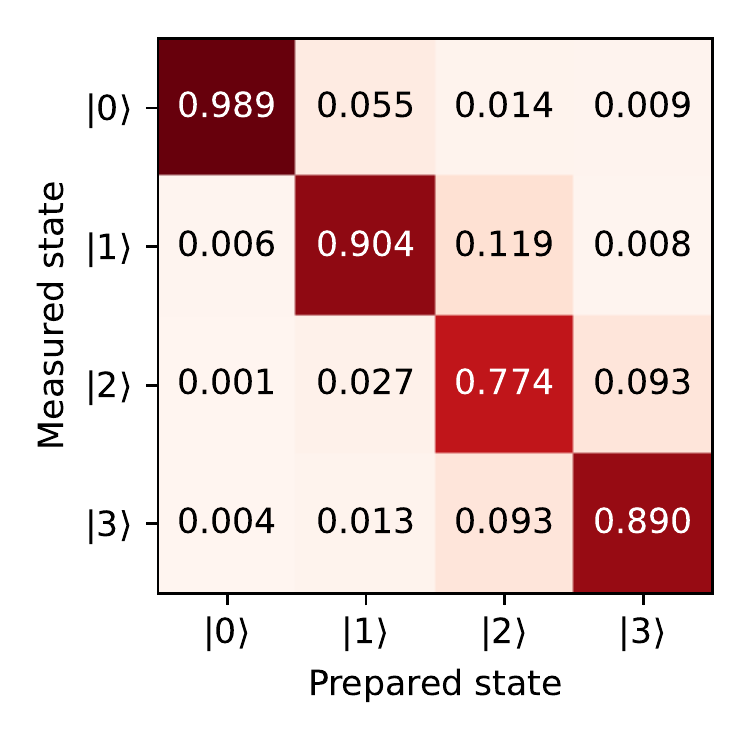}
	}
	\sidesubfloat[]{
    \includegraphics[width=.4\linewidth]{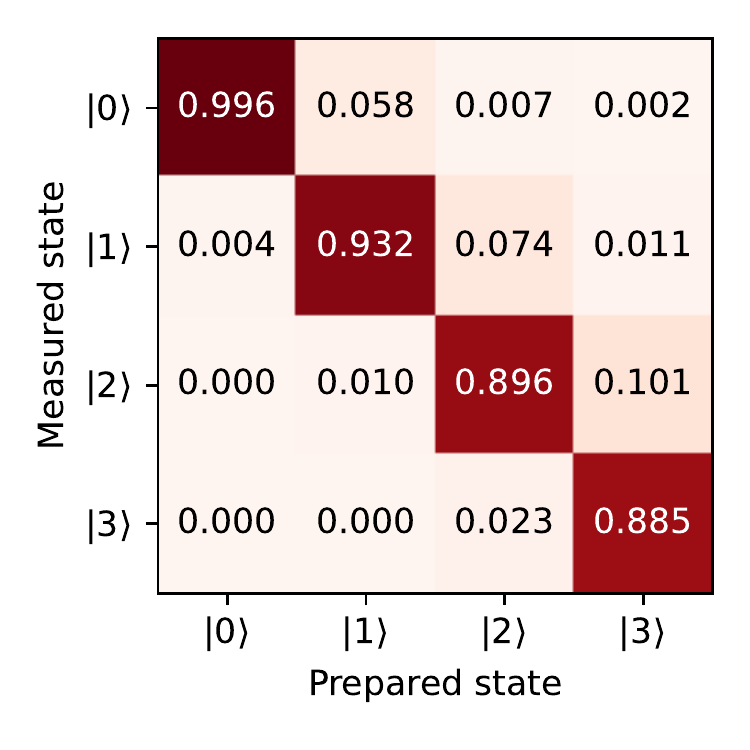}
	}\\
    \caption{The assignment matrix measured during the execution of VQE sequences. (a) The assignment matrix was measured from the experiment. \sx{(b) the simulated assignment matrix that only accounts for the T1 decay event. The measured assignment matrix was used to implement measurement mitigation.}
    }
    \label{fig:VQE_assignment_matrix}
\end{figure}

The Variational quantum eigensolver (VQE) is considered a promising application of near-term quantum devices \cite{Cerezo2021VariationalAlgorithms, Tilly2021ThePractices}, which has been used to solve electronic structure problems such as the bonding energy of a hydrogen molecule. This particular problem requires a minimum of two qubits using the symmetry-conserving Bravyi-Kitaev encoding of the Hamiltonian \cite{OMalley2016ScalableEnergies,Kandala2017Hardware-efficientMagnets,Tapering-off-Hamiltonians}. As an evaluation of the practicability of the emulator, we use a 4-level qudit system to emulate this two-qubit system and demonstrate VQE on a qudit.

To solve electronic structure problems, the VQE optimizes a wave function with a pre-assumed form, or "ansatz" $\ket{\Psi}$. The variational principle guarantees that the ground-state energy of the electronic system denoted as $E_0$, can be approximated by optimizing the parameters $\Vec{\theta}$ in the ansatz $\ket{\Psi(\Vec{\theta})}$. It can be written as $\bra{\Psi(\Vec{\theta})} H \ket{\Psi(\Vec{\theta})} \geq E_0$, where $\ket{\Psi(\Vec{\theta})}$ is the ansatz parametrized by $\Vec{\theta}$, and $H$ is the Hamiltonian. The wave function is typically generated using a parametrized quantum circuit, and in this case, we use the ansatz 
$\ket{\Psi(\Vec{\theta})} = e^{-i\theta XY}\ket{\Psi_{\mathrm{HF}}}$, which is suggested from the unitary coupled cluster theory~\cite{OMalley2016ScalableEnergies,Anand2022,Romero2018}. $\ket{\Psi_{\mathrm{HF}} = \ket{01}}$ is the Hartree-Fock state which is used as a starting point in this theory, and where it is assumed electrons occupy the lowest lying orbitals. 

\begin{figure}[]
    \centering
		\includegraphics[width=\linewidth]{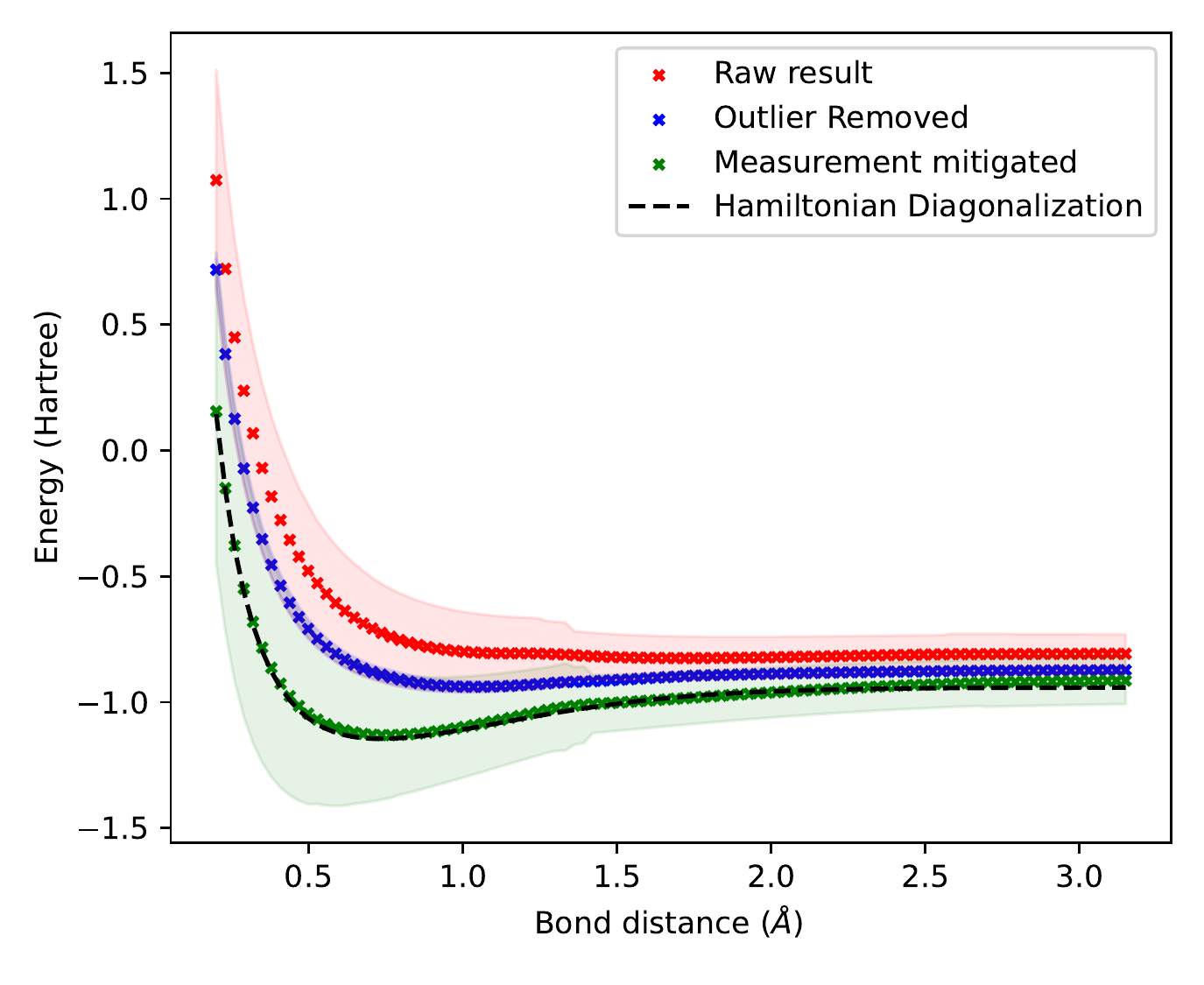}
    \caption{The energy of the hydrogen molecule as a function of bond distance, obtained by the variational quantum eigensolver (VQE) on a 4-level transmon. The crosses represent the average energy value from multiple experiments, while the coloured shadow indicates the error bar corresponding to one standard deviation. The error bar is evaluated by propagating the uncertainty of each Pauli operator expectation value to the evaluated energy. The black line represents the energy of hydrogen obtained by diagonalizing the Hamiltonian. }
    \label{fig:VQE_result}
\end{figure}

We implement the chosen ansatz on the two-qubit emulator using the gate sequence shown in Figure \ref{fig:VQE_circuit}. The energy of the hydrogen molecule is evaluated as $E(\Vec{\theta}) = \bra{\Psi(\Vec{\theta})}H\ket{\Psi(\Vec{\theta})} = \sum_i g_i \bra{\Psi(\Vec{\theta})}P_i\ket{\Psi(\Vec{\theta})}$, where $P_i \in \{IZ, ZI, ZZ, XX, YY\}$ and $g_i$ are coefficients that can be computed from the hydrogen bond distance $R$. The measurement of the quantum state gives the probability distribution of the state in $\ket{0}$, $\ket{1}$, $\ket{2}$, and $\ket{3}$, which we represent as a vector $R=\{p_{\ket{0}},p_{\ket{1}},p_{\ket{2}},p_{\ket{3}}\}$. The expectation value for an operator $M\in {IZ,ZI,ZZ}$ is evaluated as $\braket{M} = \mathrm{diag}(M)^T R$. The $\braket{XX}$ and $\braket{YY}$ are measured by swapping the $X$ or the $Y$ basis with $Z$ basis of both qubits, respectively, and measure $\braket{ZZ}$. There is only one parameter in the ansatz; in the experiment, the parameter is swept using 100 points from $-\pi$ to $\pi$. Each point is sampled 100 times, each time uses 500 shots to estimate the expectation values. The distribution of the estimated expectation value is displayed in the form of a heatmap in Figure \ref{fig:vqe}. The solved energy versus the hydrogen bond distance is given in Figure \ref{fig:VQE_result}.  

In comparison to the simulation results, the experimental data displays some noticeable differences. As shown in Figure \ref{fig:vqe}, we observe that the amplitude of the expectation value for all operators is lower than 1, whereas in the simulation it should be 1. Additionally, the $ZZ$ expectation value exhibits a larger value when the parameters are set to 0 compared to $\pi$, which is not the case in the simulation where it remains constant. Moreover, the distribution plot for each expectation value shows a "shadow," indicating that some of the measured distributions are not well-approximated by a simple Gaussian distribution. See  the supplementary materials 
for more details. These results suggest that the system is subject to a complex noise model. We consider two potential sources of error for this result: (1) amplitude damping error during the long measurement time, and (2) misassignment between the $\ket{2}$ and $\ket{3}$ states, as indicated by GST (see Figure \ref{fig:GST_SPAM}(d)). We use simulations to investigate the impact of these two types of noise and compare the simulated expectation value with the experiment result.

To address the misclassification event between the $\ket{2}$ and $\ket{3}$ states, we introduce a readout model $\Tilde{P}(\ket{2}) = (1-\epsilon)P(\ket{2}) + \epsilon P(\ket{3})$ and $\Tilde{P}(\ket{3}) = (1-\epsilon)P(\ket{3}) + \epsilon P(\ket{2})$, where $\Tilde{P}(\ket{i})$ is the measured probability for state $\ket{i}$ under the noise model. The blue lines in the simulation plots of Figure \ref{fig:vqe} represent the simulated expectation value versus parameter under the misclassification noise, with $\epsilon$ values of $0.1$, $0.25$, and $0.5$, where lighter colours indicate lower noise strength and darker colours indicate higher noise strength. The simulation results reveal that this misclassification event is likely the cause of the observed "shadow" in the distribution of the expectation values.

For the amplitude damping channel, we introduce a completely positive trace preserving (CPTP) channel operator $\mathcal{D}$ for a qudit to describe the amplitude damping error \cite{Nielsen2000QuantumInformation,Chessa2021}. Here we use $\mathcal{D}(\rho) = K_0\rho K_0^\dagger + \sum_{0\leq i < j < d-1} K_{ij} \rho K_{ij}^\dagger$, where $K_{ij} = \sqrt{\gamma_{ij}}\ket{i}\bra{j}, \forall i,j~\mathrm{s.t.}~0\leq i < j < d-1$ and $K_0 = \ket{0}\bra{0} + \sum_{1\leq j < d-1}\sqrt{1-\xi_j}\ket{j}\bra{j}$. $\gamma_{ji}$ has real value, describs the decay rate from the $j$-th to the $i$-th level ($\xi_j = \sum_{0\leq i < j < d-1}\gamma_{ji} \leq 1$). We performed simulations using the error model described above and set $\gamma_{ji} = 1 - \exp(-t \Gamma_{ji})$, where $1/\Gamma_{ji}$ is the effective $T_1$ for the $\{\ket{i},\ket{j}\}$ subspace and $t$ is the simulated waiting time for the decay of the system. The simulation results indicate that the distortion of the $ZZ$ expectation value is similar to the simulated value from the amplitude-damping channel. This suggests that the amplitude damping error during the measurement is likely to be the main source of error for the $ZZ$ expectation value.

To mitigate the effect of the misclassification error, a Gaussian fit is applied to the measured probability distribution to identify and remove the outliers that contribute to the ``shadow" in the distribution \footnote{We used the ``EllipticEnvelope'' outlier detector from the scikit-learn python package.}. We exclude $50\%$ of the population as outliers to visually eliminate all the ``shadow'', and show the remaining population in column (c) of Figure \ref{fig:vqe} (``Outlier removed"). The resulting expectation values, shown in blue in Fig. \ref{fig:VQE_result}, have significantly reduced error bars compared to the raw results. However, there is still a noticeable difference between the VQE result and the Hamiltonian diagonalization result.

To mitigate both the readout assignment error and the amplitude damping error during measurement, we applied measurement assignment mitigation using the assignment matrix $A$. This matrix describes the probability of measuring the state $\ket{j}$ when the system is prepared in state $\ket{i}$, and is used to obtain a mitigated probability distribution $\Tilde{R}$ by inverting the assignment matrix as $\Tilde{R} = A^{-1}R$. The mitigated expectation value was then evaluated as $\braket{\Tilde{M}} = diag(M)^\top  A^{-1} R$. The measured assignment matrix is shown in Fig. \ref{fig:VQE_assignment_matrix}(a). Notable misclassification between the $\ket{2}$ and $\ket{3}$ states is observed, consistent with data from gate-set tomography. \sx{For comparison, we present the simulated assignment matrix, which accounts only for the $\mathrm{T}_1$ decay event, as shown in Fig. \ref{fig:VQE_assignment_matrix}(b). This matrix is evaluated as $\Tilde{A} = (I-\Gamma^\top)^t$, where $t$ is set to $10$ microseconds (the measurement pulse length) to show a worst-case scenario.} \sx{The error-mitigated expectation values are presented in Fig.\ref{fig:vqe} (c). We observed that the mitigated expectation value of ZZ and ZI exceeds the physical bond, and we suspect this is due to the difference between the GMM (Gaussian Mixture Model) classifier used for evaluating the expectation value and the GMM classifier used to evaluate the assignment matrix, see supplementary materials for more details. The latter classifier is trained on a different dataset collected after the raw expectation values are evaluated. We employ the mitigated expectation value without a hard physical bond to evaluate the energy, and the} result is shown in green in Fig. \ref{fig:VQE_result}. By applying the measurement assignment mitigation technique, we are able to decrease the difference between the averaged solved energy and the Hamiltonian Diagonalization result to reach the chemical accuracy threshold of $1.5\times10^{-2}$ Hartree \cite{OMalley2016ScalableEnergies}. However, due to the large misclassification error, the error bar remains significant. It is worth noting that the assignment matrix already captures both misclassification errors and the amplitude damping error, the result would be overcorrected if we apply both assignment mitigation and outlier removal. 

\section{Discussion}

This work aims to explore the potential of higher levels in a transmon as a computational space for near-term quantum applications. By introducing higher levels, we increase the Hilbert space available for quantum algorithms, which can lead to more efficient and powerful algorithms. 
In this work, we demonstrate a high-coherence transmon device as a 2-qubit emulator to address these challenges.

We conduct detailed characterizations and benchmarking on both the hardware itself and the emulator, including active reset to prepare a high-fidelity initial state. We then implement variational quantum algorithms on the qudit device, confirming that charge noise at higher levels is not detrimental to the algorithm. The main source of error is the decay that occurs during the measurement process. We then mitigate this error with post-processing techniques and obtain a solution for the energy of a Hydrogen molecule with an error within chemical accuracy. 

\sx{This study has focused on developing an emulator on a superconducting qudit. A  future research direction is to implement entangling gates between ququart transmons. This direction could build upon earlier research that developed qutrit entangling gates \cite{PhysRevX.11.021010, Goss2022}. The next step would be to determine the most effective compilation method for these emulators, and demonstrate its universality.}  Additionally, further exploration of the impact of higher levels on the noise and error rates of the device could help improve its overall performance.

\begin{acknowledgments}
P.L.~acknowledges support from the EPSRC b. [EP/T001062/1, EP/N015118/1, EP/M013243/1]. M.B. acknowledges support from EPSRC QT Fellowship grant EP/W027992/1.  D.L. and I.R. acknowledge the support of the UK government department for Business, Energy and Industrial Strategy through the UK National Quantum Technologies Programme. We thank Yutian Wen for his insightful discussion and thank Jules Tilly for reviewing the manuscript and providing constructive feedback. \sx{We acknowleges the use of scQubits library \cite{Groszkowski2021scqubitspython}.} The authors would like to acknowledge the use of the University of Oxford Advanced Research Computing (ARC) facility in carrying out this work \cite{richards_2015_22558}.
\end{acknowledgments}

\bibliography{refnc}

\appendix
\begin{widetext}
\section{\sx{Intrinsic SU(4) operations and v}irtual Z gates\label{app:vz}}

\sx{\paragraph*{Physical gates} Physical drives at the transition frequency between the $\ket{i}$ and the $\ket{j}$ states implement the X or Y rotation in a qubit-like subspace of the qudit. The parametrized SU(4) rotation operator for X-like and Y-like rotations are $X_{ij}(\theta) = exp(i\frac{\theta}{2}X_{ij})$ and $Y_{ij}(\theta) = exp(i\frac{\theta}{2}Y_{ij})$, respectively. The generators are shown as follows:}

\sx{
\begin{equation}
\begin{aligned}
    &X_{01} = \begin{pmatrix}
    0 & -i & 0 & 0\\ 
    i & 0 & 0 & 0\\ 
    0 & 0 & 1 & 0\\ 
    0 & 0 & 0 & 1
    \end{pmatrix}\qquad
    &X_{12} = \begin{pmatrix}
    1 & 0 & 0 & 0\\ 
    0 & 0 & -i & 0\\ 
    0 & i & 0 & 0\\ 
    0 & 0 & 0 & 1
    \end{pmatrix}\qquad
    &X_{23} = \begin{pmatrix}
    1 & 0 & 0 & 0\\ 
    0 & 1 & 0 & 0\\ 
    0 & 0 & 0 & -i\\ 
    0 & 0 & i & 0
    \end{pmatrix} \\
\end{aligned}
\label{eq:SU4_X_gates}
\end{equation}
}

\sx{
\begin{equation}
\begin{aligned}
    &Y_{01} = \begin{pmatrix}
    0 & 1 & 0 & 0\\ 
    1 & 0 & 0 & 0\\ 
    0 & 0 & 1 & 0\\ 
    0 & 0 & 0 & 1
    \end{pmatrix}\qquad
    &Y_{12} = \begin{pmatrix}
    1 & 0 & 0 & 0\\ 
    0 & 0 &1 & 0\\ 
    0 & 1 & 0 & 0\\ 
    0 & 0 & 0 & 1
    \end{pmatrix}\qquad
    &Y_{23} = \begin{pmatrix}
    1 & 0 & 0 & 0\\ 
    0 & 1 & 0 & 0\\ 
    0 & 0 & 0 & 1\\ 
    0 & 0 & 1 & 0
    \end{pmatrix} \\
\end{aligned}
\label{eq:SU4_Y_gates}
\end{equation}
The pulses employ a 50 nanoseconds Blackman-DRAG envelope, and they are separated by a 10 nanoseconds buffer time.
}

\sx{\paragraph*{Virtual Z gates}} The Z rotation can be implemented virtually by shifting the phase of all gates in the rest of the sequence \cite{McKay2017EfficientComputing}. In this work we define the Z gate notation as follows:
\begin{equation}
\begin{aligned}
    &Z_1(\theta) = \begin{pmatrix}
    1 & 0 & 0 & 0\\ 
    0 & e^{i\theta} & 0 & 0\\ 
    0 & 0 & 1 & 0\\ 
    0 & 0 & 0 & 1
    \end{pmatrix} \qquad
    &Z_2(\theta) = \begin{pmatrix}
    1 & 0 & 0 & 0\\ 
    0 & 1 & 0 & 0\\ 
    0 & 0 & e^{i\theta} & 0\\ 
    0 & 0 & 0 & 1
    \end{pmatrix} \qquad
    &Z_3(\theta) = \begin{pmatrix}
    1 & 0 & 0 & 0\\ 
    0 & 1 & 0 & 0\\ 
    0 & 0 & 1 & 0\\ 
    0 & 0 & 0 & e^{i\theta}
    \end{pmatrix} \\
\end{aligned}
\label{eq:virtual_Z_gates}
\end{equation}

The implementation of the Z rotation for a multi-level system is with an extension of qubit virtual $Z$ gate. Suppose we would like to implement a Z gate in the following gate sequence.

\begin{equation}
    U = G_n G_{n-1} ... G_k Z_m(\theta) G_{k-1} ... G_{0} 
\end{equation}
where $G_k$ denotes a gate. Now insert an identity gate sequence $Z_m(\theta)Z_m^{-1}(\theta)$ between all the following gates, and we get 
\begin{equation}
    U = Z_m(\theta)Z_m^{-1}(\theta)G_n Z_m(\theta)Z_m^{-1}(\theta)... Z_m(\theta)Z_m^{-1}(\theta) G_k Z_m(\theta) G_{k-1} ... G_{0} 
\end{equation}
which is equivalent to the original gate sequence. Now we rewrite the $G_k^\prime = Z_m^{-1}(\theta)G_k Z_m(\theta)$ and get a new sequence 
\begin{equation}
    U = Z_m(\theta)G_n^\prime G_{n-1}^\prime ... G_k^\prime G_{k-1} ... G_{0} 
\end{equation}
where $G_k^\prime$ is implemented by shifting the phase of driving pulses. The last gate $Z^{-1}_m(\theta)$ will not make any difference if the state is measured with an operator commutes with $Z_m^{-1}(\theta)$, which is always the case for the standard dispersive readout. For a four-level qudit system, the virtual Z gate can be implemented as shown in table \ref{tab:chapter_3:virtual_z_phase_shift} .

\begin{table}[h]
\centering
    \begin{tabular}{llll}
              & $\{\ket{0},\ket{1}\} $ subspace & $\{\ket{1},\ket{2}\} $ subspace  & $\{\ket{2},\ket{3}\} $ subspace    \\\hline
$Z_1(\theta)$ & $-\theta$ & $\theta$ & 0\\
$Z_2(\theta)$ & $0$ & $-\theta$ & $\theta$\\
$Z_3(\theta)$ & $0$ & $0$       & $-\theta$ 
\end{tabular}
    \caption{Shifting the drive pulse sequence to implement multilevel transmon virtual Z gate}
    \label{tab:chapter_3:virtual_z_phase_shift}
\end{table}



\section{Dispersive readout for qudits \label{app:readout}}

The Jaynes-Cummings Hamiltonian in the dispersive limit can be used to describe the Hamiltonian of a qubit-resonator system with capacitive coupling~\cite{Bianchetti2009DynamicsElectrodynamics,Bianchetti2010ControlAtom}. The Hamiltonian takes the form:
\begin{equation}
    \mathcal{H} = \hbar \omega_r a^\dagger a + \sum_{k=0}^n \hbar\omega_k \ket{k}\bra{k} + \sum_{k=1}^{n-1} \hbar \chi_{k-1}\ket{k}\bra{k} + \sum_{j=0}^{n-1}\hbar s_j\ket{j}\bra{j}a^\dagger a
\end{equation}
where $g_j$ is the coupling strength to the transmon transition $j$ and $j+1$, and $\Delta_j$ is the detuning between the resonator frequency and the $j$ and $j+1$ transition frequency. The resonator exhibits a dispersive shift $s_j = -(\chi_j-\chi_{j-1})$ from interaction with the transmon in state $\ket{j}$. Here, we define $\chi$ as $\chi_1 = -s_1$ to maintain consistency with the conventions in most of the transmon literature. The transmon frequency also experiences a "Lamb shift" that depends on the photon number within the resonator.

\begin{figure}
    \centering
    \includegraphics[width=.8\textwidth]{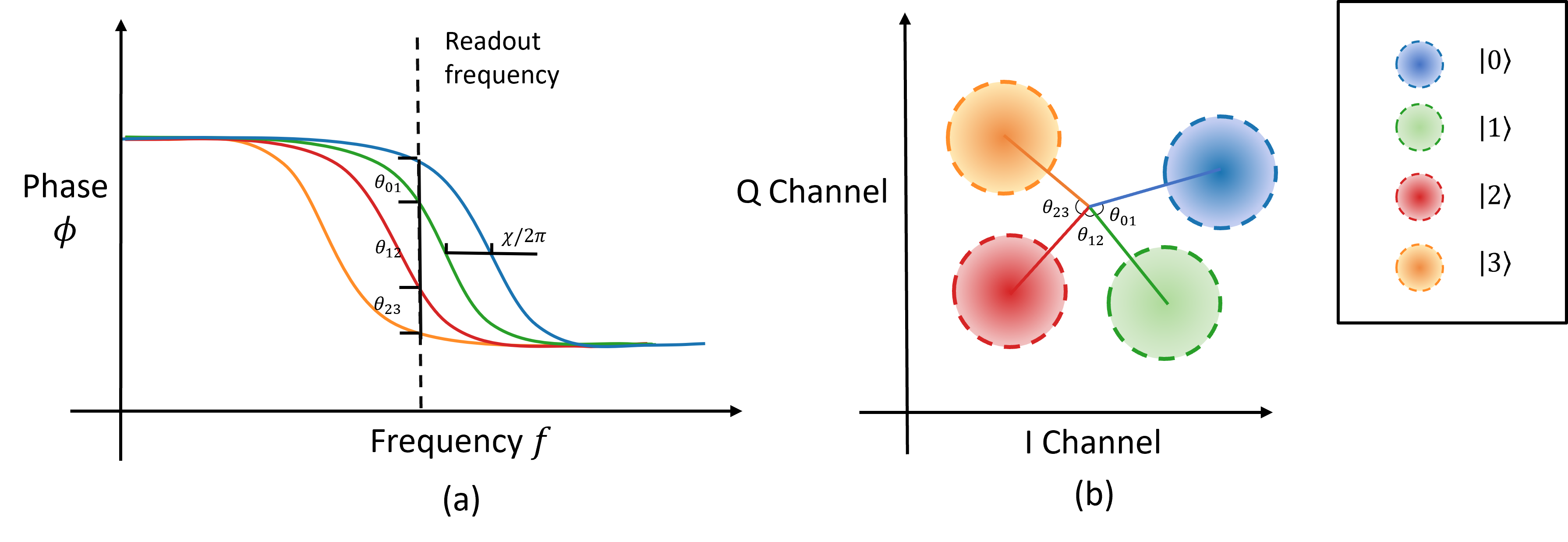}
    \caption{(a) plot shows the spectroscopy phase response of a resonator coupled to a qubit. The resonator oscillation frequency shifts depends on the transmon's states, resulting in different phase responses shown in different colors. To read all four levels simultaneously, a readout signal is sent to the resonator with a frequency that maximally distinguishes the phase response of all four different states. (b) The scatter plot shows the I channel signal and Q channel signal, often referred to as the "IQ plane." The integrated sum of the demodulated signal over time gives a point on the IQ plane. The four different states correspond to four different regions on the IQ plane, whose size is determined by the measurement noise. The separation angle $\theta_{ij}$ between region $i$ and $j$ is related to the frequency shift shown in (a).}
    \label{fig:dispersive_readout}
\end{figure}

The above analysis demonstrates that the resonance frequency of the coupled resonator is dependent on the state of the transmon qubit, allowing the transmon state to be inferred by probing the resonator's frequency. Since the dispersive shift is typically small, it is easier to distinguish the transmon state by the phase of the probing signal. A typical reflection response from a resonator is shown in Figure \ref{fig:dispersive_readout}. The resonator phase response is shifted in the left figure, and by choosing a center frequency that can distinguish more signals, a single pulse can distinguish multiple states. Typically, the device is designed to have $\chi = \kappa$ so that a readout frequency can be chosen to make $\theta_{01}=\pi$, maximizing the readout fidelity. However, for reading out four different states simultaneously, we would like to have $\theta_{01}\approx\theta_{12}\approx\theta_{23}\approx \pi / 2$, which can be approximately achieved by setting $\chi = \kappa/2$.

\section{Basic characterization \label{app:characterization}}

\begin{figure*}[!t]
    \centering
    \sidesubfloat[]{
        \includegraphics[width=.3\linewidth]{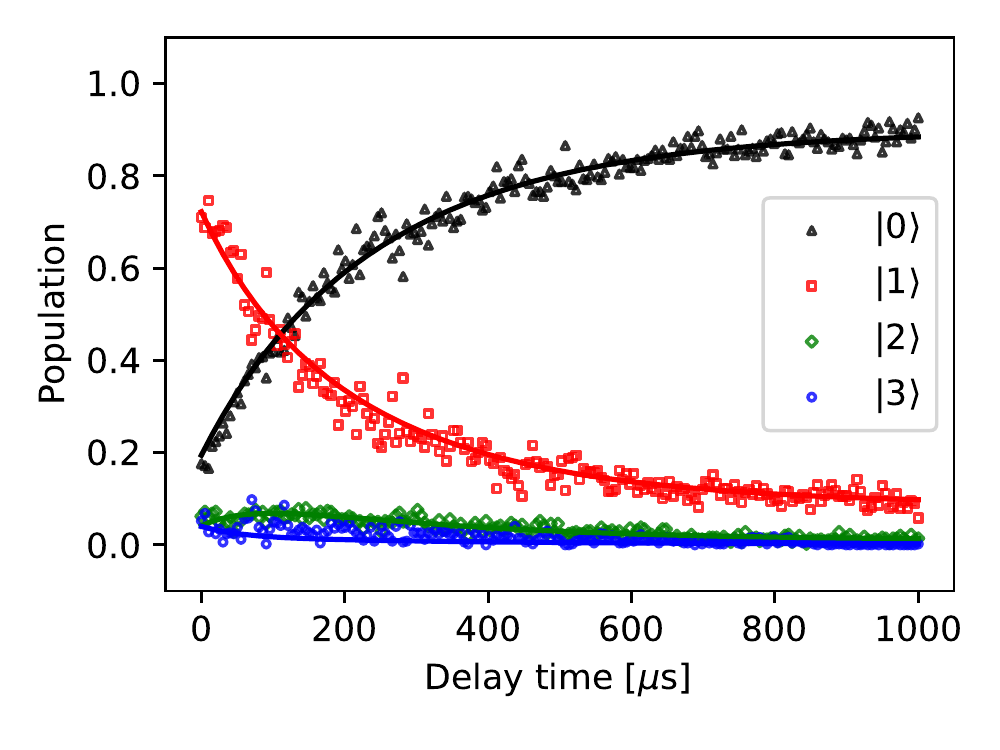}
	}
	\sidesubfloat[]{
    \includegraphics[width=.3\linewidth]{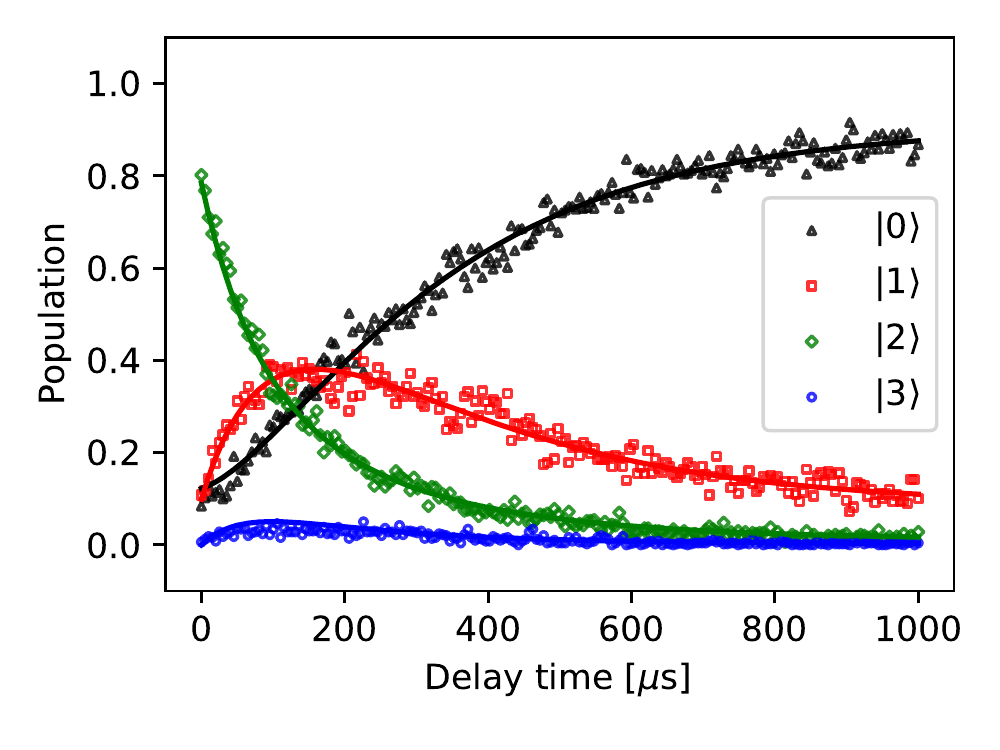}
	}
	\sidesubfloat[]{
    \includegraphics[width=.3\linewidth]{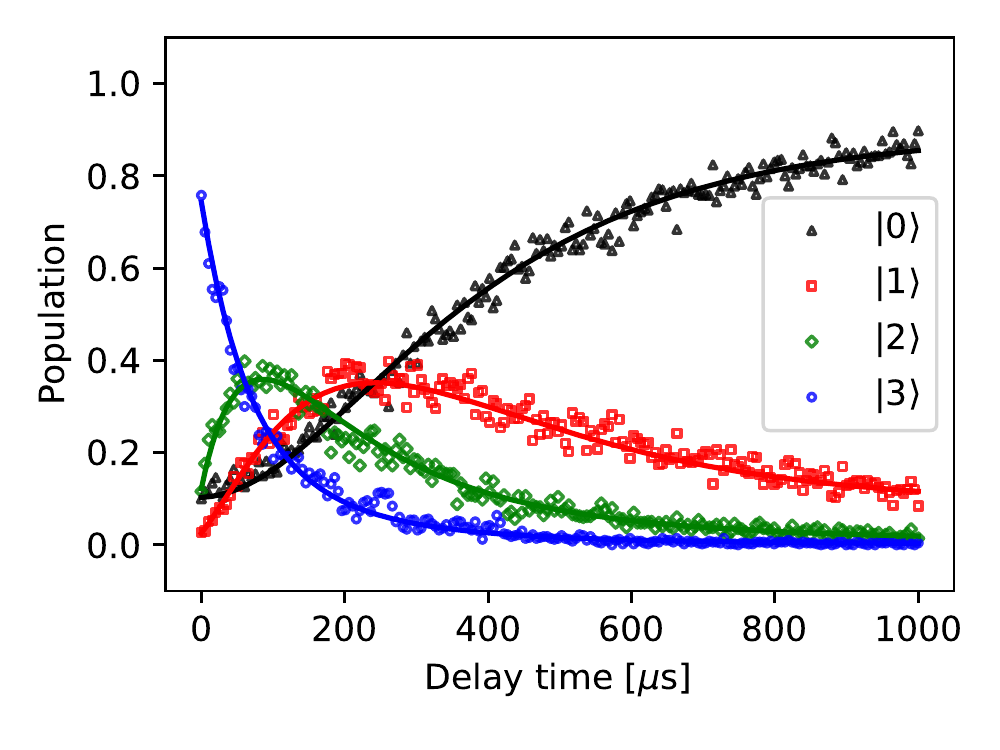}
	}\\
	\sidesubfloat[]{
        \includegraphics[width=.3\linewidth]{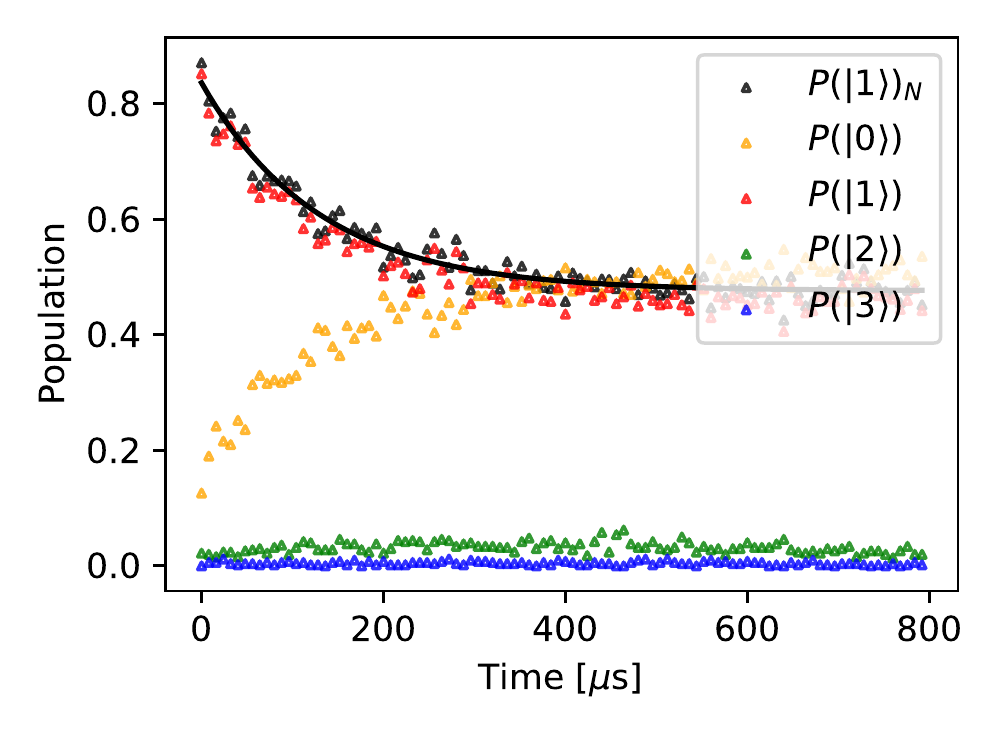}
	}
	\sidesubfloat[]{
        \includegraphics[width=.3\linewidth]{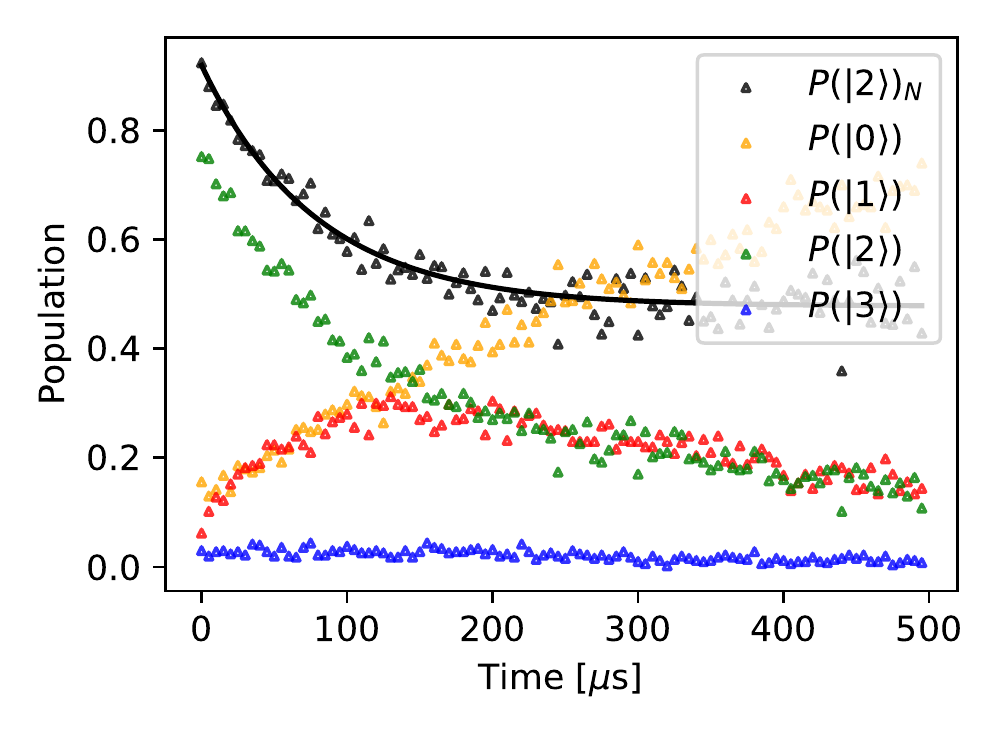}
	}
	\sidesubfloat[]{
        \includegraphics[width=.3\linewidth]{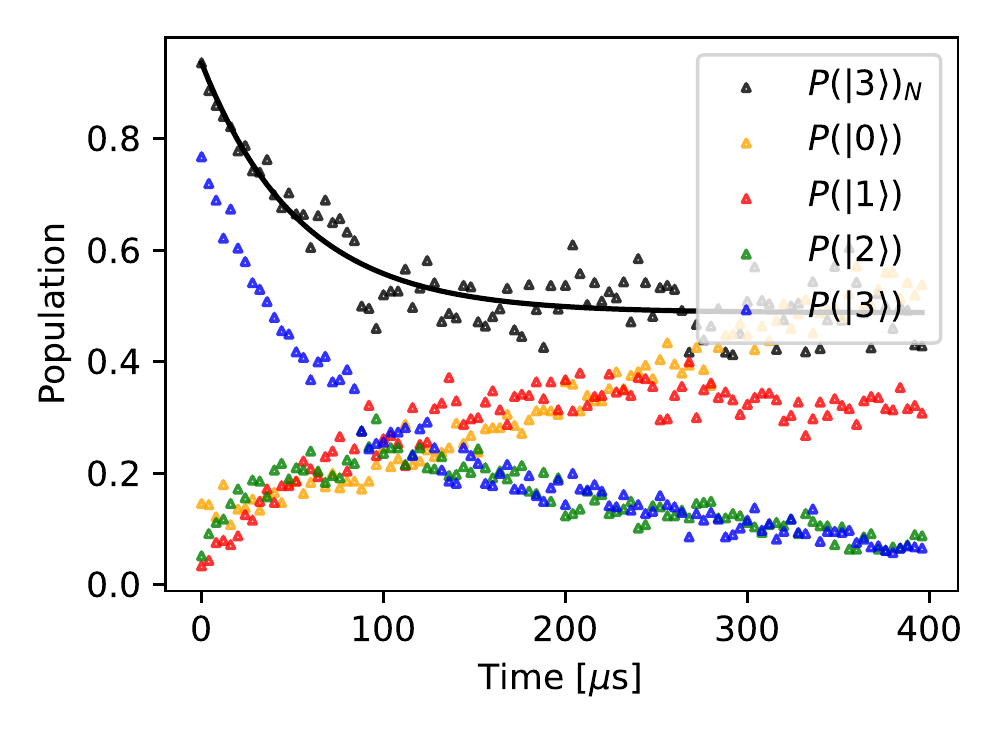}
	}\\
	\sidesubfloat[]{
        \includegraphics[width=.3\linewidth]{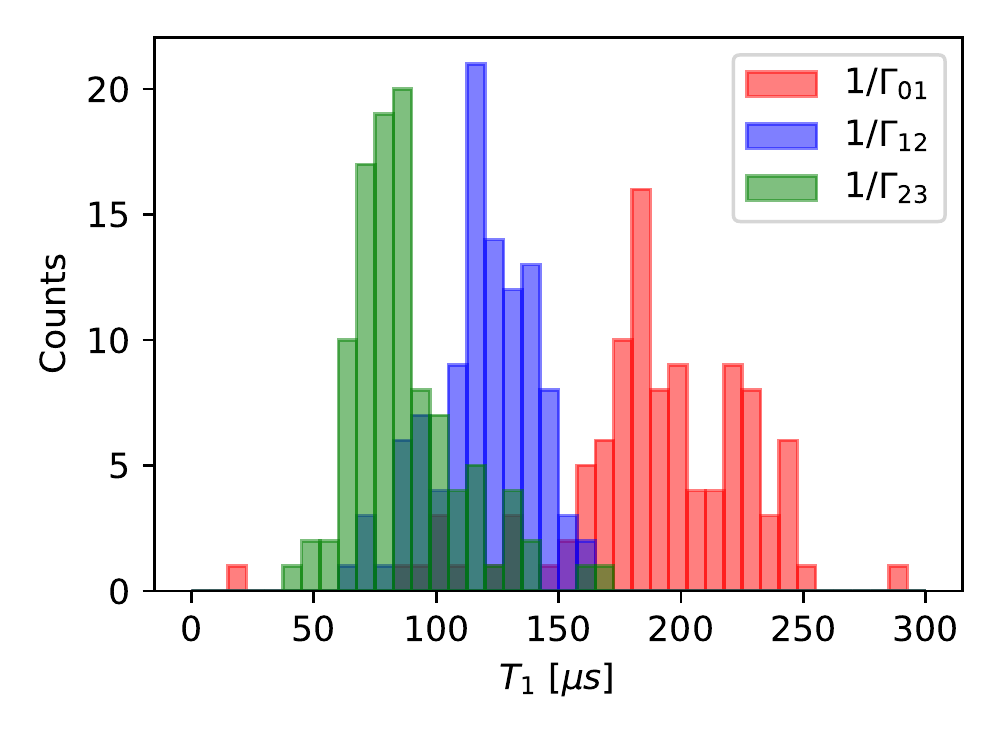}
	}
	\sidesubfloat[]{
        \includegraphics[width=.3\linewidth]{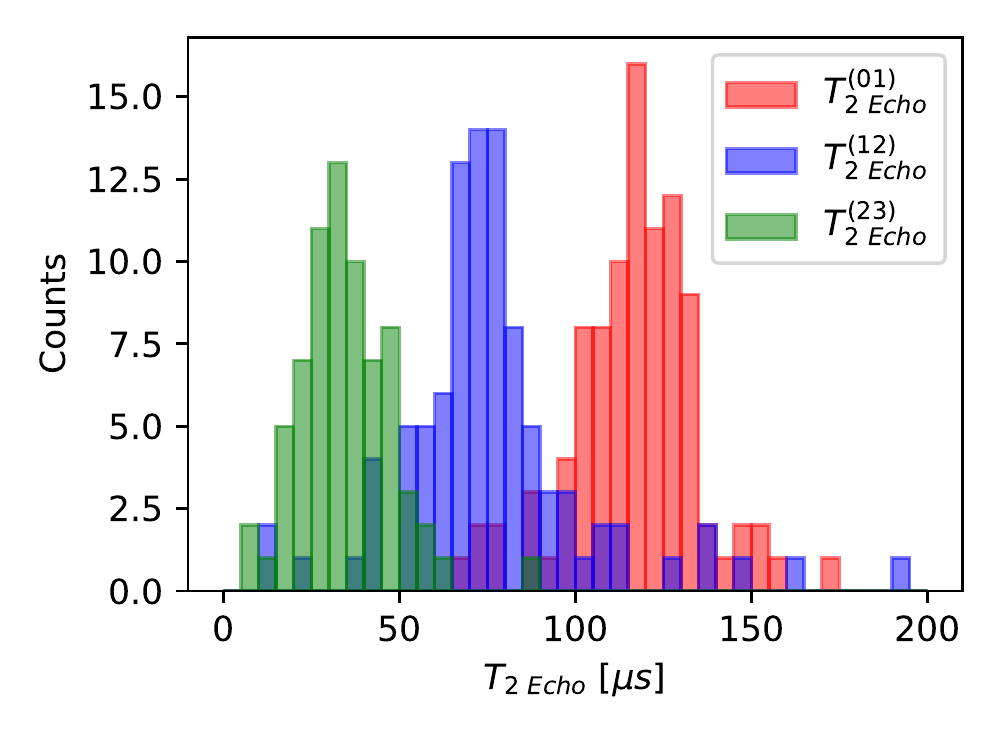}
	}
	\sidesubfloat[]{
        \includegraphics[width=.3\linewidth]{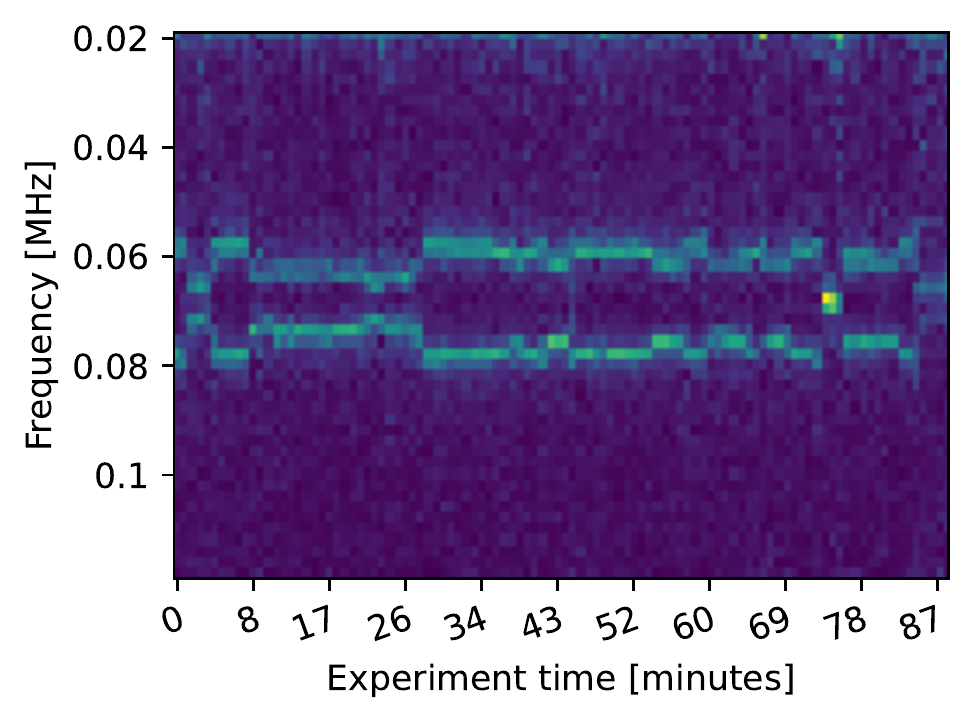}
	}
    \caption{(a) to (c) show the population in each state as a function of the waiting time, when the transmon is initially prepared in the $\ket{1}$, $\ket{2}$, $\ket{3}$ state, respectively. (d) to (f) shows a typical result Spin Echo experiment performed on the $\{\ket{0},\ket{1}\}$, $\{\ket{1},\ket{2}\}$, $\{\ket{2},\ket{3}\}$ subspaces, respectively. The black \sx{marker} in these figures represents the normalized success rate $P = P(\ket{i+1})/(P(\ket{i})+P(\ket{i+1}))$, which is used to reduce the impact of the decay of the transmon during the experiment. (g) and (h) show the distribution of the effective $T_1$ and $T_2$ echo times, respectively, with repeated measurements. These experiments are repeated for 100 times. (i) displays the Fourier spectrum of the trace of the Ramsey interferometry experiment performed on the $\{\ket{2},\ket{3}\}$ subspace over time. The frequency axis shows the measured frequency detuning from the transition drive frequency, while the experiment time axis denotes the time interval between the start of collecting the Ramsey trace and the start of the experiment. }
    \label{fig:characterization}
\end{figure*}

To model the spontaneous decay dynamics, a matrix $\Gamma$ is employed to capture the decay rate between all energy levels \cite{Peterer2015CoherenceQubit}. 
The state population as a function of waiting time is given by \sx{$\Vec{P}(t) = (I-\Gamma^\top)^t \Vec{P_0}$}, where $\Vec{P}(t)$ is the population of each state observed at time $t$, corresponding to the initial state population $\Vec{P_0}$. The effective $T_1^{(ij)} = 1/\Gamma_{ji}$ is then defined from the $\Gamma$ matrix, where $i$ and $j$ are the neighboring energy levels. \sx{The averaged value of the $\Gamma$ matrix over 100 measurements on this device is presented as follows:
\begin{equation}
\Gamma = \left[\begin{matrix}
0.00044 & -0.00044 & 0.0 & 0.0\\
-0.00599 & 0.00706 & -0.00108 & 0.0\\
-0.00055 & -0.00802 & 0.01112 & -0.00255\\
-0.00017 & -0.00078 & -0.0118 & 0.01222
\end{matrix}\right]
\end{equation}
The $T_1$ life time characterization of the device is by first evaluate the $T_1$ values for each individual measured $\Gamma$ matrix and then evaluate the statistics.} The transmon is reported to have $T_1^{(01)} = 189 \pm 41~\mathrm{\mu s}$, $T_1^{(12)} = 119 \pm 21~\mathrm{\mu s}$, and $T_1^{(23)} = 87 \pm 23~\mathrm{\mu s}$. To characterize the dephasing dynamics, spin echo experiments are performed on each neighbouring subspace to determine $T_2$, which describes the phase coherence of the subspace. The normalized survival population $P(\ket{i})N$ is defined as $P(\ket{i})_N = P(i)/(P(i) + P(i-1))$ to remove the effect of energy relaxation, where $P(\ket{i})$ is the population ratio measured in state $\ket{i}$ with $i>0$. $P(\ket{i})_N$ is fitted to $P(\ket{i})_N(t) = e^{-t/T_{2}^{(i)}} P(\ket{i})_N(0)$, where $T_{2}^{(i)}$ is the effective $T_2$ value for the $\{\ket{i},\ket{i-1}\}$ subspace. The experimentally determined transmon parameters are $T_{2}^{(01)}=118\pm 21  \mathrm{~\mu s}$, $T_{2}^{(12)}=76\pm 27 \mathrm{~\mu s}$, and $T_{2}^{(23)}=35\pm14  \mathrm{~\mu s}$.

The charge-noise-induced error on the higher levels can be a problem for executing quantum algorithms. To measure the sensitivity of the charge noise, we implement a Ramsey interferometry experiment on the ${\ket{2},\ket{3}}$ subspace, as it is the most sensitive subspace among all three neighbouring subspaces of the four lowest transmon levels \cite{Tennant2022Low-FrequencyTransmon, Martinez2022Noise-specificQubit}. Our results show that the frequency shift due to charge noise is around $20\mathrm{~kHz}$, which is significantly lower than the rabi rate of a single qudit pulse (which is \sx{$10$} MHz for a $50$ ns long $\pi$ pulse). This implies that the charge noise contribution to the error would not be detrimental to the implementation of quantum algorithms.

\section{Randomized benchmarking and gate set tomography\label{app:gst}}

\begin{figure}
    \centering
    \includegraphics[width=.7\linewidth]{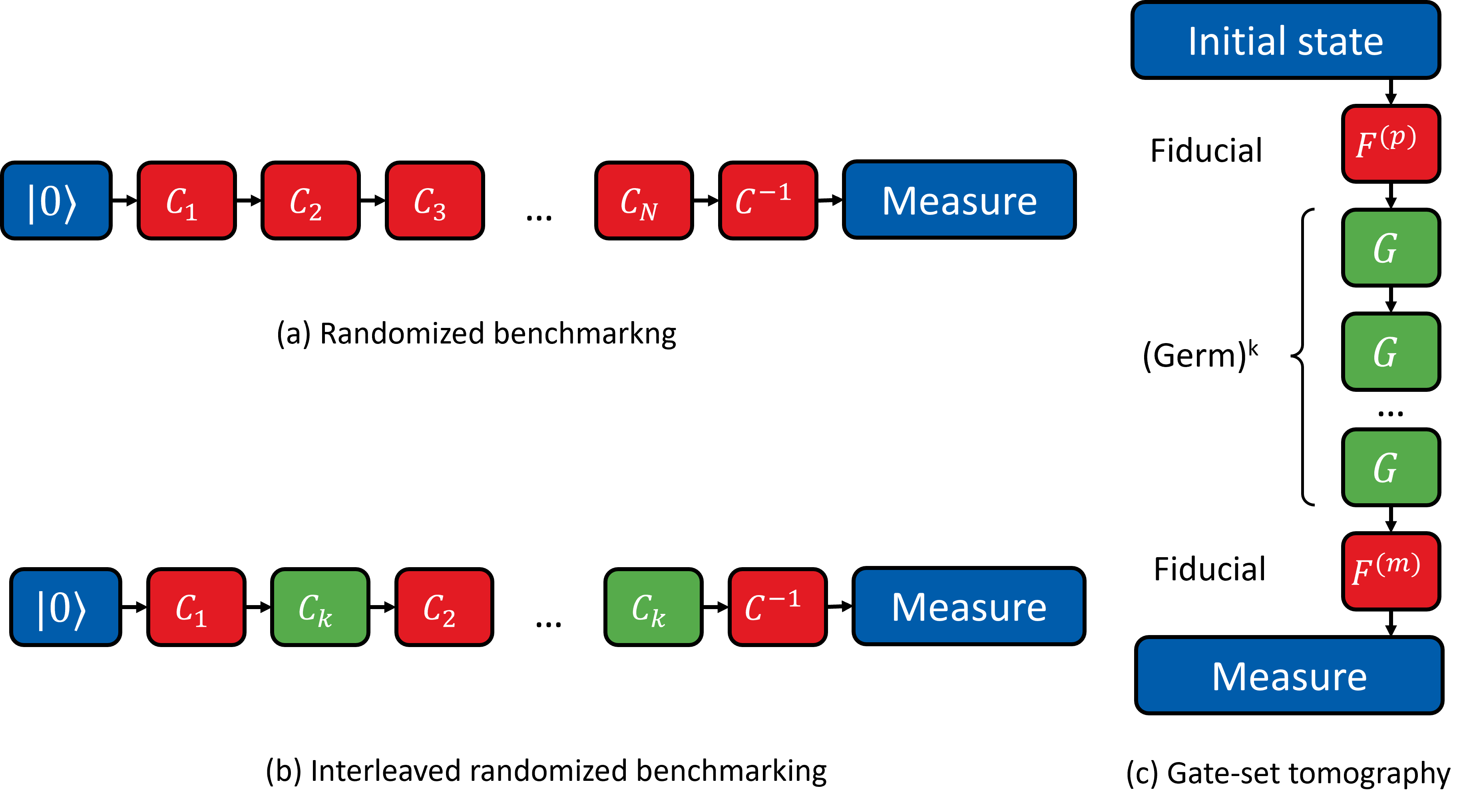}
    \caption{\sx{Experiment schemes for (a) randomized benchmarking, (b) interleaved randomized benchmarking (c) gate-set tomography. }}
    \label{fig:measurement_schemes}
\end{figure}

\sx{Randomized benchmarking (RB) and Gate Set Tomography (GST) are both protocols for Quantum Characterisation, Verification and Validation (QCVV). RB uses random sequences of Clifford groups to estimate the fidelity of gates. The experiment schemes are presented in Fig. \ref{fig:measurement_schemes} (a) and (b) for regular RB and interleaved RB.} GST performs process tomography for the full gate set, and estimates the state preparation and measurement (SPAM) error while reconstruct the CPTP map of the quantum operations~\cite{Greenbaum2015IntroductionTomography,Nielsen2021GateTomography}. The experiment scheme is shown in Fig.\ref{fig:measurement_schemes} (c).  To implement gate-set tomography on the qudit system, we need to define the Preparation and measurement fiducials to span the whole Hilbert space. The fiducials are simply the two-qubit fiducials mapped to the qudit native, gates, given in Tab.\ref{tab:qudit_fiducials}. 
The tomography result of the gate set is shown in Fig.\ref{tab:all_ptm_report}. 

\begin{table}[h]
\makeatletter\long\def\@ifdim#1#2#3{#2}\makeatother
\caption{\label{tab:qudit_fiducials} Fiducials for qudit gate set tomography. These fiducials are the two-qubit fiducials to mapped to the qudit, and can be straightforwardly implemented on the emulator. }
    \centering
        \begin{tabular}{ll}
        \hline        \hline
         \multicolumn{2}{c}{Preparation Fiducials ($F^{(p)}$)}          \rule{0pt}{3ex}    \\\hline
        1&$I$                     \\  
        2&$X_{01}(\frac{\pi}{2})$ \\ 
        3&$X_{01}(\frac{\pi}{2}).Z_{1}(\frac{\pi}{2})$ \\
        4&$X_{01}(\frac{\pi}{2}).X_{01}(\frac{\pi}{2})$ \\
        5&$X_{01}(\frac{\pi}{2}).X_{01}(\frac{\pi}{2}).X_{12}(\frac{\pi}{2})$ \\
        6&$X_{01}(\frac{\pi}{2}).X_{01}(\frac{\pi}{2}).X_{12}(\frac{\pi}{2}).Z_{2}(\frac{\pi}{2})$ \\
        7&$X_{01}(\frac{\pi}{2}).X_{12}(\frac{\pi}{2}).X_{12}(\frac{\pi}{2})$ \\
        8&$X_{01}(\frac{\pi}{2}).Z_{1}(\frac{\pi}{2}).X_{12}(\frac{\pi}{2}).X_{12}(\frac{\pi}{2})$ \\
        9&$X_{01}(\frac{\pi}{2}).X_{01}(\frac{\pi}{2}).X_{12}(\frac{\pi}{2}).X_{12}(\frac{\pi}{2})$ \\
        10&$X_{01}(\frac{\pi}{2}).X_{12}(\frac{\pi}{2}).X_{12}(\frac{\pi}{2}).X_{23}(\frac{\pi}{2}).X_{23}(\frac{\pi}{2})$                     \\  
        11&$X_{01}(\frac{\pi}{2}).Z_{1}(\frac{\pi}{2}).X_{12}(\frac{\pi}{2}).X_{12}(\frac{\pi}{2}).X_{23}(\frac{\pi}{2}).X_{23}(\frac{\pi}{2})$                     \\  
        12&$X_{01}(\frac{\pi}{2}).X_{01}(\frac{\pi}{2}).X_{12}(\frac{\pi}{2}).X_{23}(\frac{\pi}{2}).X_{23}(\frac{\pi}{2})$                     \\  
        13&$X_{01}(\frac{\pi}{2}).X_{01}(\frac{\pi}{2}).X_{12}(\frac{\pi}{2}).Z_{2}(\frac{\pi}{2}).X_{23}(\frac{\pi}{2}).X_{23}(\frac{\pi}{2})$                     \\  
        14&$X_{01}(\frac{\pi}{2}).X_{01}(\frac{\pi}{2}).X_{12}(\frac{\pi}{2}).X_{12}(\frac{\pi}{2}).X_{23}(\frac{\pi}{2})$                     \\  
        15&$X_{01}(\frac{\pi}{2}).X_{01}(\frac{\pi}{2}).X_{12}(\frac{\pi}{2}).X_{12}(\frac{\pi}{2}).X_{23}(\frac{\pi}{2}).Z_{3}(\frac{\pi}{2})$                     \\  
        16&$X_{01}(\frac{\pi}{2}).X_{01}(\frac{\pi}{2}).X_{12}(\frac{\pi}{2}).X_{12}(\frac{\pi}{2}).X_{23}(\frac{\pi}{2}).X_{23}(\frac{\pi}{2})$                     \\

        \hline        \hline
         \multicolumn{2}{c}{Measurement Fiducials($F^{(m)}$)}    \rule{0pt}{3ex}\\        \hline
        1&$I$              \\
        2&$Z_{1}(\frac{\pi}{2}).Z_{3}(\frac{\pi}{2}).X_{01}(\frac{\pi}{2}).X_{23}(\frac{\pi}{2})$                     \\  
        3&$X_{01}(\frac{\pi}{2}).X_{23}(\frac{\pi}{2})$ \\ 
        4&$X_{12}(\frac{\pi}{2}).X_{12}(\frac{\pi}{2}).Z_{1}(\frac{\pi}{2}).Z_{2}(\frac{\pi}{2}).Z_{1}(\frac{\pi}{2}).Z_{3}(\frac{\pi}{2}).X_{01}(\frac{\pi}{2}).X_{23}(\frac{\pi}{2}).$\\
          &$X_{12}(\frac{\pi}{2}).X_{12}(\frac{\pi}{2}).Z_{1}(\frac{\pi}{2}).Z_{2}(\frac{\pi}{2})$	\\
        5&$X_{12}(\frac{\pi}{2}).X_{12}(\frac{\pi}{2}).Z_{1}(\frac{\pi}{2}).Z_{2}(\frac{\pi}{2}).Z_{1}(\frac{\pi}{2}).Z_{3}(\frac{\pi}{2}).X_{01}(\frac{\pi}{2}).X_{23}(\frac{\pi}{2}).$\\
        &$X_{12}(\frac{\pi}{2}).X_{12}(\frac{\pi}{2}).Z_{1}(\frac{\pi}{2}).Z_{2}(\frac{\pi}{2}).Z_{1}(\frac{\pi}{2}).Z_{3}(\frac{\pi}{2}).X_{01}(\frac{\pi}{2}).X_{23}(\frac{\pi}{2})$	\\
        6 & $X_{12}(\frac{\pi}{2}).X_{12}(\frac{\pi}{2}).Z_{1}(\frac{\pi}{2}).Z_{2}(\frac{\pi}{2}).Z_{1}(\frac{\pi}{2}).Z_{3}(\frac{\pi}{2}).X_{01}(\frac{\pi}{2}).X_{23}(\frac{\pi}{2}).$\\
        &$X_{12}(\frac{\pi}{2}).X_{12}(\frac{\pi}{2}).Z_{1}(\frac{\pi}{2}).Z_{2}(\frac{\pi}{2}).X_{01}(\frac{\pi}{2}).X_{23}(\frac{\pi}{2})$\\
        7 & $X_{12}(\frac{\pi}{2}).X_{12}(\frac{\pi}{2}).Z_{1}(\frac{\pi}{2}).Z_{2}(\frac{\pi}{2}).X_{01}(\frac{\pi}{2}).X_{23}(\frac{\pi}{2}).X_{12}(\frac{\pi}{2}).X_{12}(\frac{\pi}{2}).$\\
        &$Z_{1}(\frac{\pi}{2}).Z_{2}(\frac{\pi}{2})$\\
        8 & $X_{12}(\frac{\pi}{2}).X_{12}(\frac{\pi}{2}).Z_{1}(\frac{\pi}{2}).Z_{2}(\frac{\pi}{2}).X_{01}(\frac{\pi}{2}).X_{23}(\frac{\pi}{2}).X_{12}(\frac{\pi}{2}).X_{12}(\frac{\pi}{2}).$\\
        &$Z_{1}(\frac{\pi}{2}).Z_{2}(\frac{\pi}{2}).Z_{1}(\frac{\pi}{2}).Z_{3}(\frac{\pi}{2}).X_{01}(\frac{\pi}{2}).X_{23}(\frac{\pi}{2})$ \\
        9 & $X_{12}(\frac{\pi}{2}).X_{12}(\frac{\pi}{2}).Z_{1}(\frac{\pi}{2}).Z_{2}(\frac{\pi}{2}).X_{01}(\frac{\pi}{2}).X_{23}(\frac{\pi}{2}).X_{12}(\frac{\pi}{2}).X_{12}(\frac{\pi}{2}).$\\
        &$Z_{1}(\frac{\pi}{2}).Z_{2}(\frac{\pi}{2}).X_{01}(\frac{\pi}{2}).X_{23}(\frac{\pi}{2})$ \\
        \hline        \hline
        \end{tabular}
\end{table}

We found the estimated gate fidelity for the $\{\ket{0},\ket{1}\}$ subspace is lower than $\{\ket{1},\ket{2}\}$ and $\{\ket{2},\ket{3}\}$ subspace, despite it has better coherence. The control pulses RF signal is generated by mixing an IF signal generated by a 2 Gsps DAC and a fixed LO signal. The LO frequency is selected as 3.904 GHz to cover the transition frequencies of the qubit states $\{\ket{0},\ket{1}\}$, $\{\ket{1},\ket{2}\}$, and $\{\ket{2},\ket{3}\}$. This choice of LO frequency results in an IF frequency of 230 MHz for the $\{\ket{0},\ket{1}\}$ transition, which is at the upper limit of our DAC's frequency bandwidth. Therefore, we hypothesize that this could be a contributing factor to the lower gate fidelity observed in the $\{\ket{0},\ket{1}\}$ subspace compared to the $\{\ket{1},\ket{2}\}$ and $\{\ket{2},\ket{3}\}$ subspace.

\newcommand{\iptm}[1]{\begin{tabular}{c} \includegraphics[width=4.5cm,height=4.5cm]{#1}\end{tabular}}

\begin{figure}[h]
    \centering
    \begin{tabular}{c|c|c|c}
         Gate name & $X_{01}(\frac{\pi}{2})$ & $X_{12}(\frac{\pi}{2})$ & $X_{23}(\frac{\pi}{2})$  \\ &&& \\ 
         \begin{tabular}{c}Average\\infidelity\end{tabular}  & $0.00498(11)$ & $0.002966(52)$ & $0.002906(68)$\\&&& \\
         \begin{tabular}{c}PTM\\\end{tabular} & \iptm{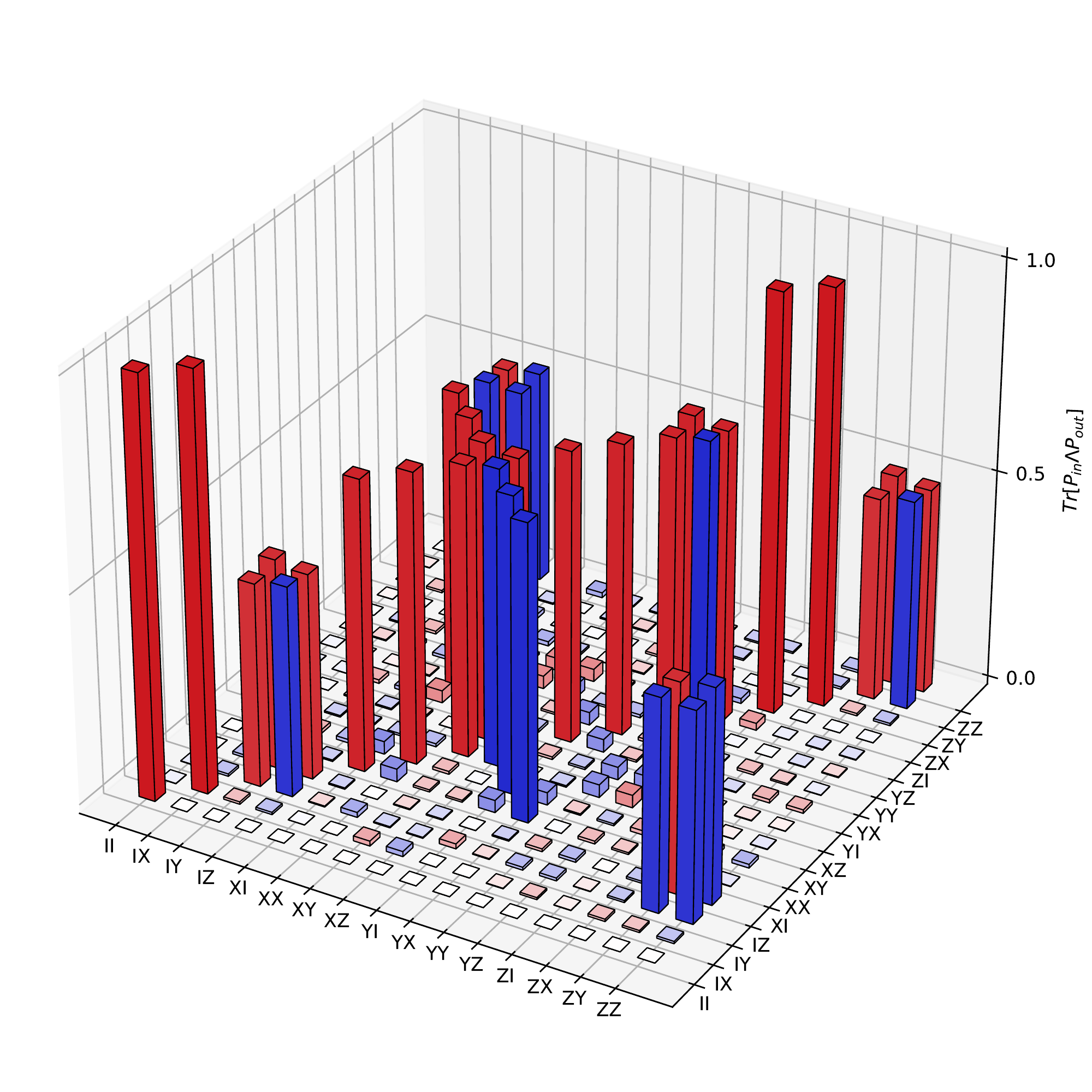} & \iptm{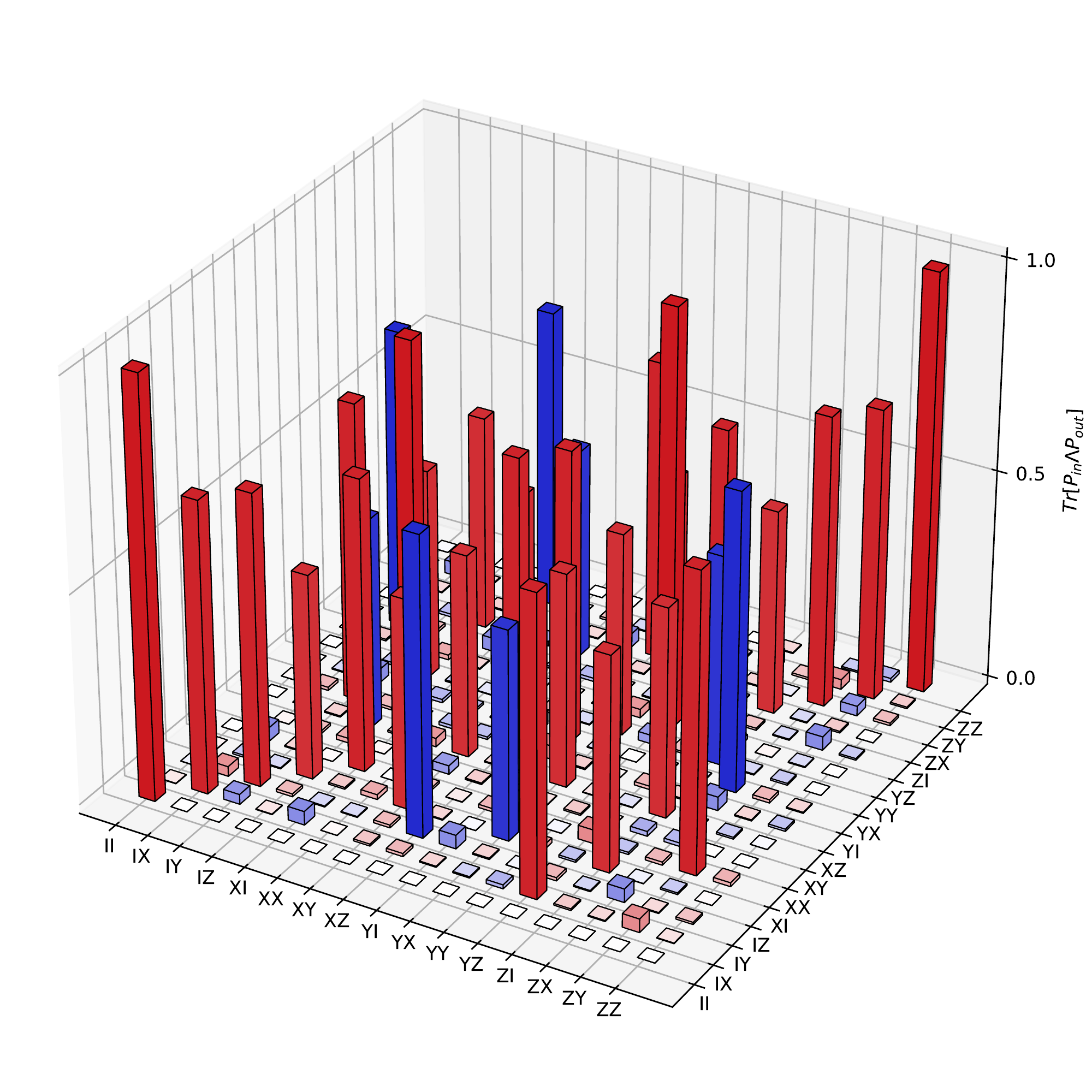} & \iptm{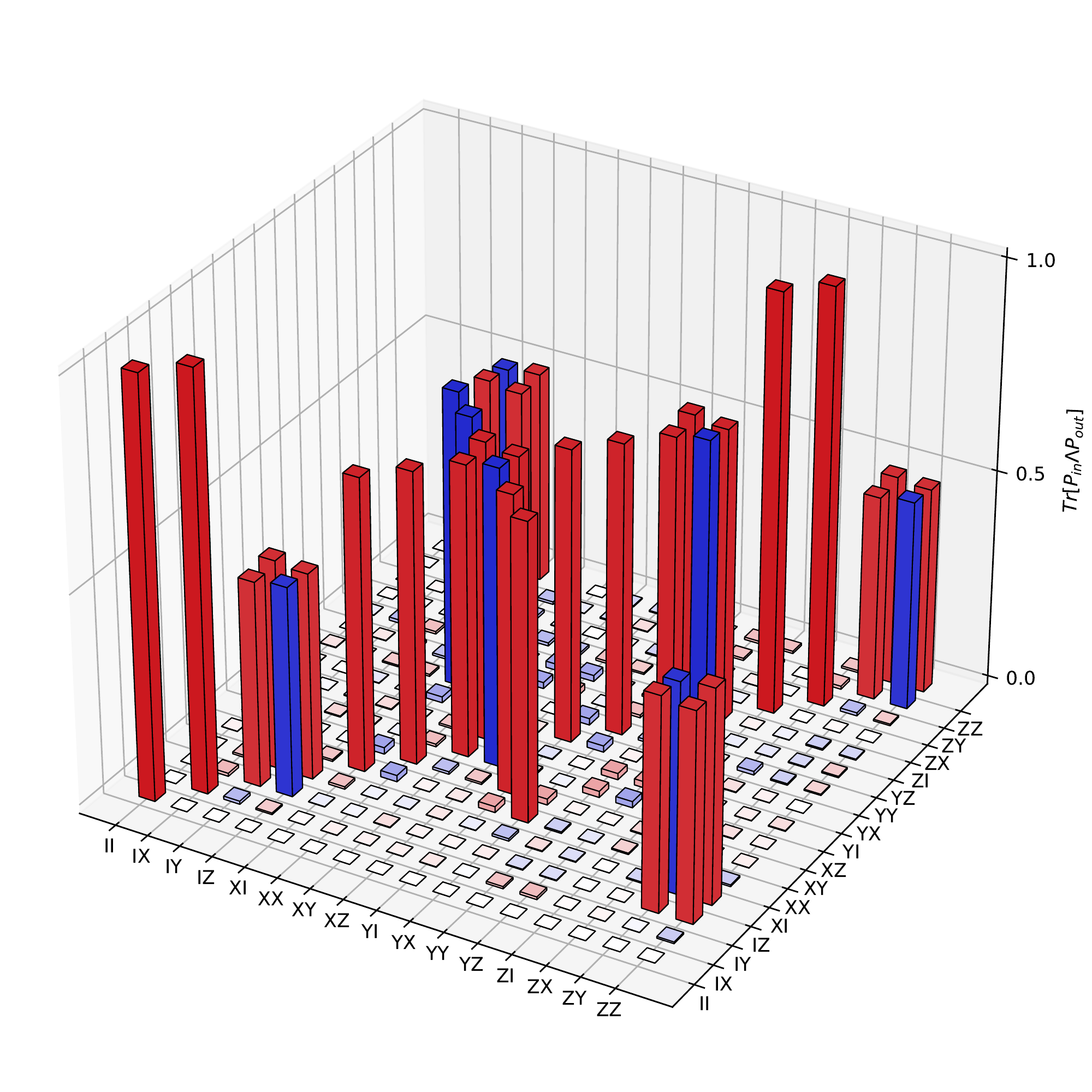} \\ &&&\\
         \begin{tabular}{c}Error\\generator\end{tabular} & \iptm{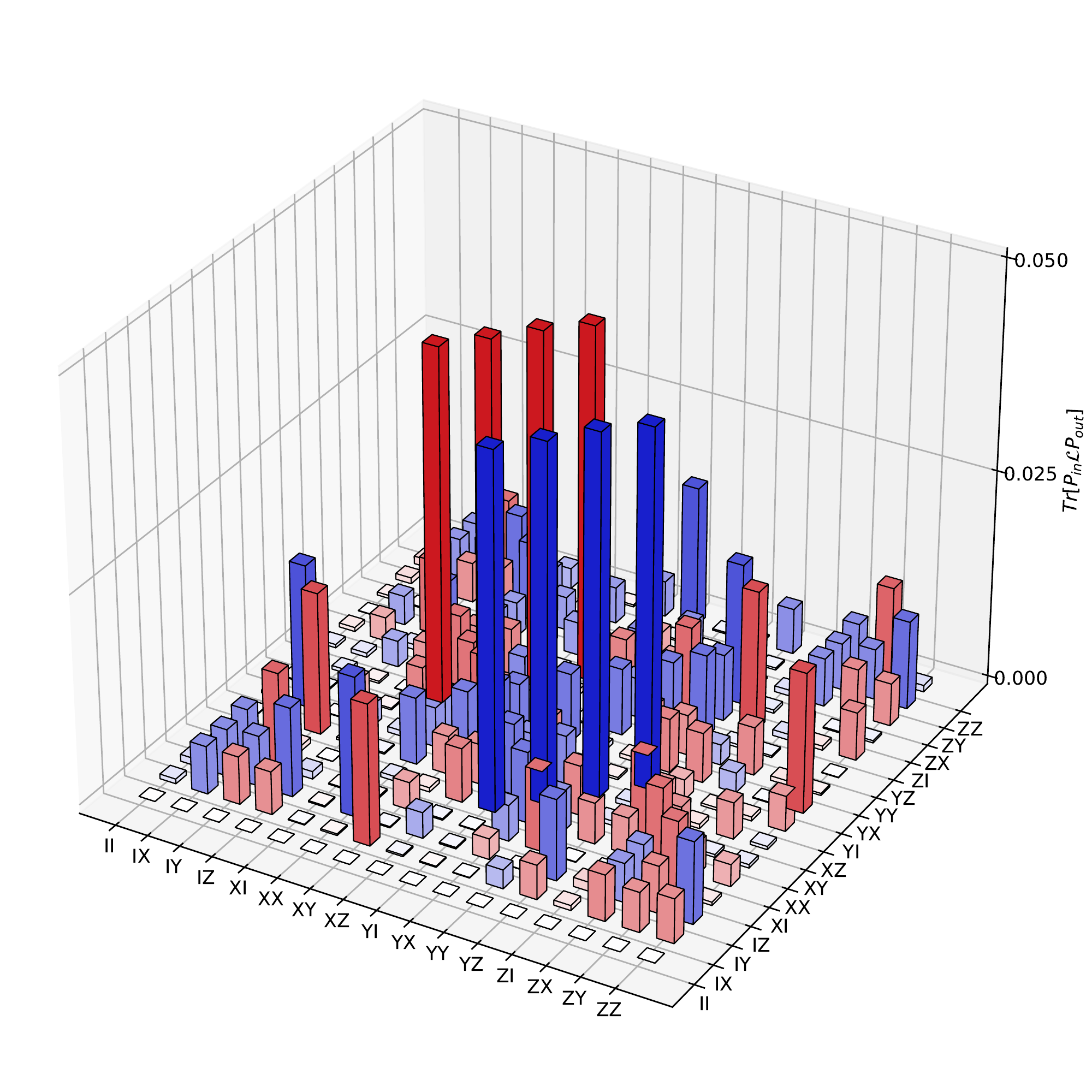} & \iptm{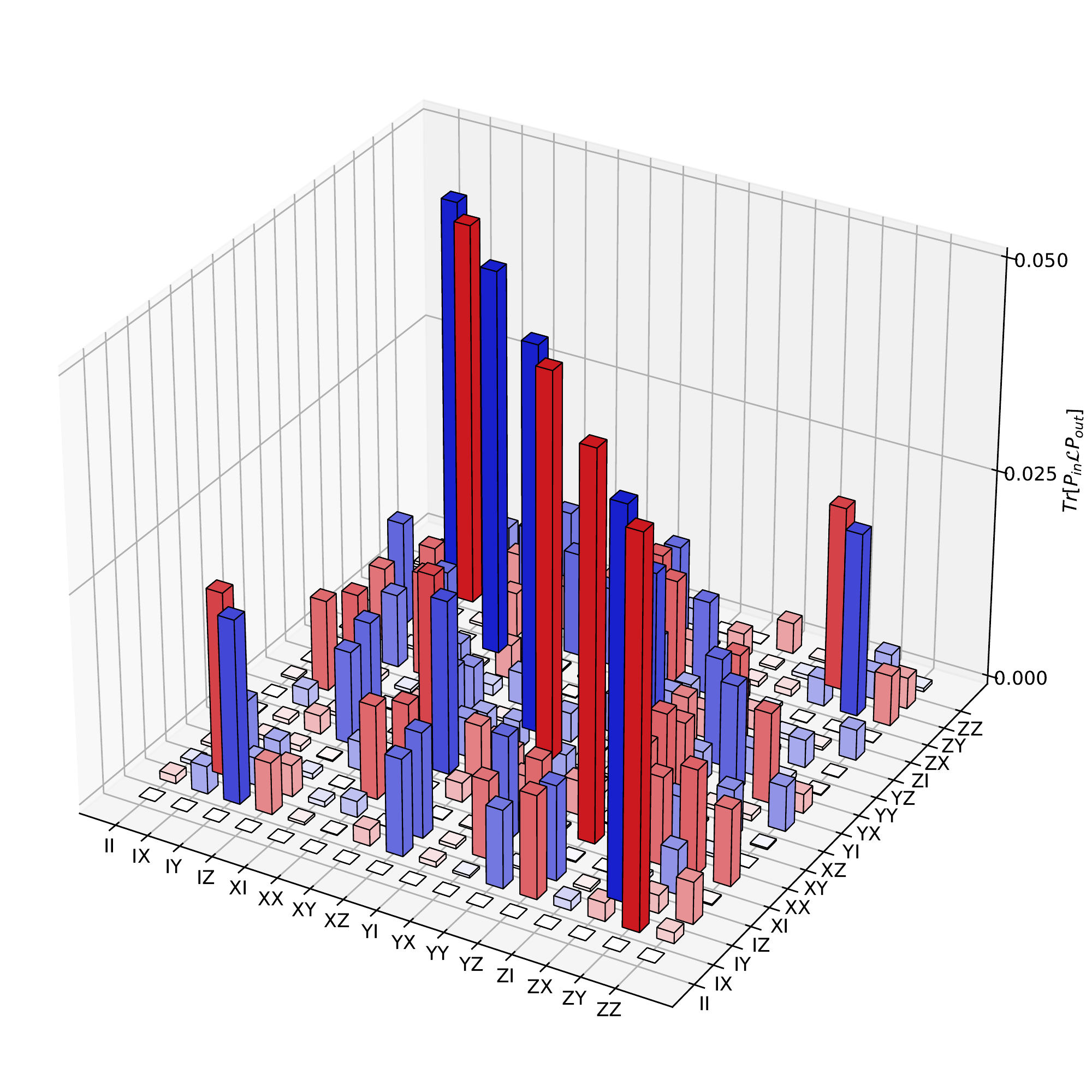} & \iptm{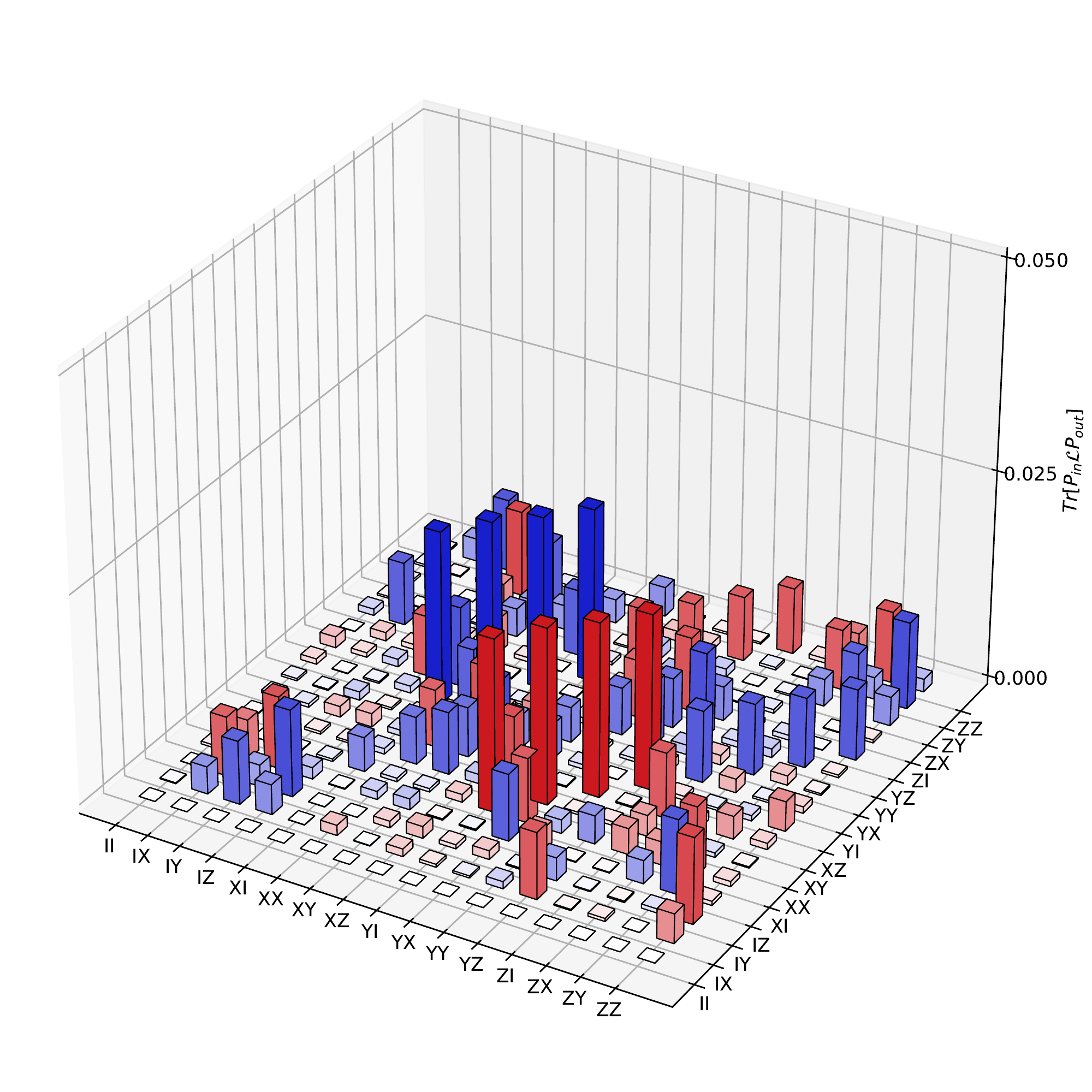}\\ \hline&&& \\
         Gate name & $Z_{1}(\frac{\pi}{2})$ & $Z_{2}(\frac{\pi}{2})$ & $Z_{3}(\frac{\pi}{2})$  \\&&& \\
         \begin{tabular}{c}Average\\infidelity\end{tabular}  & $0.000085(15)$ & $0.000218(70)$ & $0.000113(80)$ \\&&& \\
         \begin{tabular}{c}\\PTM\\\end{tabular} & \iptm{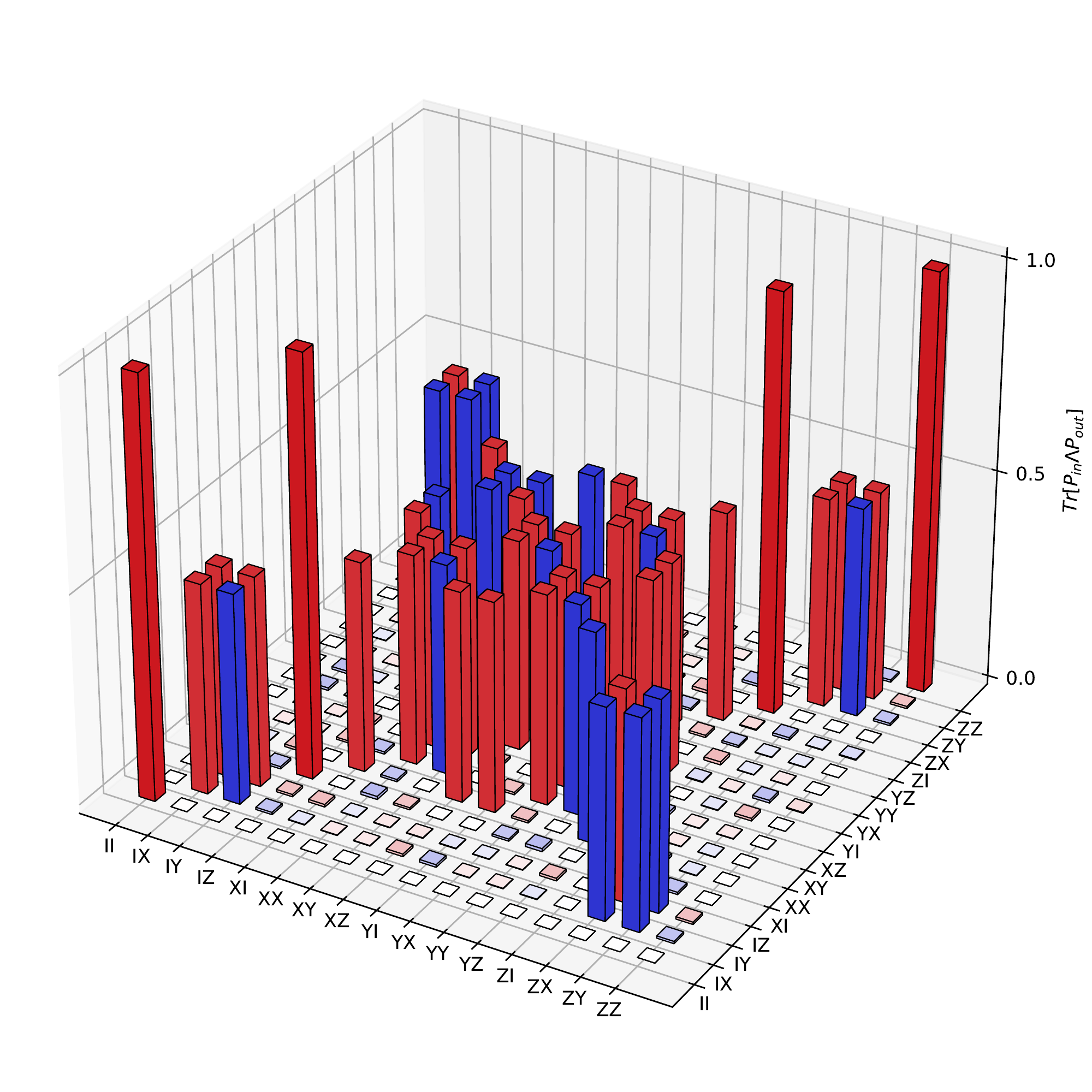} & \iptm{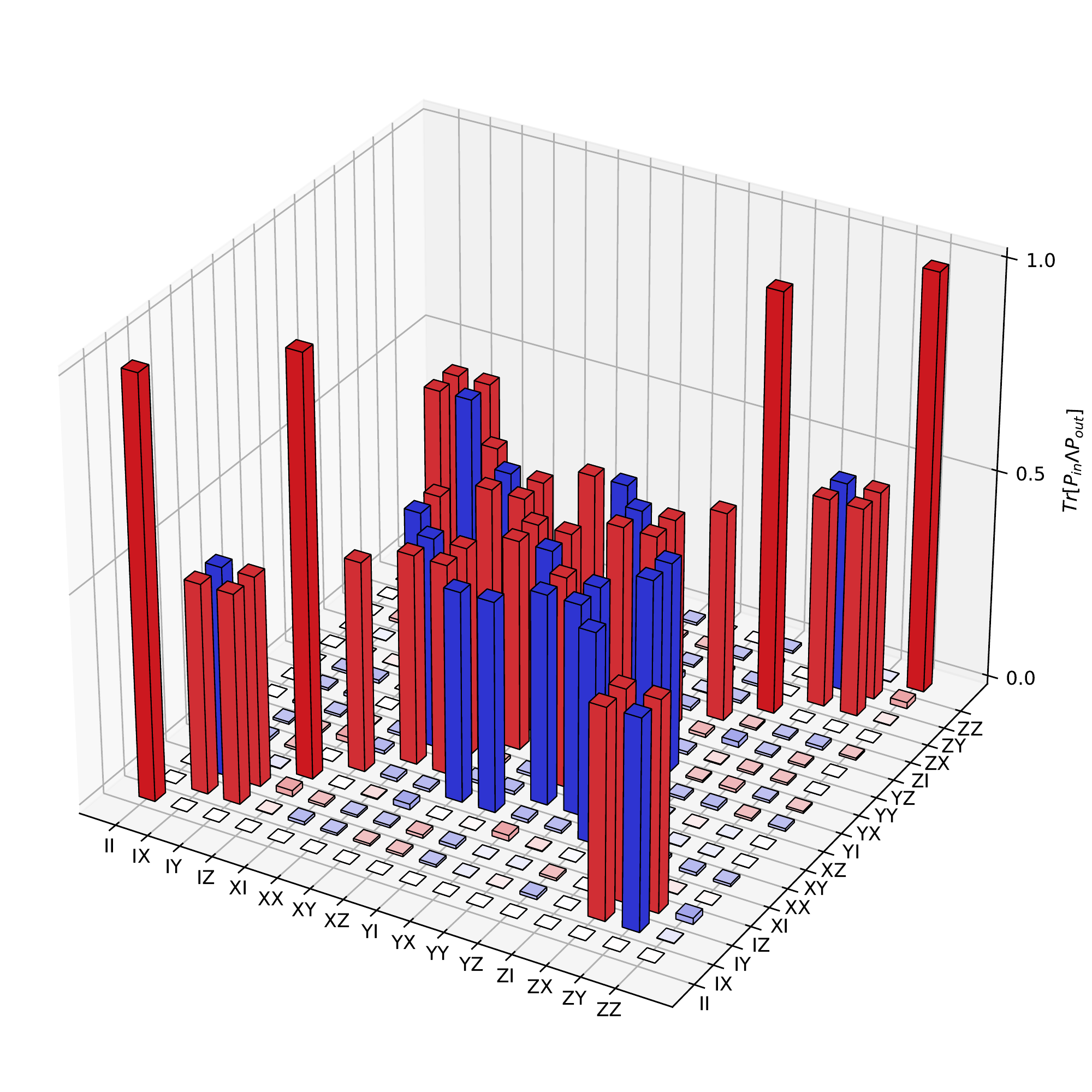} & \iptm{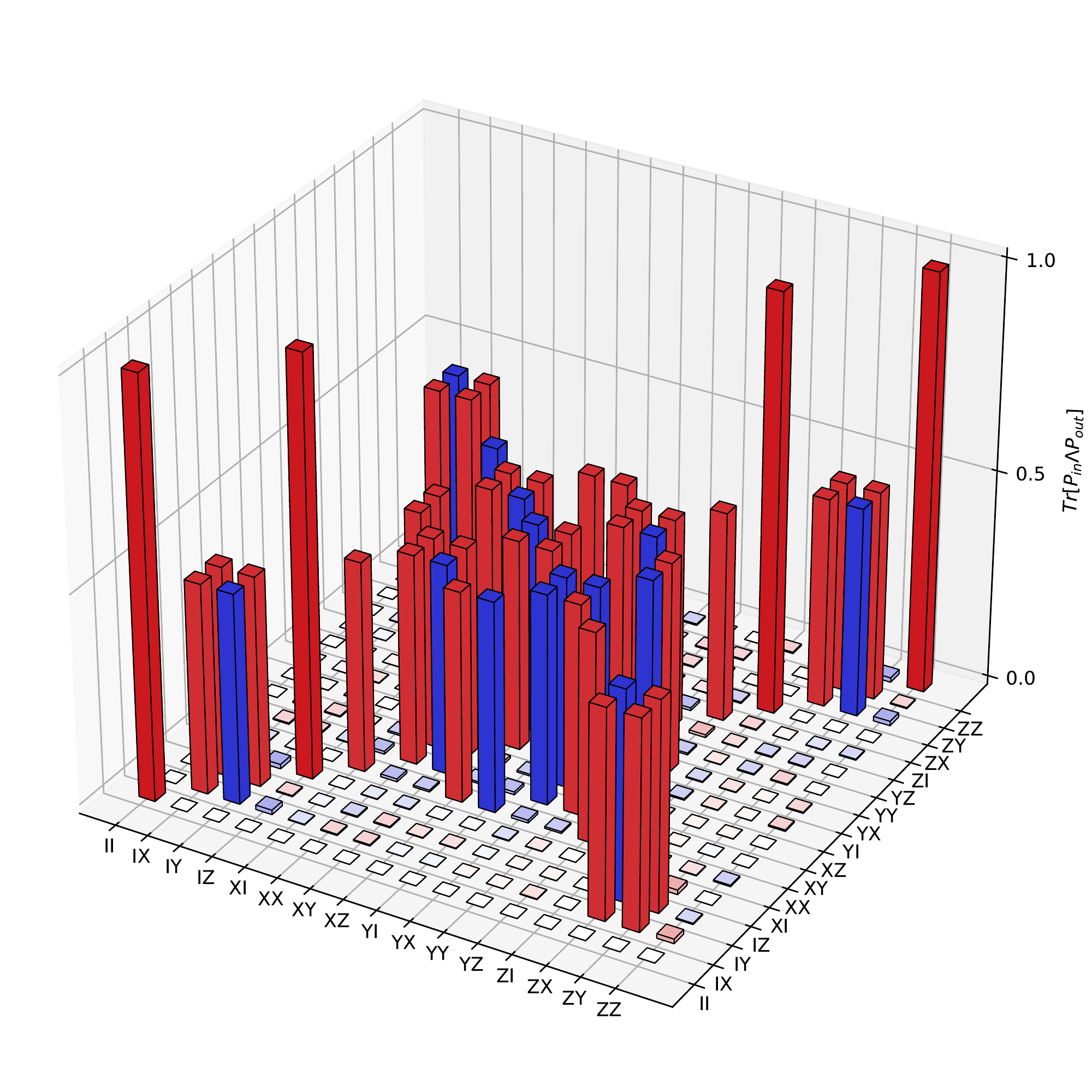} \\ &&&\\
         \begin{tabular}{c}Error\\Generator\end{tabular} & \iptm{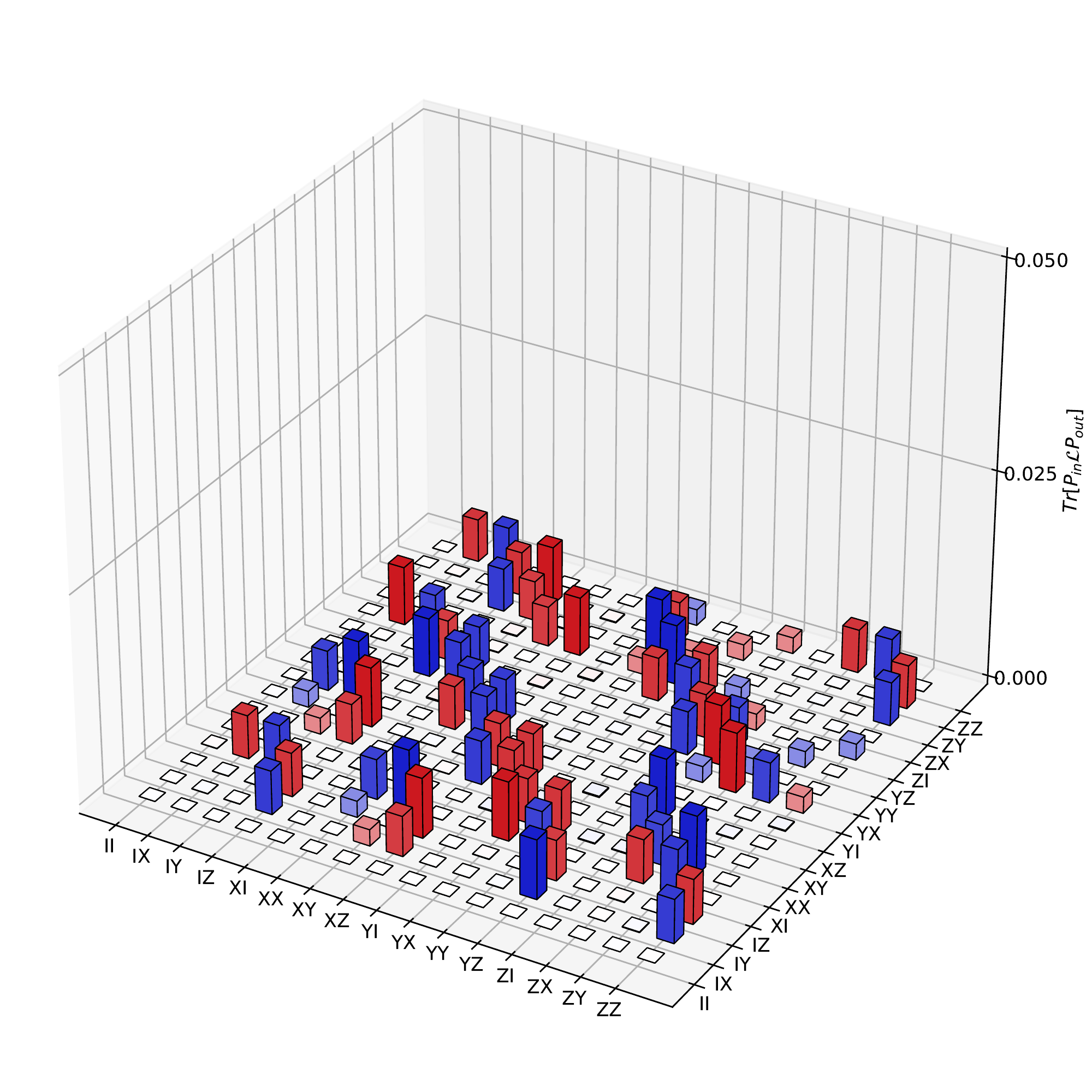} & \iptm{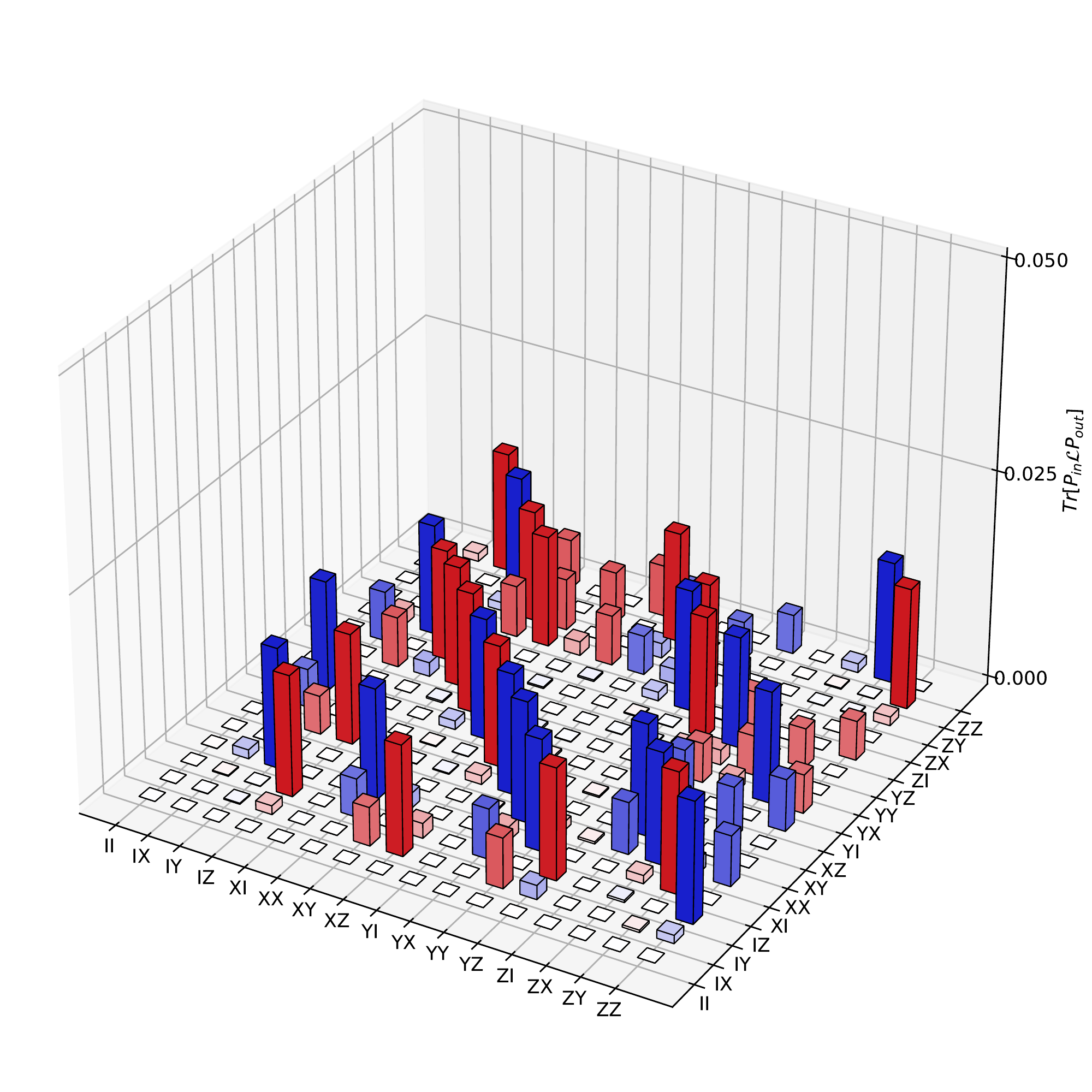} & \iptm{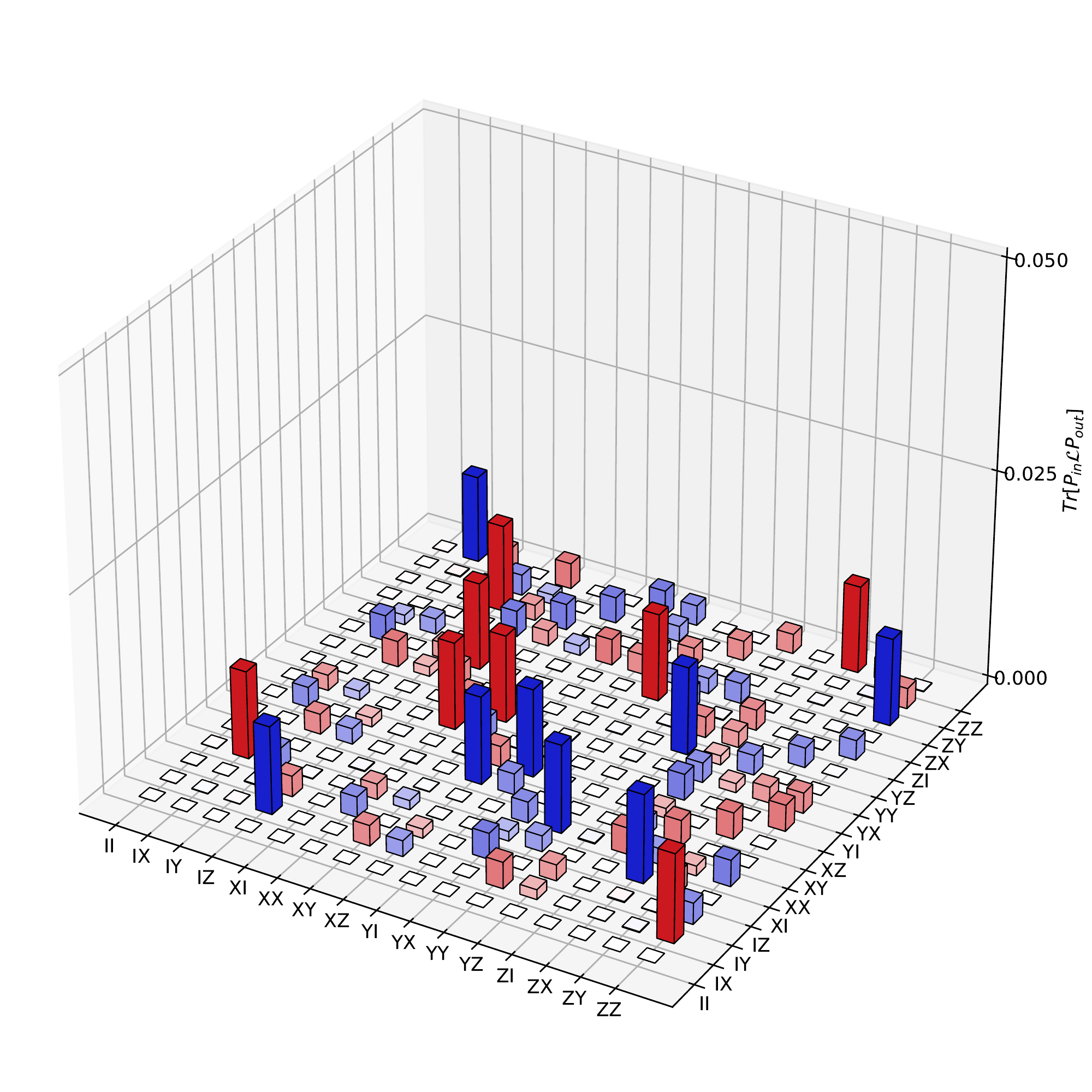} \\ 
         
    \end{tabular}
    \caption{Averaged gate infidelities, reconstructed process matrix and error generators of the gate set. The result is generated by pyGSTi maximum likelihood estimation.} 
    \label{tab:all_ptm_report}
\end{figure}

\section{VQE Ansatz\label{app:ansatz}}

The ansatz used in our approach is identical to that of \cite{OMalley2016ScalableEnergies}. However, in contrast to the use of CNOTs and Z gates in the original ansatz, we use $ZZ(\theta)$ gates to directly implement two-qubit interactions. The expanded ansatz is shown in Figure \ref{fig:ansatz_expanded}.

\begin{figure*}
    \centering    \includegraphics[width=\linewidth]{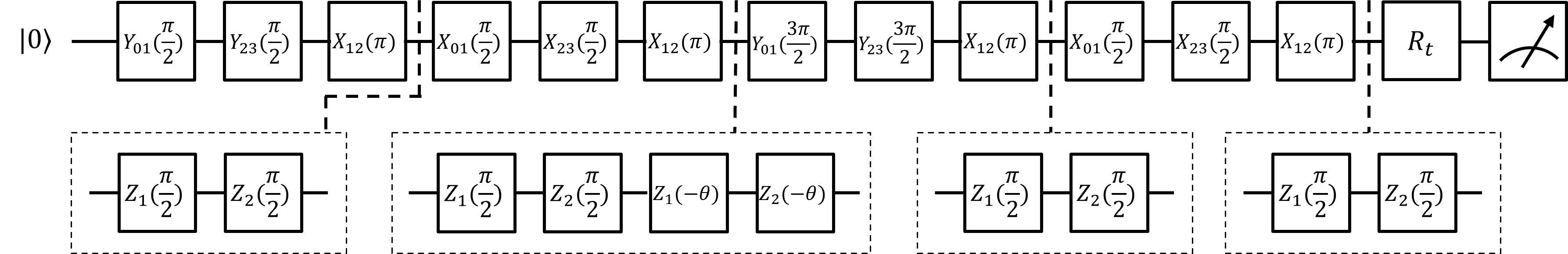}
    \caption{The ansatz presented in the native gate set. The first row in our implementation denotes all the physical gates that have been used, where $R_t$ represents the fiducials used for preparing the measurement basis. The dashed lines in the figure denote virtual Z gates that have been inserted into the circuit. The virtual Z gate sequences are described in the second row.}
    \label{fig:ansatz_expanded}
\end{figure*}

\section{``Shadow'' of the expectation valuez\label{app:shadow}}

The measured distribution of the expectation values is not a Gaussian distribution, resulting in a "shadow" in the heat map plot. To further illustrate this effect, we present a 2D histogram of the distribution where the shadow exists, shown in Figure \ref{fig:plot_slice}.

\begin{figure*}[!t] 
    \centering
	\sidesubfloat[]{
    \includegraphics[width=.3\linewidth]{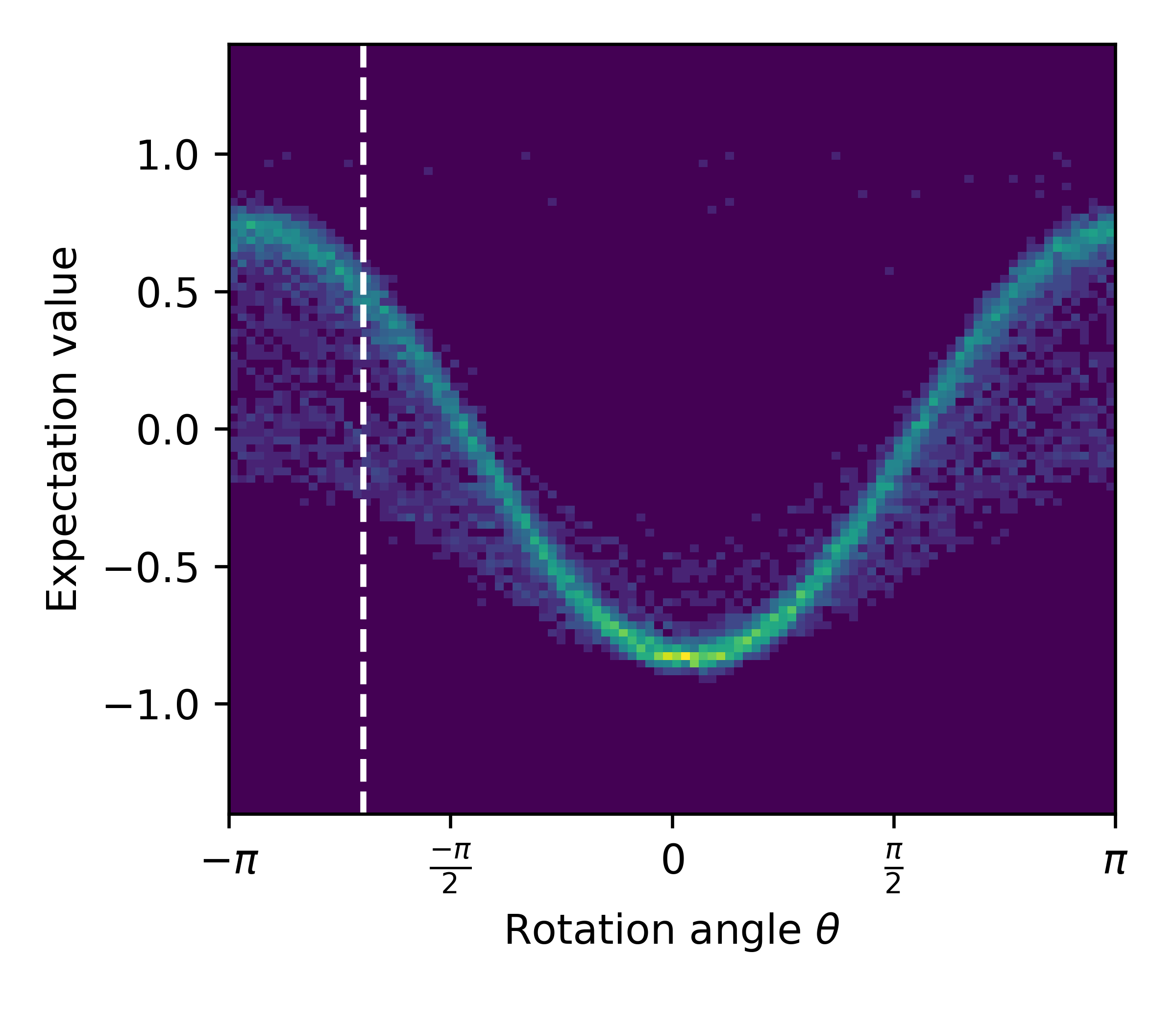}
	}
	\sidesubfloat[]{
        \includegraphics[width=.3\linewidth]{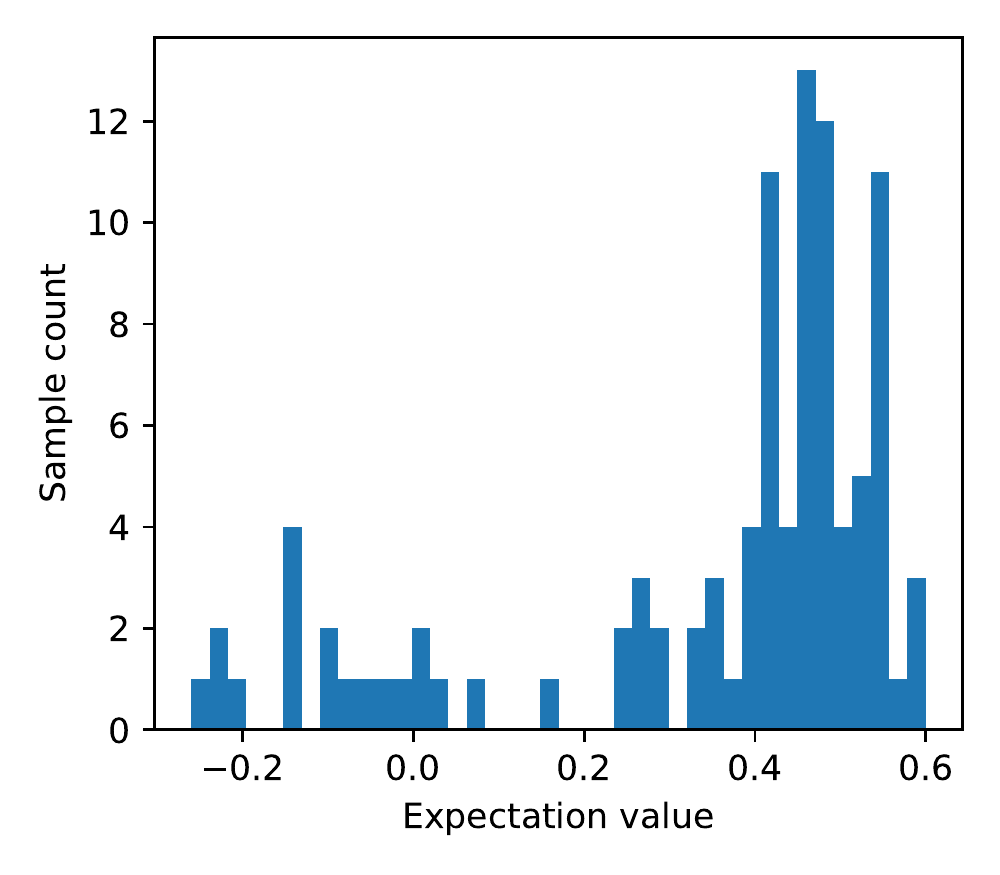}
	}\\
	\sidesubfloat[]{
        \includegraphics[width=.3\linewidth]{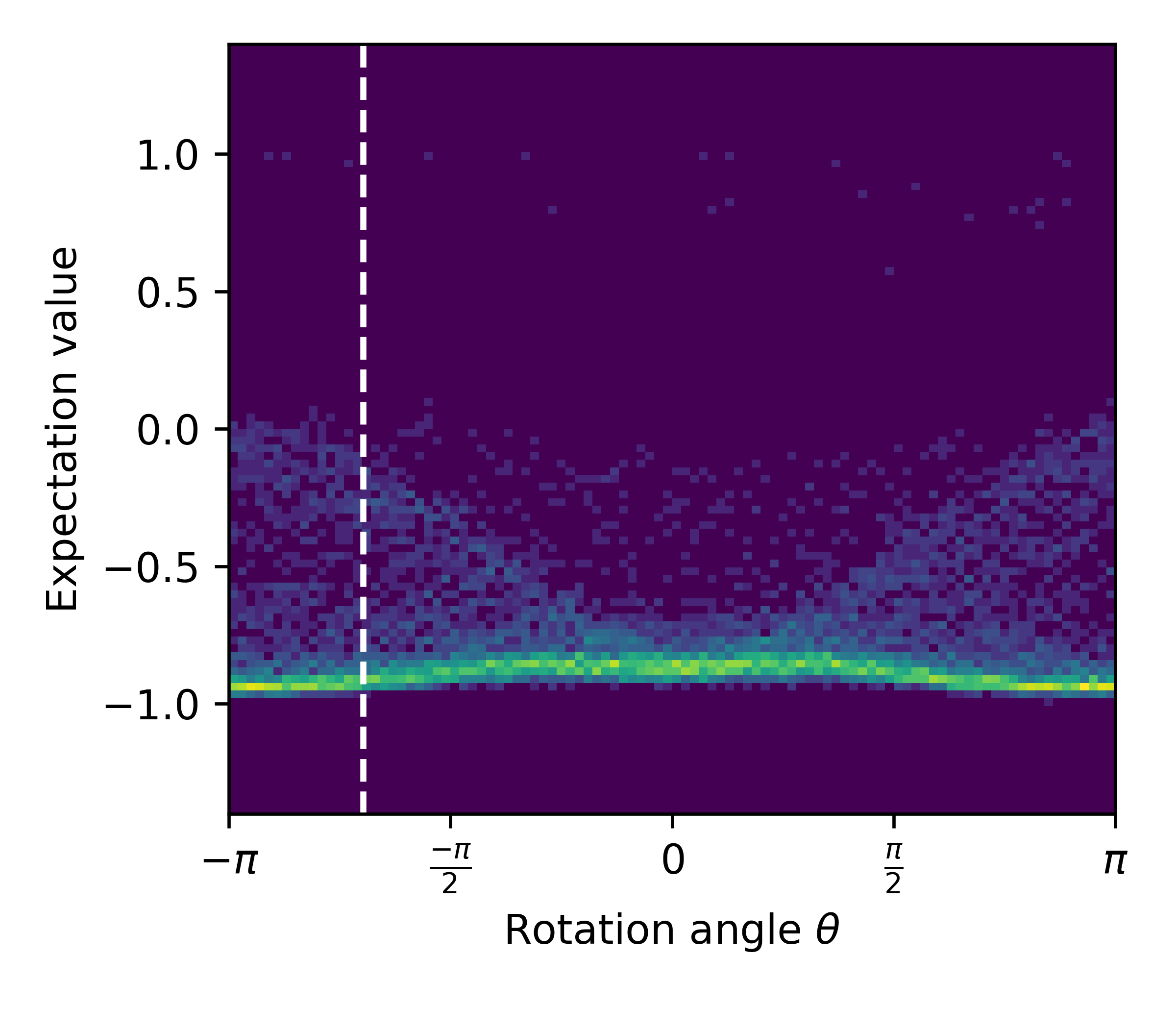}
	}
	\sidesubfloat[]{
        \includegraphics[width=.3\linewidth]{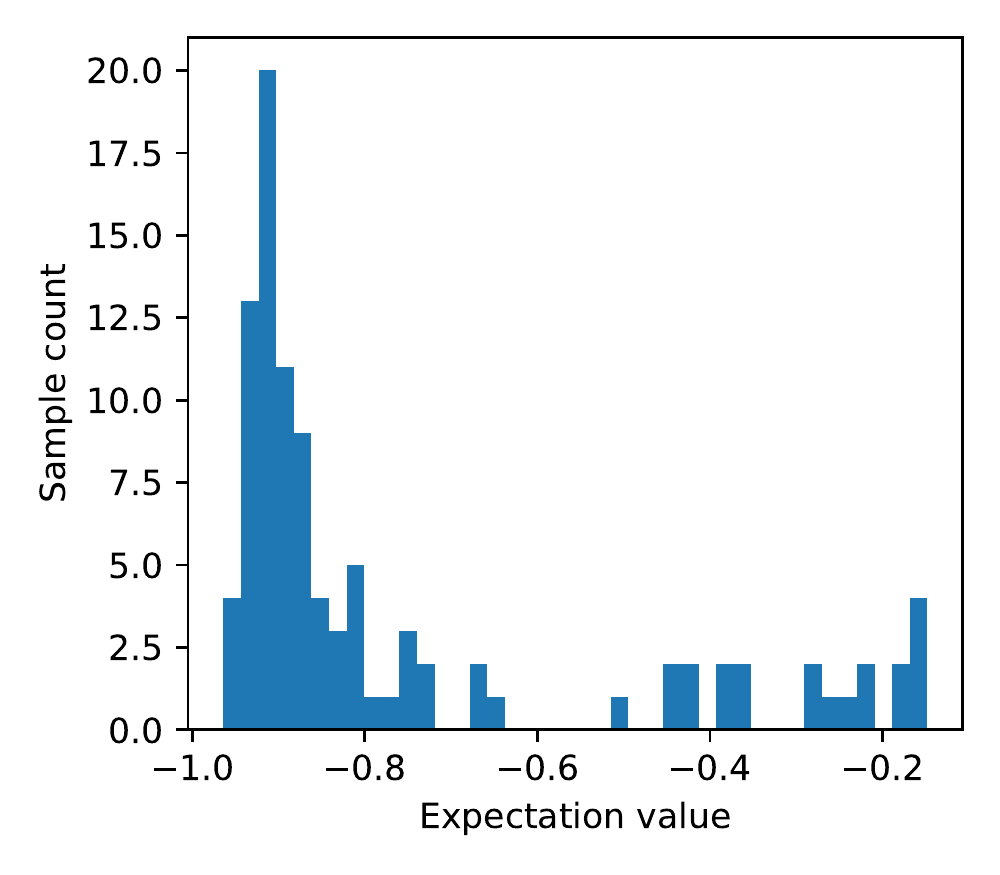}
	}
    \caption{(a) and (c) show heat maps of the distribution of the expectation values of the $IZ$ and $ZZ$ operators, respectively, plotted against the ansatz parameter $\theta$. Figures (b) and (d) display histograms of the distribution of expectation values along the white dashed line shown in figures (a) and (c).
}
    \label{fig:plot_slice}
\end{figure*}

\sx{\section{Spherical Gaussian Mixture model as readout classifier \label{app:spherical_gmm}}
The Gaussian Mixture Model is a type of machine learning algorithm that posits data points are derived from a combination of several Gaussian distributions. Its training involves determining parameters for each of these distributions. Specifically, each Gaussian component, denoted as \( \mathcal{N}(x | \mu_k, \Sigma_k) \), is characterized by its mean \( \mu_k \) and covariance \( \Sigma_k \). The Probability Distribution Function of a Gaussian component is expressed as:
\[
\mathcal{N}(x | \mu_k, \Sigma_k) = \frac{1}{\sqrt{(2\pi)^d |\Sigma_k|}} \exp\left(-\frac{1}{2}(x-\mu_k)^T\Sigma_k^{-1}(x-\mu_k)\right),
\]
where $d$ represents the dimensionality of each data point, and $k$ is the index of a distribution. Generally, the variance within each Gaussian component is characterized by its covariance matrix \( \Sigma_k \), which allows the distribution has correlations across different dimensions. In the context of qubit state discrimination, the measurement signals are aggregated and summed using FPGA hardware. It's assumed that the Gaussian distributions originate from signal noise, which is independent in both the $I$ and $Q$ channels and has the same variance. Therefore for this work, the covariance matrices \( \Sigma_k \) are selected to be diagonal and all the diagonal element are the same, ensuring the model to fit distributions that appear 'spherical' in the IQ plane. The implementation of the Gaussian mixture model is provided by the scikit-learn library \cite{pedregosa2011scikit}.}

\section{\sx{Experiment setup \label{app:exp_setup}}}

\begin{figure}
    \centering
    \includegraphics[width=\linewidth]{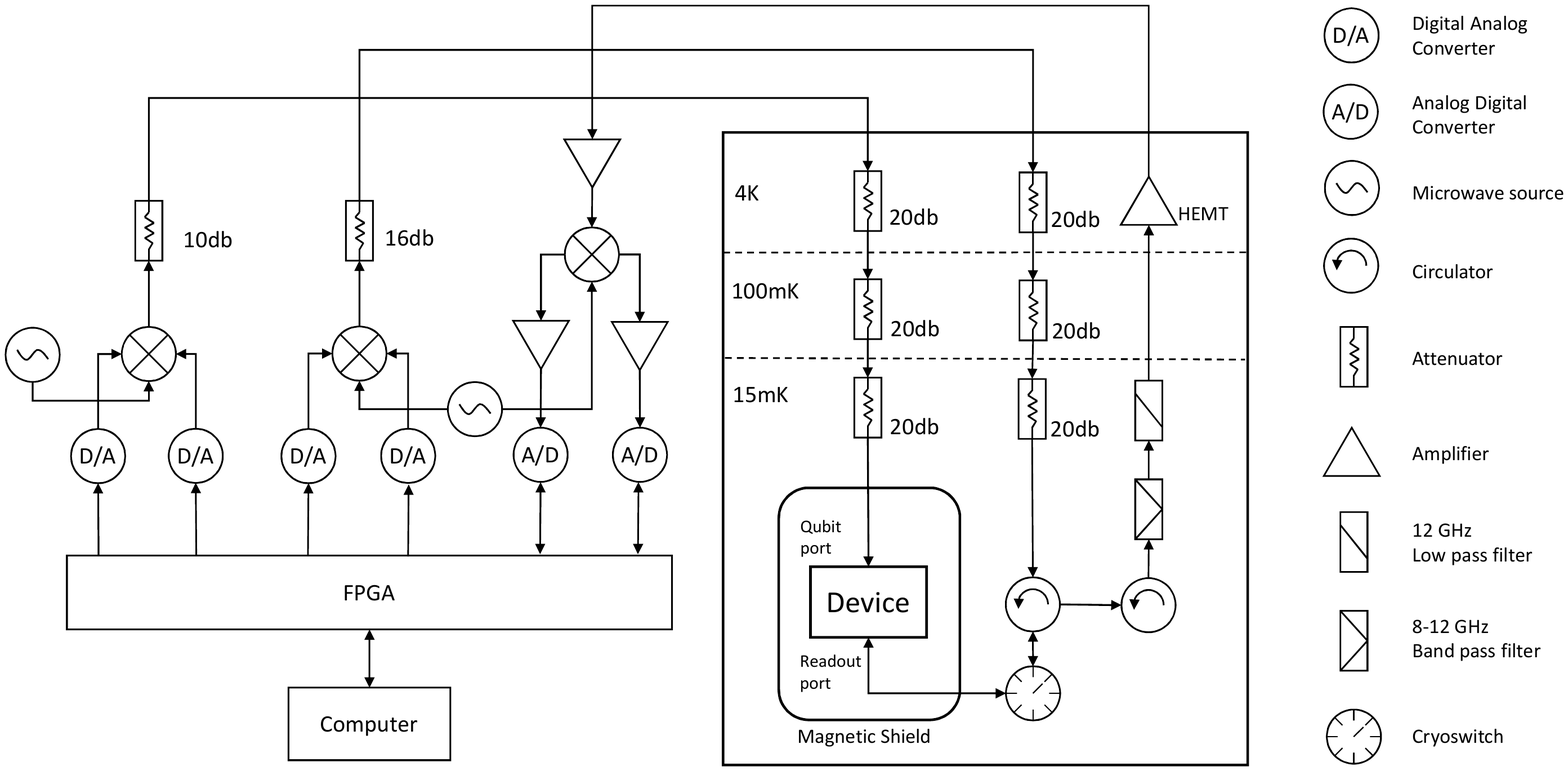}
    \caption{The wiring diagram for the experiment setup. The setup does not include a parametric amplifier on the readout output line. The cryoswitch is installed to connect to the readout port among multiple qubits on the same chip, which enables the use of a single readout line for characterizing different qubits.}
    \label{fig:exp_setup}
\end{figure}

\end{widetext}

\end{document}